\documentclass[11pt,a4paper]{article}

\usepackage{jheppub, hyperref}
\usepackage[capitalise]{cleveref} % automatic referencing, always capital
\usepackage{amsfonts, amsmath, amssymb, latexsym}
\usepackage{enumitem}
\usepackage{colortbl,bm,booktabs,multirow}
\usepackage{comment}
 
\usepackage{xspace}
\usepackage[normalem]{ulem}
\usepackage[dvipsnames]{xcolor}
\definecolor{alizarin}{rgb}{0.82, 0.1, 0.26}

\newcommand{\mpl}{M_\text{\tiny P}}
\renewcommand{\d}{{\rm d}}

\newcommand{\calP}{{\cal P}}

\newcommand{\SIGWAY}{\texttt{SIGWAY}\xspace}

\newcommand{\vk}{{\boldsymbol k}}
\newcommand{\vp}{{\boldsymbol p}}

\newcommand{\Pz}{\mathcal P_\zeta}
\newcommand{\Ph}{\mathcal P_h}
\newcommand{\tauNL}{\tau_{\text{NL}}}

\newcommand{\s}{\mathfrak s}
\newcommand{\tc}{\mathfrak t}
\newcommand{\uc}{\mathfrak u}

\definecolor{rosewood}{rgb}{0.4, 0.0, 0.04}

\definecolor{midori}{HTML}{008000}

\definecolor{TARDISBlue}{RGB}{0, 21, 129}

\newcommand{\Om}[1]{{\Omega_{\rm GW}} }

\preprint{CERN-TH-2024-217}

\title{
Reconstructing Primordial Curvature Perturbations via Scalar-Induced Gravitational Waves with LISA}

\author[a]{
Jonas {E}l~{G}ammal\footnote{Corresponding author: \href{mailto:jonas.elgammal@hotmail.com}{jonas.elgammal@hotmail.com}},}
\author[b]{Aya Ghaleb,}
\author[c]{Gabriele Franciolini\footnote{Project coordinator: \href{mailto:gabriele.franciolini@cern.ch}{gabriele.franciolini@cern.ch}},}
\author[def]{Theodoros Papanikolaou,}
\author[gh]{Marco Peloso,}
\author[gh]{Gabriele Perna\footnote{Corresponding author: \href{mailto:gabriele.perna@phd.unipd.it}{gabriele.perna@phd.unipd.it}},}
\author[c]{Mauro Pieroni,}
\author[ij]{Angelo Ricciardone,}
\author[k]{Robert Rosati\footnote{Project coordinator: \href{mailto:robert.j.rosati@nasa.gov}{robert.j.rosati@nasa.gov}, NASA Postdoctoral Program Fellow},}
\author[blm]{Gianmassimo Tasinato\vspace{.3cm},}
\author[]{\qquad \qquad \qquad \qquad \qquad \qquad \qquad \qquad \qquad}
\author[n]{Matteo Braglia,}
\author[o]{Jacopo Fumagalli,}
\author[ghr]{Jun'ya Kume,}
\author[pq]{Enrico Morgante,}
\author[a]{Germano Nardini,}
\author[st]{Davide Racco,}
\author[u]{S\'ebastien Renaux-Petel,}
\author[v]{Hardi Veerm\"ae,}
\author[u]{Denis Werth,}
\author[b]{and Ivonne Zavala}

\author[]{\\ \vspace{.2cm}
\centering \texttt{(For the LISA Cosmology Working Group)}
\vspace{.2cm}
}

\affiliation[a]{Department of Mathematics and Physics, University of Stavanger, NO-4036 Stavanger, Norway}
\affiliation[b]{Department of Physics, Faculty of Science and Engineering, Swansea University, Singleton Park, SA2 8PP, Swansea, United Kingdom}
\affiliation[c]{CERN, Theoretical Physics Department,
Esplanade des Particules 1, Geneva 1211, Switzerland}
\affiliation[d]{Scuola Superiore Meridionale, Largo San Marcellino 10, 80138 Napoli, Italy}
\affiliation[e]{Istituto Nazionale di Fisica Nucleare (INFN), Sezione di Napoli, Via Cinthia 21, 80126 Napoli, Italy}
\affiliation[f]{National Observatory of Athens, Lofos Nymfon, 11852 Athens, 
Greece}
\affiliation[g]{Dipartimento di Fisica e Astronomia ``G. Galilei'', Universit\`a degli Studi di Padova, via Marzolo 8, I-35131 Padova, Italy}
\affiliation[h]{INFN, Sezione di Padova, via Marzolo 8, I-35131 Padova, Italy}
\affiliation[i]{Dipartimento di Fisica “Enrico Fermi”, Università di Pisa, Largo Bruno Pontecorvo 3, Pisa I-56127, Italy}
\affiliation[j]{INFN, Sezione di Pisa, Largo Bruno Pontecorvo 3, Pisa I-56127, Italy}
\affiliation[k]{NASA Marshall Space Flight Center, Huntsville, AL 35812, USA}
\affiliation[l]{Dipartimento di Fisica e Astronomia, Universit\`a di Bologna}
\affiliation[m]{INFN, Sezione di Bologna, I.S. FLAG, viale B. Pichat 6/2, 40127 Bologna,   Italy}
\affiliation[n]{Center for Cosmology and Particle Physics, New York University, 726 Broadway, New York, NY 10003, USA}
\affiliation[o]{Departement de F\'isica Qu\`antica i Astrofisica and Institut de Ci\`encies del Cosmos (ICC), Universitat de Barcelona, Mart\'i i Franqu\`es 1, 08028 Barcelona, Spain}
\affiliation[p]{Dipartimento di Fisica, Università di Trieste, Strada Costiera 11, I-34151 Trieste, Italy}
\affiliation[q]{INFN, Sezione di Trieste, Via Valerio 2, 34127 Trieste, Italy}
\affiliation[r]{Research Center for the Early Universe (RESCEU), Graduate School of Science, The University of Tokyo, Hongo 7-3-1
Bunkyo-ku, Tokyo 113-0033, Japan}
\affiliation[s]{Institut f\"ur Theoretische Physik, ETH Z\"urich,Wolfgang-Pauli-Str.\ 27, 8093 Z\"urich, Switzerland}
\affiliation[t]{Physik-Institut, Universit\"at Z\"urich, Winterthurerstrasse 190, 8057 Z\"urich, Switzerland}
\affiliation[u]{Institut d’Astrophysique de Paris, UMR 7095 du CNRS et de Sorbonne Universit\'e, 98 bis bd Arago, 75014 Paris, France}
\affiliation[v]{Keemilise ja bioloogilise f\"u\"usika instituut, R\"avala pst. 10, 10143 Tallinn, Estonia}

\abstract{
Many early universe scenarios predict an enhancement of scalar perturbations at scales currently unconstrained by cosmological probes. These perturbations source gravitational waves (GWs) at second order in perturbation theory, leading to a scalar-induced gravitational wave (SIGW) background. The LISA detector, sensitive to mHz GWs, will be able to constrain curvature perturbations in a new window corresponding to scales $k \in [10^{10}, 10^{14}] \,{\rm Mpc}^{-1}$, difficult to probe otherwise.
In this work, we forecast the capabilities of LISA to constrain the source of SIGWs using different approaches: {\it i)} agnostic, where the spectrum of curvature perturbations is binned in frequency space; {\it ii)} template-based, modeling the curvature power spectrum based on motivated classes of models;
{\it iii)} ab initio, starting from first-principles model of inflation featuring an ultra-slow roll phase.
We compare the strengths and weaknesses of each approach. We also discuss the impact on the SIGW spectrum of non-standard thermal histories affecting the kernels of SIGW emission and non-Gaussianity in the statistics of the curvature perturbations. Finally, we propose simple tests to assess whether the signal is compatible with the SIGW hypothesis.
The pipeline used is built into the
\href{https://github.com/jonaselgammal/SIGWAY}{\SIGWAY} code. 
}

\begin{document}
\begin{figure}
\begin{flushright}
{\includegraphics[width = 0.2 \textwidth]{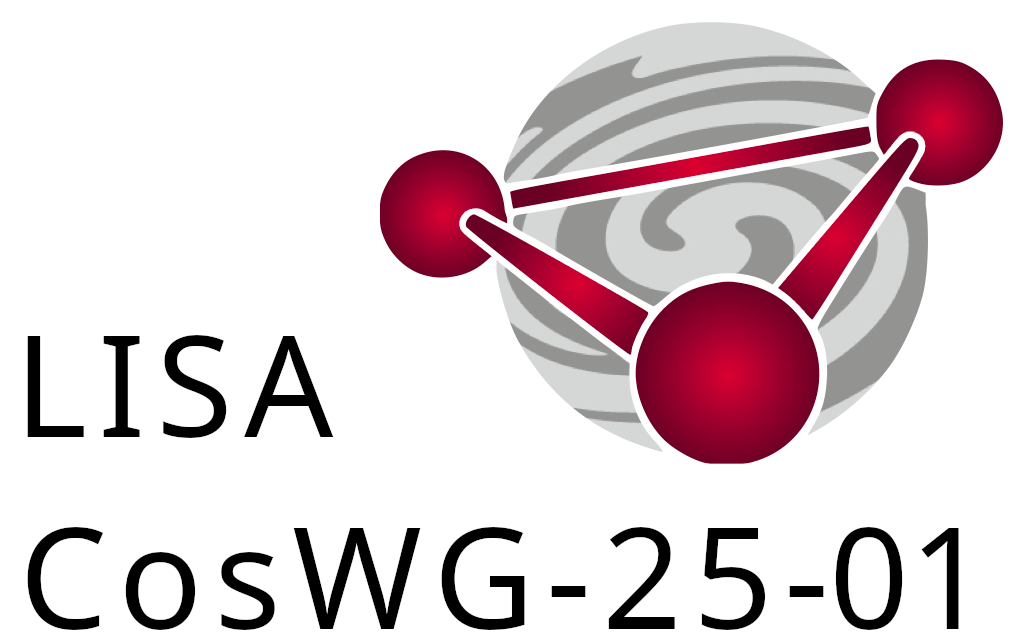}}\\[5mm]
\end{flushright}
\end{figure}

\maketitle

\section{Introduction}
\label{sec:intro}

The Laser Interferometer Space Antenna (LISA)~\cite{LISA:2017pwj} represents a groundbreaking gravitational wave (GW) observatory aimed to probe and impact our understanding of fundamental physics, astronomy, and cosmology~\cite{LISACosmologyWorkingGroup:2022jok, LISA:2022kgy, LISA:2022yao}. With the first-ever direct probe of the stochastic gravitational-wave background (SGWB) in the millihertz frequency range, LISA provides the opportunity to unveil processes that occurred in the first stages of the Universe, including inflation~\cite{Bartolo:2016ami, LISACosmologyWorkingGroup:2024hsc}. 
Probing a primordial SGWB at millihertz frequencies corresponds to exploring comoving scales that lie between those accessible to ground-based GW interferometers and those probed by pulsar-timing arrays, cosmic microwave background (CMB), or large-scale structure surveys.
These scales correspond to comoving wavenumbers in the range \( k \sim  [10^{10}, 10^{14}] \, \mathrm{Mpc}^{-1} \).
Several cosmological models (see e.g.~refs.~\cite{Maggiore:2007ulw, Guzzetti:2016mkm, Cai:2017cbj, Caprini:2018mtu} for reviews) predict a measurable SGWB at these scales, often without any other complementary distinctive signature, placing LISA in a unique position to test these scenarios. 

Inflationary models exhibiting amplified scalar fluctuations are one of the candidates for sourcing an SGWB in the LISA frequency band. 
Enhanced scalar fluctuations generate scalar-induced gravitational waves (SIGWs)
 at second order in perturbations~\cite{Tomita:1975kj, Matarrese:1992rp,Matarrese:1993zf,Matarrese:1997ay, Acquaviva:2002ud, Mollerach:2003nq,Carbone:2004iv, Ananda:2006af, Baumann:2007zm, Domenech:2021ztg}, resulting in a potentially large SGWB.
Amplified scalar fluctuations naturally arise in single-field inflationary scenarios 
with features in the potential like those leading to ultra-slow-roll (USR) phases, multi-field setups, and mechanisms such as preheating or early matter-dominated eras. Intriguingly, the same perturbations that seed the SIGWs can also trigger the formation of primordial black holes (PBHs)~\cite{Zeldovich:1967lct,Hawking:1971ei,Carr:1974nx,Carr:1975qj,Chapline:1975ojl}. SIGWs in the millihertz frequency band arise in correspondence with the asteroidal-mass window for PBHs, a viable candidate for addressing the dark matter puzzle~\cite{Bartolo:2018evs,Bartolo:2018rku, Yuan:2021qgz,LISACosmologyWorkingGroup:2023njw,Balaji:2022rsy}. 
By detecting or setting upper bounds on SIGWs, LISA would not only shed light on the inflationary epoch but also on dark matter and non-astrophysical black hole formation channels. This makes SIGWs a high-gain, well-motivated target for LISA. 

SIGWs are also powerful tools to investigate non-Gaussianity (NG) in the early universe since their production is highly sensitive to the statistical properties of scalar curvature fluctuations~\cite{Gangui:1993tt, Matarrese:2000iz, Bartolo:2001cw, Maldacena:2002vr, Bartolo:2004if,Chen:2010xka,Byrnes:2010em,Wands:2010af,Renaux-Petel:2015bja,Achucarro:2022qrl}. NG is typically characterized by parameters such as $f_{\rm NL}$, which quantifies NG at the bispectrum level (three-point correlation function), and $\tau_{\rm NL}$, which appears in the trispectrum (four-point correlation function). Earlier analysis have focused on contributions to the SIGW involving $f_{\rm{NL}}$,~\cite{Nakama:2016gzw, Garcia-Bellido:2017aan, Unal:2018yaa, Cai:2018dig, Cai:2019amo, Ragavendra:2020sop, Yuan:2020iwf, Adshead:2021hnm, Davies:2021loj, Abe:2022xur, Garcia-Saenz:2022tzu, Garcia-Saenz:2023zue}. More recent studies have extended this analysis to higher-order NG terms~\cite{Yuan:2020iwf, Li:2023xtl,Yuan:2023ofl, Perna:2024ehx}. Recently, a Fisher forecast analysis for LISA about NG has been performed in~\cite{Perna:2024ehx}.
The tensor power spectrum of SIGWs is directly related to the four-point correlation function of the curvature fluctuations. Such a correlation function has both connected and disconnected contributions. While the latter contributes only to the Gaussian SIGW power spectrum, the former is directly linked to the trispectrum through the $\tau_{\rm NL}$ parameter~\cite{Garcia-Saenz:2022tzu}, which is the key observable to constrain NG from SIGWs.

The detection and characterization of the primordial SGWB is one of the most challenging objectives of the LISA mission~\cite{Colpi:2024xhw}. LISA is a signal-dominated detector, where a multitude of transient or quasi-monochromatic events overlap in time and frequency with the stochastic superposition of all unresolved astrophysical events and, potentially also with a significant primordial SGWB. Additionally, the stationary component of the instrumental noise can mimic a SGWB to some extent. Completing the LISA science program for the SGWB therefore requires: 
\begin{enumerate}[label=\textit{\roman*})]
    \item Determining whether a primordial SGWB is present in the data.
    \item Reconstructing the SGWB frequency shape and, if possible, its statistical properties.
    \item Setting upper limits on cosmological sources of SGWB not supported by the data.
    \item Constraining the parameters of the most likely SGWB source candidates.
\end{enumerate}
ESA and NASA plan to address these tasks through the so-called ``global fit", a data analysis procedure where modules fitting each class of sources (galactic binaries, supermassive black hole binaries, SGWB, etc.) iterate until convergence~\cite{Colpi:2024xhw}. Recently, successful prototype global fit analyses became available in the literature \cite{Littenberg:2023xpl,Katz:2024oqg,Strub:2024kbe}, tested on the \textit{Sangria} LISA Data Challenge (LDC) dataset \cite{le_jeune_2022_7132178}, which contains no primordial SGWB. It is still an open question how the global fit should support a primordial SGWB search,  and how the SGWB properties should be represented in the detection catalogs that the space agencies will publish. A recent study \cite{Rosati:2024lcs} has attempted to perform an SGWB search directly on the global fit residual. 

In this work, we aim to bridge these gaps by providing elements of the global fit SGWB module useful for the tasks \emph{i)} - \emph{iv)}  in the presence of a SGWB due to SIGWs. To develop and test our rationale, we work in the limit that all resolvable events have been precisely reconstructed\footnote{Although this optimistic working hypothesis may seem unrealistic, it is the correct one to use in a global fit module. All current implementations of the global fit are based on a blocked Gibbs sampling scheme, where each source type is sampled independently, assuming a perfect subtraction of the other source types. By periodically alternating which source type is being sampled over, imperfect source subtraction and source type confusion are properly included and fully modeled in the resulting posterior.
When working after the global fit, with one sample of the residual as we assume in this work, these possible degeneracies are not fully modeled. Additionally, if the global fit has experienced a convergence failure (the MCMC is still ``burning in''), unmodeled source power may still be in the data and lead to false detections of an SGWB. Ref. \cite{Rosati:2024lcs} studies how this convergence failure affects stochastic background recovery in the available prototype global fit residuals.}, leaving us with data containing the stationary component of the noise, the SIGW background, and the foregrounds from the unresolved galactic and extragalactic binaries.

As we focus on SIGW sources, we perform an analysis starting from the properties of the source curvature power spectra $\mathcal{P}_\zeta(k)$ of the source, instead of the GW energy density $\Omega_{\rm GW}(f)$ generated by these power spectra.

Concerning \emph{i)} and \emph{ii)}, we prototype a model-agnostic method that reconstructs the power spectrum $\mathcal{P}_\zeta(k)$ by binning it in frequency space. This approach allows for maximal flexibility in capturing unknown SIGW features. It is however not as agnostic as other generic SGWB searches~\cite{Karnesis:2019mph, Caprini:2019pxz, Flauger:2020qyi, Pieroni:2020rob, Baghi:2023qnq, Pozzoli:2023lgz, Dimitriou:2023knw} since it requires, by construction, a SIGW source, i.e.~a SGWB that can be derived as the proper convolution of a generic $\mathcal{P}_\zeta(f)$.  Due to this additional information, the method is expected to be more sensitive to SIGW signals than other fully-agnostic approaches. It can be particularly useful for placing upper bounds on the SIGW amplitude if no signal is detected in the LISA data, or act as a key ingredient for SIGW model selection if a signal is present. 

On the other hand, if the computational resources available to the mission allow running the global fit for every SGWB template,  several modules for the template-based SIGW reconstruction have been conceived since the first iterations of the global fit.\footnote{See ref.~\cite{LISACosmologyWorkingGroup:2024hsc} for other template-based reconstructions suitable for inflationary models.} The advantage of such a possibility is clear: the more signal characterization is included in the search, the higher the sensitivity to that signal. This process also reduces the risk of SGWB misreconstructions that the global fit might absorb into the parameter estimation of other sources. To address the points \emph{i)} and \emph{ii)} above within this framework, we collect several well-motivated inflationary models with known  $\mathcal{P}_\zeta(k)$ predictions, we design template classes that effectively parameterize these $\mathcal{P}_\zeta(k)$, and we prototype the SIGW template-based searches for them.

Accurately performing \emph{i)} and \emph{ii)} enables LISA to identify the most favored SIGW models and then proceed with tasks \emph{iii)} and \emph{iv)}. Accordingly, we implement a prototype data analysis pipeline, choosing the USR inflationary setup as a representative example. 
In particular, we develop a fast numerical algorithm determining $\mathcal{P}_\zeta(k)$ once the inflationary model parameters are known. Thanks to its speed, the algorithm allows for rapid likelihood evaluations in the fundamental-parameter space, enabling direct inference on the USR model parameters from GW data.

As a proof of concept, we further perform inference on ${\cal P}_\zeta(k)$ in cosmological scenarios where standard assumptions on Gaussianity of curvature perturbations and on the standard thermal history of the Universe are relaxed. We evaluate the SIGW energy density sourced by non-Gaussian contributions parametrized by $\tauNL$, resorting to the local ansatz emerging from a perturbative expansion of scalar fluctuations. This contribution is known to modify its frequency profile compared to the Gaussian counterpart~\cite{Cai:2018dig, Unal:2018yaa, Adshead:2021hnm, Garcia-Saenz:2022tzu, Yuan:2023ofl, Perna:2024ehx}. Since
for models of inflation with local type NG, where the curvature perturbation is dominated by one degree of freedom, $\tau_{\rm NL}$ and $f_{\rm NL}$ are related, we take advantage of such a relation to forecast the ability of LISA of probing NG, focusing the analysis on $f_{\rm NL}$ and discussing its implications for $\tau_{\rm NL}$. 
It is worth mentioning that, as recently argued by \cite{Iovino:2024sgs}, the effects of NG on the SIGW background may not always be accurately captured by an expansion around a Gaussian field. Properly accounting for the full impact of intrinsic non-linearities may significantly suppress or enhance the spectrum compared to the predictions based on the local ansatz. Achieving this would require the development of fully non-perturbative approaches to compute the SIGW spectrum, which are beyond the scope of this work.
Finally, we perform some diagnostic tests to assess whether the reconstructed signal is consistent with the SIGW hypothesis. Such tests could help to rule out a scalar-induced origin as a viable explanation for some SGWB spectral shapes.

\begin{figure}[t]
\centering
\includegraphics[width=1\textwidth]{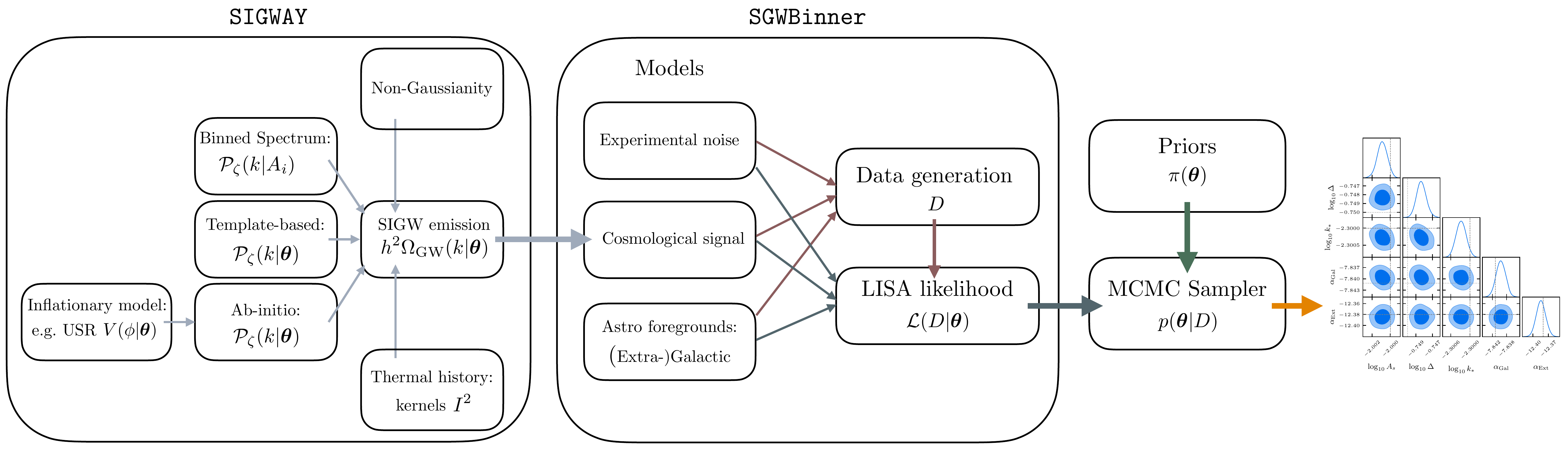}
\caption{ Flowchart describing the inputs and modules of the\SIGWAY\ code developed for this project, alongside the \texttt{SGWBinner} adopted to make forecasts for LISA. 
}
\label{fig:flowc}
\end{figure}

The core of our numerical analysis is implemented in the \SIGWAY\  code.\footnote{\href{https://github.com/jonaselgammal/SIGWAY}{https://github.com/jonaselgammal/SIGWAY}} 
This stand-alone \texttt{Python} code addresses tasks \textit{i) - iv)} for SIGW signals by offering the following functionalities:
\begin{itemize}
    \item A fast vectorized numerical integrator for computing the SGWB resulting from any spectrum of primordial curvature fluctuations of modes reentering the Hubble radius during radiation domination or a phase of early matter domination.
    \item An integration algorithm for computing the SGWB assuming a binned spectrum of $\mathcal{P}_\zeta(k)$ for agnostic reconstructions of $\mathcal{P}_\zeta$.
    \item Solvers for the background- and perturbation equations of motion for the inflaton in a single-field scenario that can be called by the SIGW integrator starting from the inflaton Lagrangian.
    \item Capabilities for computing the SGWB including non-Gaussian contributions for a lognormal shape of $\mathcal{P}_\zeta$.
    \item Functionality for pairing to the \texttt{SGWBinner} pipeline \cite{Caprini:2019pxz, Flauger:2020qyi} for computing the LISA likelihood, and performing inference on the parameters governing the primordial curvature fluctuations.
\end{itemize}
In Fig.~\ref{fig:flowc} we report a flowchart that schematically describes the inputs and outputs of the \SIGWAY\ code developed for this project, alongside the \texttt{SGWBinner} adopted to make forecasts for LISA. 

The paper is organized as follows. In Sec.~\ref{sec:models} we review some representative models predicting enhanced power spectra of curvature perturbations. In Sec.~\ref{sec:modellingPzeta} we identify functional forms that describe the shapes of the aforementioned spectra in terms of effective parameters. In Sec.~\ref{sec:computation} we describe the analytic and numerical tools that we implement in the \SIGWAY\  code. The functionality of the \texttt{SGWBinner} code that is relevant for performing inference is briefly described in Sec.~\ref{sec:tempbinnerpipe}. Sec.~\ref{sec:results} illustrates, for representative benchmark signals, how the elements built in this work help tackle the key tasks  \emph{i)} -- \emph{iv)}, 
including testing whether the SIGW hypothesis is compatible with the putative signal in Sec.~\ref{sec:tests}. Finally, Sec.~\ref{sec:conclusions} presents our main conclusions, while App.~\ref{app:SIGWBinner} and App.~\ref{App:largeNbinning} discuss technicalities and subtleties regarding the proposed SIGW signal reconstruction and interpretation.
\paragraph{Notation.}
We indicate with $k$ the comoving wavenumber 
while with $f$ the associated frequency $f = k / 2 \pi$.
For scales relevant to LISA, we translate wavenumbers in units of Hz using $c = {\rm Hz\, Mpc} /(1.023 \times 10^{14})$. 
For presentation purposes, we differentiate between frequencies and momenta arbitrarily denoting them with units of Hz and $\mathrm{s}^{-1}$, respectively.
As usually done in the literature, we report the GW spectral energy density $\Omega_{\rm GW}$ multiplied by the rescaled Hubble rate $h=H_0 /(100\, \mathrm{km/s/Mpc})$ squared.
Finally, we indicate vectors with bold symbols (e.g.~$\vec x \equiv {\bm x}$) while their magnitude with the lower-case letter.

\section{Early universe models leading to enhanced curvature spectra}\label{sec:models}

In this section, we summarise the main classes of models predicting enhanced curvature power spectra at small scales, which can lead to potential GW signatures in LISA. 

\subsection{Single field inflation}\label{sec:models_inflation}

In the simplest models of inflation, a single scalar field known as the inflaton moves gradually down its potential under the influence of Hubble friction, resulting in a slow-roll (SR) phase. The generated fluctuations are nearly scale-invariant, Gaussian, and adiabatic: they freeze on super-Hubble scales, producing a universe that is statistically homogeneous and isotropic~\cite{ Starobinsky:1980te, Guth:1980zm, Linde:1981mu, Albrecht:1982wi,Akrami:2018odb}.

In certain models, however, the inflaton potential in the Einstein frame can contain features such as a flat region or a mild bump that causes the field velocity to decrease rapidly in a brief USR\footnote{\label{foot:SR}We consider a slightly broader definition of USR, which is often characterized by $\eta_H \equiv - \ddot{H}/(2H \dot H) = 3$ (or $\eta \equiv \dot{\epsilon}/(H \epsilon ) = -6$) and can be realized on a flat plateau. However, models considered in the literature often exhibit a small bump instead, which results in $\eta_H > 3$ due to the curvature of the potential. We will include such deviations in our definition of USR.} phase~\cite{Ivanov:1994pa,Kinney:1997ne,Inoue:2001zt,Kinney:2005vj,Martin:2012pe,Motohashi:2017kbs} followed by another SR or a constant-roll phase~\cite{Ozsoy:2019lyy, Karam:2022nym,Balaji:2022dbi}.\footnote{See e.g. \cite{Ashoorioon:2019xqc,Ballesteros:2021fsp} for scenarios based on the EFT of inflation.} 
The shape of the enhanced spectral features is determined by the amplification of the curvature perturbations during the USR phase, as well as by the specifics of the transition into the USR era. If the field accelerates significantly during that transt epoch, it can cause sizeable spectral modulation, and even be the dominant source of amplification. The resulting perturbations deviate significantly from scale invariance, exhibiting the strongest amplification for wavelengths exiting the horizon around the USR era.

Generically, models of this kind can be subdivided into four categories: 
{\it i) }
Quasi-inflection points and plateaus~\cite{Garcia-Bellido:2017mdw, Kannike:2017bxn, Ballesteros:2017fsr, Germani:2017bcs, Motohashi:2017kbs, Ezquiaga:2017fvi, Di:2017ndc, Hertzberg:2017dkh, Rasanen:2018fom, Cicoli:2018asa, Ozsoy:2018flq, Gao:2018pvq, Atal:2019cdz, Atal:2019erb, Mishra:2019pzq, Ballesteros:2019hus, Dalianis:2018frf, Bhaumik:2019tvl, Drees:2019xpp, Dalianis:2019asr, Ballesteros:2020qam, Ragavendra:2020sop, Nanopoulos:2020nnh, Iacconi:2021ltm, Stamou:2021qdk, Wu:2021zta, Ng:2021hll, Rezazadeh:2021clf, Wang:2021kbh, Gu:2022pbo, Frolovsky:2022qpg, Cicoli:2022sih,Ghoshal:2023wri,Barker:2024mpz};
{\it ii)} 
Upward~\cite{Cai:2021zsp, Inomata:2021tpx}
or downward~\cite{Inomata:2021tpx, Inomata:2021uqj, Kefala:2020xsx, Dalianis:2021iig} steps;
{\it iii)}
Models in which the inflaton rolls through a global minimum/double-well potentials~\cite{Yokoyama:1998pt, Saito:2008em, Bugaev:2008bi, Fu:2020lob, Briaud:2023eae, Karam:2023haj}; and
{\it iv)} 
Potentials with stacked features/oscillating potentials~\cite{Cai:2019bmk,Tasinato:2020vdk,Zhou:2020kkf,Peng:2021zon,Inomata:2022yte,Fumagalli:2023loc,Caravano:2024tlp}. It was also suggested that models going beyond a non-minimally coupled inflaton, e.g.~within modified gravity theories, can introduce features in terms other than the inflaton potential or the non-minimal coupling~\cite{Frolovsky:2022ewg, Ballesteros:2018wlw, Fu:2019ttf, Fu:2019vqc, Heydari:2021gea, Heydari:2021qsr, Kawai:2021edk, Arya:2019wck, Ashoorioon:2019xqc, Bastero-Gil:2021fac, Ozsoy:2020kat, Solbi:2021wbo, Solbi:2021rse, Teimoori:2021pte, Correa:2022ngq, Kawaguchi:2022nku,Poisson:2023tja}. See e.g.~\cite{Ozsoy:2023ryl} for a model-building review. Even though the literature on these models is quite vast, many scenarios predict a curvature power spectrum that is enhanced at small scales with similar properties. In particular, most models, especially those in category {\it i)}, produce a single peak in the power spectrum that is approximately captured by a broken power law. However, models with sharp features can produce spectral oscillations at and after the peak in the power spectrum.
Such enhancement mechanisms are mainly in the categories {\it ii)} and {\it iii)}. 
Non-standard potentials in category {\it iv)} can also deviate strongly from this picture.

We finish this section with a note on the theoretical consistency of these enhancing mechanisms. There has been significant debate about the potential impact of loop corrections, induced by the enhanced modes during an USR phase, on long-wavelength scales, with some even challenging the validity of perturbative computations in these scenarios. This discussion is also relevant for the possible interpretation of a stochastic signal as originating from SIGW.  
The question of whether these corrections can become sufficiently large to undermine the predictive power of inflationary models related to PBH and SIGW has been also explored in Res.~\cite{Kristiano:2022maq,Riotto:2023hoz,Kristiano:2023scm,Riotto:2023gpm,Firouzjahi:2023aum,Firouzjahi:2023ahg,Franciolini:2023lgy,Cheng:2023ikq,Maity:2023qzw,Davies:2023hhn,Ballesteros:2024zdp}, while other studies have questioned 
% (and disproven) 
the very existence of these corrections and proposed an argument against their presence \cite{Fumagalli:2023hpa,Tada:2023rgp,Inomata:2024lud,Kawaguchi:2024rsv,Fumagalli:2024jzz}.

\subsection{Multi-field inflation}\label{sec:models_multi-field}

\paragraph{Hybrid/multi-field inflation.} 

Given the high number of degrees of freedom within multi-field inflationary setups, we can separate and control more efficiently two stages responsible respectively for the generation of nearly scale-invariant primordial curvature perturbations on large CMB scales, and for enhanced curvature perturbations on small scales which lead  to the production of SIGWs. Hybrid/multi-field models of inflation tend to generate slowly-growing lognormal-like peaks in the curvature power spectrum~\cite{Garcia-Bellido:1996mdl, Kawasaki:2015ppx, Pi:2017gih, Kallosh:2022vha, Braglia:2022phb, Tada:2023pue, Tada:2023fvd,Wang:2024vfv,Iacconi:2024hmg} while strong deviations from a geodesic trajectory in field space may lead to sharp peaks and features such as spectral oscillations in $\Pz$~\cite{Palma:2020ejf,Fumagalli:2020adf,Fumagalli:2020nvq,Braglia:2020eai,Braglia:2020taf,Bhattacharya:2022fze,Aragam:2023adu,Aragam:2024nej}.

\paragraph{Curvaton models}

Within curvaton scenarios~\cite{Lyth:2001nq}, one can realize setups with the curvaton field being characterized by a steep blue spectrum either due to interactions with the inflaton or other degrees of freedom during inflation~\cite{Chen:2019zza, Gow:2023zzp} or due to a non-trivial
kinetic term~\cite{Liu:2020zzv,Pi:2021dft,Meng:2022low}. In particular,  within axion-like curvaton setups, the curvaton field is identified with the phase of a complex field whose modulus decreases rapidly during inflation~\cite{Kawasaki:2012wr,Ando:2017veq,Ando:2018nge,Inomata:2020xad}. We should also highlight that in any curvaton model the curvature perturbations on small scales originate from non-adiabatic curvaton field fluctuations during inflation, leading to a non-Gaussian probability distribution function for the primordial curvature perturbations~\cite{Bartolo:2003jx,Bartolo:2005fp,Sasaki:2006kq,Enqvist:2008gk} with important consequences at the level of the SIGW signal~\cite{Kawasaki:2013xsa,Cai:2018dig,Ferrante:2023bgz,Perna:2024ehx} (but see also \cite{Chen:2024pge}).

\paragraph{Axion-gauge field coupling}

Enhanced scalar~\cite{Barnaby:2010vf} and tensor~\cite{Barnaby:2010vf,Sorbo:2011rz} modes can be produced by gauge fields amplified by their pseudo-scalar $\phi F {\tilde F}$ coupling with a rolling inflaton or spectator~\cite{Namba:2015gja,Shiraishi:2013kxa} axion during inflation. The gauge field amplitude is exponentially sensitive to the axion velocity, thus providing naturally blue signals. These enhanced curvature modes can lead to PBH and SIGW~\cite{Linde:2012bt,Bugaev:2013fya,Garcia-Bellido:2016dkw,Garcia-Bellido:2017aan}. The precise shape of these signals is sensitive to the axion evolution, which is significantly impacted by the backreaction of the amplified gauge fields, which is recently being explored via lattice simulations~\cite{Caravano:2022epk,Figueroa:2023oxc,Caravano:2024xsb,Figueroa:2024rkr}.

\subsection{Other classes of models}\label{sec:other_models}

\paragraph{Preheating.} 

During the preheating phase following inflation, as the inflaton field undergoes coherent oscillations around the minimum of its potential, a striking phenomenon emerges: the resonant amplification of quantum inflaton field fluctuations, which drives particle production~\cite{Kofman:1994rk,Kofman:1997yn}. These enhanced quantum fluctuations are accompanied by a resonant amplification of the scalar metric fluctuations (usually quoted as metric preheating~\cite{Finelli:1998bu,Bassett:1999mt,Jedamzik:1999um,Bassett:1999cg}), or, in other terms, with enhanced curvature perturbations, responsible for the generation of SIGWs~\cite{Jedamzik:2010hq}
and potentially for PBH formation~\cite{Jedamzik:2010dq,Martin:2019nuw,Martin:2020fgl} (see however~\cite{Ballesteros:2024hhq} for an assessment on the role of non-linearities and anharmonicities). Most studies have focused on multi-field inflationary setups since in such scenarios the enhancement of entropic (isocurvature) fluctuations can give rise to the enhancement of the adiabatic/curvature fluctuations in the broad resonance regime~\cite{Bassett:1998wg,Green:2000he,Bassett:2000ha,Suyama:2004mz,Torres-Lomas:2013uzl,Torres-Lomas:2014bua}. This leads to a notable amplification of the primordial curvature power spectrum, deviating from the standard scale-invariant behavior at small scales~\cite{Kou:2019bbc,Joana:2022uwc,Adshead:2023mvt}. Interestingly, recent works also suggest that a parametric amplification of the curvature perturbations can occur even in the narrow regime in the case of single field inflation~\cite{Jedamzik:2010dq,Easther:2010mr,Martin:2019nuw,Martin:2020fgl,Ballesteros:2024hhq}.
We should also note that these gravitational waves are typically peaked at MHz or GHz frequencies far above those of LISA, although some scenarios do allow for a peak in LISA's range \cite{Cui:2021are}.

\paragraph{Matter Bouncing Scenarios.} In non-singular matter bouncing cosmological models \cite{Brandenberger:2012zb}, the matter contracting phase inevitably amplifies super-horizon curvature perturbations. This enhancement can lead to an enhanced primordial curvature power spectrum on small scales compared to the ones probed by CMB. As these perturbations cross the cosmological horizon, either during the contracting phase~\cite{Cai:2012va,Cai:2013kja,Chen:2016kjx} or the expanding Hot Big Bang phase~\cite{Banerjee:2022xft,Papanikolaou:2024fzf}, they can lead to the abundant production of SIGWs.

\paragraph{Early PBH domination.}
At distances much larger than the mean PBH separation length, a population of PBHs can be viewed as an effective pressureless fluid. One can then treat this PBH fluid within the context of cosmological perturbation theory showing that the PBH energy fluctuations are isocurvature in nature~\cite{Papanikolaou:2020qtd,Domenech:2020ssp} and can convert to adiabatic curvature perturbations in an early matter-dominated era driven by light PBHs ($m_\mathrm{PBH}<10^9\mathrm{g}$) occurring before BBN. Interestingly enough, these PBH-induced curvature perturbations can source abundant SIGWs detectable by GW observatories~\cite{Papanikolaou:2020qtd,Domenech:2020ssp,Domenech:2021wkk}. Notably, these PBH associated SIGWs~\cite{Domenech:2021and,Domenech:2023jve} can serve as a novel portal to probe primordial non-Gaussianities (NGs)~\cite{Papanikolaou:2024kjb,He:2024luf} at small scales ($k>\mathrm{Mpc}^{-1}$) as well the underlying gravity theory~\cite{Papanikolaou:2021uhe,Papanikolaou:2022hkg,Tzerefos:2023mpe} and Hawking evaporation \cite{Dvali:2020wft,Domenech:2023mqk,Franciolini:2023osw,Balaji:2024hpu,Dvali:2024hsb,Kohri:2024qpd,Bhaumik:2024qzd,Domenech:2024wao}.

\subsection{A simple benchmark scenario: single field USR}\label{sec:benchmodelUSR}

In this work, we consider one of the simplest realizations of the single-field scenarios discussed in the previous section.
This class of models is described by the inflaton potential in the Einstein frame, $V(\phi)$. The corresponding action can be written as 
\begin{equation}\label{eq:S_standard}
    {\cal S}=\int \d^4 x \sqrt{-g} \left( \frac{1}{2} \mpl^2R-\frac{1}{2}(\partial^{\mu} \phi)^2-V(\phi) \right)\,,
\end{equation} 
where $R$ is the Ricci scalar and $\mpl$ is the reduced Planck mass.
Assuming a flat FLRW background geometry $\d s^2= -\d t^2+a^2 \d x_i^2$, where $a$ is the scale factor, the background evolution is governed by the Friedmann equation (dots indicate time derivatives) 
\begin{align}\label{eq:Friedmann}
    3\mpl^2 H^2= \dot{\phi}^2/2+V(\phi)\,,
\end{align}
with $H= \dot a/a$, and the Klein-Gordon equation (a prime denotes a derivative with respect to the field)
\begin{align}\label{eq:klein-gordon}
    \ddot{\phi} + 3 H \dot{\phi} + V'(\phi) = 0 \,.
\end{align}
In order to produce an enhancement of perturbations at LISA scales, and at the same time comply with CMB bounds at large scales, the inflationary potential should feature a shallower region or an inflection point, which breaks the SR evolution exponentially decelerating the field velocity. 

In the class of models considered here, the dynamics can be understood in relatively simple terms. In SR, the inflaton evolves with a negligible acceleration, and the SR solution gives $\dot\phi = -V'/(3H)$.
As the inflaton begins to approach the inflection point, the SR conditions are violated primarily due to a rapid change in the second SR parameter. Having almost reached the local maximum $\phi_*$, the inflaton will spend $\mathcal{O}(10)$ $e$-folds crossing it and its evolution is thus dictated by $\ddot \phi + 3 H \dot \phi  + \eta_V(\phi_*) H^2 (\phi - \phi_*) \simeq 0$, where $\eta_V \equiv \mpl^{2} V''/V$ denotes the second potential SR parameter. The two solutions of this equation describe two phases: First, a USR-like phase in which the inflaton rapidly decelerates, which leads to an amplification of the power spectrum. Second, a subsequent constant roll or a SR phase that is dual to the initial USR-like phase~\cite{Wands:1998yp,Karam:2022nym}. 

The linear superposition of these solutions describes a smooth transition between these epochs. The second SR parameter $\eta_V(\phi_*)$ determines the spectral slope after the peak $n_s-1 = 3(1-\sqrt{1 - (4/3)\eta_V(\phi_*)})$. Thus, as exact USR ($\eta_V(\phi_*) = 0$) would produce a scale-invariant spectrum at scales \textit{above} the spectral peak, the violation of scale invariance in the UV is directly related to the deviation from an exact USR.

As an example, we consider the potential given by the rational function proposed in~\cite{Garcia-Bellido:2017mdw} (see also \cite{Germani:2017bcs,Cole:2023wyx})
\begin{equation}
    V(\phi) =
    \frac{\lambda}{12}
    \phi^2
    (v \mpl)^2
    \left (6 - 4 b_l \frac{\phi}{v\mpl} + 3\frac{\phi^2}{(v\mpl)^2} \right )
    \left(1 + b\frac{\phi^2}{(v\mpl)^2}\right)^{-2}\,.
\label{eq:Germani-potential}
\end{equation}
The presence of an inflection point is enforced by setting 
\begin{equation}
    b = (1+ b_f)\left [
    1-\frac{b_l ^2}{3} +\frac{b_l ^2}{3}
    \left ( \frac{9}{2 b_l ^2} -1 \right )^{2/3}\right]\,,
\end{equation}
where we included a tuning parameter $b_f$ allowing for deviation from perfect inflection points (with $b_f>0$ the inflection point becomes a shallow minimum). The field $\phi$
appearing in the action \eqref{eq:S_standard} is canonically normalized, and minimally coupled to gravity.
This is a proxy for more realistic models in which the inflaton field has a quartic potential and couples non-minimally to gravity via a $\xi R \phi^2$ term, see e.g.~\cite{Kannike:2017bxn, Ballesteros:2017fsr, Drees:2019xpp, Madge:2023dxc}. After moving to the Einstein frame, the factor in the denominator appears, which flattens the potential at large field values. In this case, one would further need to canonically normalize the field, and possibly add logarithmic corrections to the coefficients of the monomials in Eq.~(\ref{eq:Germani-potential}).
We do not discuss the origin of such a potential, as our goal is solely to provide a simple representative model to work with.

\begin{figure}[t]
\centering
\includegraphics[width=0.49\textwidth]{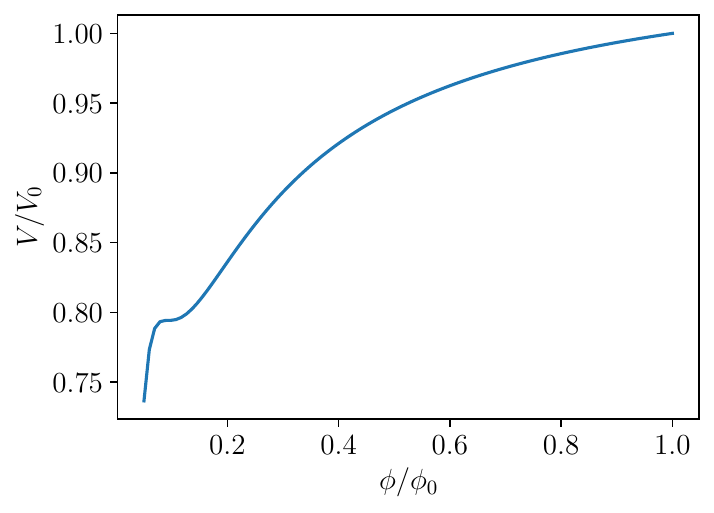}
\caption{Example of a single field model \eqref{eq:Germani-potential}, with our benchmark parameters~\eqref{eq:benchUSRvals}, leading to an early SR phase consistent with CMB data as well as an USR phase which leads to an enhancement of perturbations within the LISA frequency range. 
}
\label{fig:USR_potential}
\end{figure}

We define our benchmark potential by choosing the following parameters (close to the ones used in \cite{Cole:2023wyx})
\begin{equation}\label{eq:benchUSRvals}
\begin{gathered}
\lambda = 1.4731\times  10^{-6},
\qquad
v = 0.19688,
\\
b_l  = 0.71223,
\qquad
b_f = 1.87\times 10^{-5}\,,
\end{gathered}
\end{equation}
leading to good agreement with CMB (within 3$\sigma$ of current Planck 2018 data \cite{Akrami:2018odb}) and at the same time to a peak of the curvature spectrum: ${\cal P}_\zeta (k_{\rm peak }) \simeq 10^{-3}$ at LISA scales.
In Eq.~\eqref{eq:benchUSRvals} we report 5 significant digits because of the required tuning of the USR potential \cite{Cole:2023wyx}.
 In Sec.~\ref{sec:USR_pzeta_computation}
 we describe how to compute the spectrum of curvature perturbations in detail.
The benchmark potential is depicted in Fig.~\ref{fig:USR_potential}, where we have arbitrarily normalized the axes using the initial values $\phi_0 = 3 \mpl$ and $V_0 = 2.3 \cdot 10^{-10} \mpl^4$, which are set well before the SR phase that governs the CMB scales. 

\section{Modeling the curvature power spectrum}
\label{sec:modellingPzeta}

Throughout this paper, we describe the metric as a small perturbation of the FLRW metric in the longitudinal (conformal Newtonian) gauge
\begin{equation}
\label{eq:metric newtonian}
\d s^2 = g_{\mu\nu}\d x^\mu \d x^\nu = -a^2(1+2\Phi) \d \eta^2 
+ a^2 \left[ (1-2\Phi)\delta_{ij}+\frac{1}{2}h_{ij} \right] \d x^i \d x^j,
\end{equation}
where $\eta$ is the conformal time, and we neglect vector perturbations and  anisotropic stress (and so we can identify the two scalar Bardeen potentials, $\Phi = \Psi$).

At linear order in perturbation theory, the time and momentum dependence of the scalar potential $\Phi$ can be factored out by introducing the transfer function $T(\eta,k)$: the Bardeen's potential is related to the gauge-invariant comoving curvature perturbation $\zeta(\vk)$ by
\begin{equation}
\Phi(\eta,\vk)=\frac{3+3w}{5+3w} T(\eta,k) \zeta(\vk) \,
\label{eq: Phi to zeta}
\end{equation}
where $w$ is the equation of state characterizing the fluid dominating the energy density in the universe. In a radiation-dominated universe, $w = 1/3$, and the prefactor becomes $2/3$. 
As long as a mode $k$ is super-Hubble ($k\eta<1$), its evolution is frozen and the transfer function tends to 1.
For a homogeneous and isotropic universe, the two-point function of curvature perturbations reads
\begin{equation}
\left\langle \zeta(\vk_1)\zeta(\vk_2)\right\rangle \equiv
 (2\pi)^3 \delta(\vk_1+\vk_2) P_\zeta(k_1)
 \equiv
(2\pi)^3 \delta(\vk_1+\vk_2)\frac{2\pi^2}{k_1^3} \Pz(k_1) ,
\label{eq: def P zeta}
\end{equation}
where we introduced the dimensionful power spectrum $P_\zeta(k)$, and the dimensionless one $\Pz(k)$.
We adopt different methodologies to model the primordial curvature power spectrum $\Pz(k)$ generated in various inflationary models, by:
\begin{itemize}
\item Developing a model-independent parametrization of the spectrum, enabling an agnostic reconstruction approach;
\item Constructing analytical templates for $\calP_\zeta(k)$. These templates are motivated by specific scenarios and are consequently less flexible than the previous case. However, given the lower number of parameters controlling the templates as well as their simple analytical descriptions, they alleviate the computational challenges posed by models which would require expensive computation of the curvature power spectrum;
\item Establishing a robust pipeline for the computation of primordial curvature power spectra $\calP_\zeta(k)$ for the benchmark model described in Sec.~\ref{sec:models}.
\end{itemize}

\subsection{Model independent parameterization: binned spectrum approach}\label{sec:binnedPzeta}

In this section, we discuss a parameterization of the power spectrum $\mathcal{P}_\zeta$ in terms of a \textit{binned spectrum approach},  which is useful in developing a template-independent procedure for computing the SIGW. We
 represent the curvature power spectrum as a sum of discrete components defined within small momentum intervals. Specifically,
we decompose  ${\cal P}_\zeta$ as a sum of $N -1$ top hat functions as 
\begin{equation}\label{eq:PzetaBinning}
    {\cal P}_\zeta (p) =
    \sum_{i= 1}^{N-1}
    A_i \Theta (p - p_i)\Theta (p_{i+1} - p)\,.
\end{equation}
The coefficients $A_i$ -- which can be approximated as constants -- represent the amplitude of the spectrum within the $i$-th bin. The latter is defined within the momentum boundaries $p_i$ and 
$p_{i+1}$, using the Heaviside step functions $\Theta(p)$.\footnote{One could extend this approach to include a tilt of the spectrum at each bin and enforcing continuity for adjacent bins, at the cost of doubling the parameters controlling the power in each bin. We leave this, and other possible extensions for future work. } 
Any curvature spectrum ${\cal P}_\zeta$ can then be associated
with a specific vector  $A_i$ of the coefficients appearing in Eq.~\eqref{eq:PzetaBinning}. Therefore, our computation does not require an {\it a priori} choice of a particular template for the momentum dependence of  ${\cal P}_\zeta(p)$.

The SIGW depends quadratically
on ${\cal P}_\zeta$ using 
convolution integrals. Hence, we can expect that
the ansatz
\eqref{eq:PzetaBinning} allows us
to express the SIGW
in terms of the double sum of (model-dependent) vector components $A_i$, contracted over a general, model-independent matrix kernel. In  Sec.~\ref{sec:OmegaGWijk} we elaborate a simple procedure aimed at performing this task.

\subsection{Analytical templates for curvature spectra}\label{sec:analyticaltemplates}

We divide the list of templates into classes depending on their properties. We can define ``smooth shapes'' and shapes with features.

\subsubsection{Smooth templates}

We divide this class of templates into two cases:
\paragraph{Lognormal (LN).} 
A typical class of spectral peaks can be characterized by a LN-shape 
\begin{equation}\label{eq:PLN}
\mathcal{P}^{\rm LN}_{\zeta}(k) = 
  \frac{A_s}{\sqrt{2\pi\Delta^2}}
  \, \exp\left[ -\frac{1}{2\Delta^2} \log^2(k/k_*) \right]\, .
\end{equation}
Such spectra appear e.g.~in a subset of hybrid inflation and curvaton models, as well as from axion-gauge field coupling, see the discussion in Sec.~\ref{sec:models}.
This template can describe scenarios in which the peak is typically narrow and symmetric in log space. 
 Interestingly,  this template allows for an analytic derivation of the GW power spectrum in the broad/narrow peak approximations $\Delta\gg1$ or $\Delta\ll 1$ (see Ref.~\cite{Pi:2020otn} and improvements in~\cite{Dandoy:2023jot}).
We will consider the following benchmark scenario choosing
\begin{equation}
\log_{10} A_s=-2.50, \quad\log_{10}\Delta=\log_{10}(0.5)\approx-0.301, \quad \log_{10} \left(k_*/\mathrm{s}^{-1}\right)=-2.00\,.
\label{eq:PLN injected}
\end{equation}

\paragraph{Broken power law (BPL).} 
Another broad class of spectra is encountered, for instance, in single field inflation and curvaton models and can be described by a BPL
\begin{equation}
\label{eq:BPL}
\mathcal{P}^{\rm BPL}_{\zeta}(k)
    =A_s \frac{\left(\alpha+\beta\right)^{\gamma}}{\left(\beta\left(k / k_*\right)^{-\alpha/\gamma}+\alpha\left(k / k_*\right)^{\beta/\gamma}\right)^{\gamma}},
\end{equation}
where $\alpha, \beta>0$ describe respectively the growth and decay of the spectrum around the peak, while $\gamma$ is an ${\cal O}(1)$ model dependent parameter. The normalization is such that $\mathcal{P}^{\rm BPL}_{\zeta}(k) = A_s$ at the peak.  This template provides a very close approximation to the shape of $\mathcal{P}_\zeta$ obtained from single-field USR scenarios. There, one typically finds $\alpha \simeq 5-|1-2\eta_H| \approx 4$ \cite{Byrnes:2018txb,Ozsoy:2019lyy,Karam:2022nym}, where $\eta_H$ is the second Hubble SR parameter (see footnote \ref{foot:SR}) \textit{before} the USR phase, which is typically close to SR, that is, $\eta_H \ll 1$. 
For this template, it is possible to find an analytic GW spectrum with $\gamma=1$, see \cite{Li:2024lxx}.
We will use the following benchmark with parameters
\begin{equation}\label{eq:BPL_bench}
\begin{gathered}
\log_{10}A_s = -2.71,
\qquad
\log_{10} (k_*/ \mathrm{s}^{-1}) = -1.58,
\\
\alpha = 3.11,
\qquad
\beta = 0.221,
\qquad
\gamma = 1.25. 
\end{gathered}
\end{equation}
which provides a curvature power spectrum whose peak is within LISA's sensitivity and that fits 
the spectrum produced in the ab initio USR scenario described below (see Sec.~\ref{sec:USR_pzeta_computation}).

\subsubsection{Templates with oscillations}\label{sec:temposc}

Oscillations in the primordial power spectrum have been extensively studied to seek for deviations from the standard SR inflationary paradigm mainly at scales relevant for CMB or large-scale structure (see e.g.~\cite{Chen:2010xka,Chluba:2015bqa,Slosar:2019gvt,Achucarro:2022qrl}). These oscillations, denoted as primordial features, typically arise due to a sudden transition during inflation~\cite{Starobinsky:1992ts}, occurring over a short time-scale of the order of one $e$-fold. SIGW provides an opportunity to probe such primordial features at small-scales ($\ll \mathrm{Mpc}$)~\cite{Fumagalli:2020nvq,Braglia:2020taf,Fumagalli:2021cel,Witkowski:2021raz,Fumagalli:2021dtd}. Specifically, if the mechanism responsible for the enhancement in $\mathcal{P}_{\zeta}$ is active when modes are sufficiently deep inside the horizon, the resulting spectrum exhibits oscillations of order one. Due to their large amplitude, these oscillations could potentially leave their imprints and be detected in the SGWB. This phenomenon can occur in both single-field and multi-field inflationary models. As benchmarks for these large amplitude oscillations, we will consider a small-scale feature induced by a genuine multi-field mechanism and, secondly, weaker oscillations arising from a rapid transition between SR and USR phases in single-field inflation.

\paragraph{Turns in multi-field inflation.}
In multi-field inflationary setups, a common phenomenon is the presence of turns in the field space, being equivalent to the inflationary trajectory deviating from a geodesic in field space. This bending is quantified through the parameter $\eta_\perp$, measuring the acceleration of the inflationary trajectory perpendicular to its direction~\cite{GrootNibbelink:2000vx,GrootNibbelink:2001qt} or equivalently the deviation of the trajectory from a geodesic in the target field space.  One then can show that sharp and strong (large $\eta_\perp$) turns can lead
to the following curvature power spectrum, modulated by order-one rapid oscillations\footnote{Equation \eqref{eq:Pzeta-strong-sharp-turn} assumes the entropic field to be massless during the turn. Generalized expressions for other mass choices can be found in \cite{Fumagalli:2020nvq}, where the qualitative features remain analogous.}~\cite{Palma:2020ejf,Fumagalli:2020adf,Fumagalli:2020nvq}
\begin{align}
\label{eq:Pzeta-strong-sharp-turn}
{\cal P}_\zeta^{\rm ST}(\kappa) = 
\ & \mathcal{P}_\zeta^\textrm{env}(\kappa) \bigg[1 + (\kappa-1) \cos \Big(2 e^{-\frac{\delta}{2}} \eta_\perp \kappa \Big) + \sqrt{(2-\kappa)\kappa} \, \sin \Big(2 e^{-\frac{\delta}{2}} \eta_\perp \kappa \Big) \bigg] 
\Theta(2-\kappa) \, , 
\end{align}
and the envelope
\begin{align}\label{eq:Pzeta_ST_env}
\mathcal{P}_\zeta^\textrm{env}(\kappa) = A_s
\exp\left(-2 \eta_\perp \delta \right )
\exp\left [2 \sqrt{(2-\kappa)\kappa} \, \eta_\perp \delta\right ]
/\left [4(2-\kappa) \kappa\right] \, ,
\end{align} 
where 
$A_s$
denotes the amplitude of the power spectrum in the absence of transient instability and $\kappa \equiv k/k_*$ with $k_*$ being associated with the maximally enhanced scale, deep inside the cosmological Hubble sphere at the time of the sharp turn. 
The parameter $\delta$ is the duration in $e$-folds of the turn. $\delta\gtrsim \log \eta_\perp$ stands for broad turns and $\delta\lesssim \log \eta_\perp $ stands for sharp turns.
Finally, $\Theta(x)$ denotes the Heaviside theta function.

We define a parameterization in which the oscillations are switched off via the parameter $F \in[0,1]$, continuously interpolating between~\eqref{eq:Pzeta-strong-sharp-turn} when $F = 1$ and~\eqref{eq:Pzeta_ST_env} when $F = 0$
\begin{equation}
    {\cal P}_{\zeta}(k) 
    = 
    F
    {\cal P}_{\zeta}^{\rm ST}(k) 
    + 
    (1-F)
    {\cal P}_{\zeta}^{\rm env}(k).
\end{equation}
We consider the benchmark scenario whose parameters are given by
\begin{equation}\label{eq:benchES}
\begin{gathered}
\log_{10}A_s= -1.5,
\quad
\log_{10}{(k_*/s^{-1})} = -1.5,
\\
{\delta} = 0.5,
\quad
\eta_\perp = 14,
\quad 
F = 1.
\end{gathered}
\end{equation}

\noindent
\paragraph{Rapid transitions between SR and USR phases.}

In most USR/inflection-point scenarios, the transition from the initial SR phase to the USR-like phase depends on the properties of the potential.
In the presence of sufficiently sharp transitions between these phases, the spectrum of curvature perturbations deviates from a BPL profile due to oscillatory features.

A fully analytic spectrum can be obtained when the initial SR to USR transition is instantaneous. In that case, the peak can be described as (Eq.~(3.8) in~\cite{Karam:2022nym}) 
\begin{align}\label{eq:USR_osc}
    {\cal P}_\zeta(k) 
    = \frac{\kappa^2}{4\pi}
    \Bigg|
    &\!-\! \frac{\Gamma\left(1+\nu_{\rm II}\right)}{\zeta_1}\left(\kappa/2\right)^{-\nu_{\rm II}+ \frac{1}{2}} \! J_{\nu_{\rm II}}\left(\kappa\right)H_{\nu_{\rm I}}\left(\kappa\right) 
    \\ & \!+\! 
    \frac{\Gamma\left(\nu_{\rm II}\right)}{\zeta_2}\left(\kappa/2\right)^{-\nu_{\rm II} + \frac{3}{2}}\!
    \left[J_{\nu_{\rm II}}\left(\kappa\right)H_{\nu_{\rm I}-1}\left(\kappa\right) 
    + J_{\nu_{\rm II}+1}\left(\kappa\right) H_{\nu_{\rm I}}\left(\kappa\right)\right]
    \Bigg|^2, \nonumber
\end{align}
where, $\kappa = k/k_*$, $H_{\nu}(x)$ and $J_{\nu}(x)$ denote the Hankel- and Bessel functions of the first kind, respectively, $k_{*}$ is the scale of the mode that exits the Hubble sphere during the SR to USR transition, and $\nu_{\rm I}$ and $\nu_{\rm II}$ are related to the spectral slopes in the attractor phases before and after the transition as $n_{s} = 2(2-\nu)$. 
The first line does not contribute to the peak and can be omitted in case there is no sensitivity to spectral features away from the peak. The parameters $\zeta_{1}$, $\zeta_{2}$ control the amplitude at the IR  (or CMB) scales and the peak, respectively.

This spectrum resembles a broken power law with a modulation around the peak. This modulation has a period of $2k_*$ and is damped as $1/k$. It is generated in an instantaneous SR to USR transition and is typically suppressed or removed when the transition is non-instantaneous~\cite{Cole:2022xqc, Karam:2022nym, Franciolini:2022pav, Briaud:2025hra}. In this way, such oscillations carry information about the evolution of the inflaton during the transition. Moreover, spectral oscillations can be greatly enhanced in some cases. For instance, when the inflaton rolls through a deep minimum before entering the USR phase~\cite{Yokoyama:1998pt, Saito:2008em, Bugaev:2008bi, Fu:2020lob, Briaud:2023eae, Karam:2023haj}.

We can test for the sensitivity of LISA to resolve these oscillations. To this aim, we only consider the second line of the spectrum~\eqref{eq:USR_osc}, which describes the peak, and consider a generalized form\footnote{For notational simplicity, the template uses a different normalization and scaling of the argument than \eqref{eq:USR_osc}.}
\begin{equation}
\label{eq:Pzeta_sharpfeature}
   {\cal P}_{\zeta}^{\rm osc}(k) 
    = 
    F
    {\cal P}_{\zeta}^{\rm osc,B}(k) 
    + (1-F)
    {\cal P}_{\zeta}^{\rm osc,BPL}(k),
\end{equation}
where $F\in[0,1]$ and with 
\begin{align}
{\cal P}_{\zeta}^{\rm osc,B}(k) 
&=
\pi^2 \kappa^{5 - 2\nu_{II}}
A_s
\left| J_{\nu_{II}}(2\kappa) H_{\nu_I - 1}^{(1)}(2\kappa) + J_{\nu_{II} + 1}(2\kappa) H_{\nu_I}^{(1)}(2\kappa) \right|^2,
\label{eq:USRSR1}
\\
{\cal P}_{\zeta}^{\rm osc,BPL}(k) 
&= 
A_s
\left[ \left( \kappa^{7/2 - \nu_I} \frac{(\nu_I + \nu_{II}) \Gamma(\nu_I - 1)}{\Gamma(\nu_{II} + 2)} \right)^{-\gamma} 
+ \left( \kappa^{3/2 - \nu_{II}} \right)^{-\gamma} \right]^{-2/\gamma}.
\label{eq:USRSR2}
\end{align}
The rewriting in terms of the two contributions in Eqs.~\eqref{eq:USRSR1} and \eqref{eq:USRSR2} is equivalent to \eqref{eq:USR_osc} when imposing $F=1$, but it is done to separate the smooth BPL contribution from the one including the oscillations. This way, changing $F$ smoothly transitions between an oscillating ($F = 1$) and a non-oscillating $F = 0$ power spectra.
 
The shape of the BPL template is recovered using \eqref{eq:BPL} with $\nu_I = (7-\alpha)/2$ and $\nu_{II} = (3+\beta)/2$ when $\alpha \leq 5$. 
To obtain an exact match with \eqref{eq:BPL} the normalization and the momenta must also be rescaled so that ${\cal P}_{\zeta}^{\rm BPL}(k) = n{\cal P}_{\zeta}^{\rm osc,BPL}(bk)$, where 
\begin{equation}
    n \equiv b^{\beta} (1 + \beta/\alpha)^{\gamma} \,, 
    \qquad
    b \equiv \left[\frac{4 (\alpha/\beta)^{\gamma} \Gamma \left((\beta +7)/2\right)^2 }{(-\alpha +\beta +10)^2 \Gamma \left((5-\alpha)/2\right)^2}\right]^{\frac{1}{\alpha+\beta}}\,
\end{equation}
or, equivalently, as $A_s \to n A_s$ and $k_* \to k_*/b$.
The benchmark scenario corresponds to
\begin{equation}\label{eq:bench_sharpfeature}
\begin{gathered}
\log_{10}A_s = -2.58 \,,\quad
\log_{10} (k_*/\mathrm{s}^{-1}) = -2.02\\ 
\nu_{I} = 1.95 \,,\quad
\nu_{II} = 1.61\,,\quad
\gamma = 1.67 \,,\quad
F = 1.
\end{gathered}
\end{equation}
We fix the parameters to match the BPL template in the limit $F=0$.

\subsection{Computation of $\calP_\zeta$ in 
single field  USR scenarios}\label{sec:USR_pzeta_computation}

The benchmark model we introduced in Sec.~\ref{sec:benchmodelUSR} is based on a single-field model of inflation featuring a phase of USR. We now describe in detail how to compute the spectrum of curvature perturbations using linear perturbation theory.

As a warm-up, let us define the system of equations in terms of dimensionless variables rescaled to the corresponding relevant quantities. This typically improves the numerical stability of a code computing
the spectrum of curvature fluctuations.  
We define
\begin{align}\label{eq:adim_phi_V}
    x \equiv \phi / \mpl,
    \qquad 
    U(x)\equiv V(\phi) / V_0,
\end{align}
where we introduce the suffix $0$ to indicate the quantities evaluated at the initial conditions of the background evolution. 
We can also define the dimensionless time and Hubble rate using
\begin{align}
    \tau \equiv t V_0 ^{1/2}/\mpl, 
    \qquad 
    h \equiv H \mpl /  V_0^{1/2}\label{eq:hh}.
\end{align}
We can now change the evolution variable from time to the number of $e$-folds $N$, defined as ${\rm d} N \equiv H \d t = h \d \tau$, setting $N = 0$ at $\tau_0$.
In this way, the background equations of motion \eqref{eq:Friedmann} and \eqref{eq:klein-gordon} become
\begin{subequations}\label{eq:adim_EomBkgUSR}
\begin{align}
    y^{\prime} 
    +  3 \left [1-\frac{y^2}{6} \right ] y + 
    \frac{U_{,x}}{h^2} &= 0,
    %\nonumber 
    \\
    x^\prime - y &= 0 ,
   % \nonumber
    \\
    h^\prime + \frac{(x^\prime)^2}{2} h &= 0,
\end{align}
\end{subequations}

\noindent where prime denotes derivatives with respect to $N$, while $U_{,x} \equiv \d U(x)/\d x$.
The initial conditions for the inflaton and the Hubble parameters can be found assuming that initially, SR is satisfied, which for a given initial $x_0$ leads to 
\begin{equation}
    y_0= - \frac{3}{U_{0,x}} 
    \left ( \sqrt{1+\frac{2}{3} U_{0,x}^2} -1\right ),
    \qquad
    h_0 \equiv \frac{1}{\sqrt{6}}
    \left ( \sqrt{1+\frac{2}{3} U_{0,x}^2} -1\right )^{1/2}.
\end{equation}
Finally, in the code, we keep track of the equation of state during inflation $w_{\rm inf} \equiv p_\phi / \rho_\phi$, which can be written in our notation as $w =[ (x')^2 -3 ]  /3$.
As inflation stops when the equation of state becomes larger than $w>-1/3$, we stop the evolution when $(x')^2  = 2$.

The resulting inflationary background can be described by the evolution of the Hubble rate $H$.
This is dictated by dynamical equations relating $H$ to the SR parameters, which are defined as
\begin{align}
\epsilon_H
&\equiv -\frac{\dot{H}}{H^2} = 
\frac{1}{2 \mpl^2} \left( \frac{ d \phi }{d N} \right )^2\,
\equiv \frac{y^2}{2},
 \\
\eta_H 
&\equiv
-\frac{\ddot{H}}{2H\dot{H}} = \epsilon_H - \frac{1}{2}\frac{d\log\epsilon_H}{dN}
=
\frac{y^2}{2} - \frac{y'}{y}
\,, \label{eq:SR parameters}
\end{align} 
where we introduced $y = x'$.
Notice that our definition of $\eta_H$ differs from another definition often used in the literature, which is expressed as $\dot\epsilon_H/(\epsilon_H H)$. 
During the USR phase, this definition equals approximately $- 2 \eta_H$ neglecting $\epsilon_H$ corrections.

As long as the SR approximation is valid, i.e.~$\epsilon_H \ll 1$ and $\eta_H \ll 1$, 
the power spectrum of curvature perturbations (see e.g.~\cite{Ellis:2023wic}) is given by
\begin{equation}
\calP_\zeta (k) = \frac{H^2_k}{8\pi^2 \mpl^2 \epsilon_{H,k}}
=
\frac{V_0}{\mpl^4} \left (\frac{h_k}{2 \pi y_k }\right)^2\,,
\label{eq:PzetaSR}
\end{equation} 
where the suffix $k$ indicates that these quantities are evaluated at Hubble crossing $k = aH$. 
Consequently, one can also show that the scalar spectral index $n_s-1\equiv \d \ln {\cal P}_\zeta/\d \ln k$ and the tensor-to-scalar ratio $r \equiv {\cal P}_h/{\cal P}_\zeta$ are given by the well-known expressions
\begin{align}
n_s &\approx 1 - 4\epsilon_H + 2\eta_H ,
    \\
r & \approx 16\epsilon_H ,\label{eq:r_slow}  
\end{align}
at leading order in the SR parameters.

The  approximations above can not be used in the context of USR scenarios, as the large deceleration of the inflaton velocity causes $\eta_H \sim {\cal O} (1)$
($\eta_H = 3$ in the limit of exponential deceleration). 
We go beyond the SR approximation by solving the Mukhanov-Sasaki (MS) equation \cite{Sasaki:1986hm, Mukhanov:1988jd}.
Introducing momentarily the conformal time $\eta$ such that $\d \eta \equiv \d t / a$, we can define the MS variable $v\equiv a \delta \phi $ in terms of the inflaton perturbations in the spatially flat gauge, which satisfies the EoM
\begin{equation}\label{eq:MS equation}
    \frac{\partial ^2 v}{\partial \eta^2} + 
    \left ( k^2 - \frac{1}{z} \frac{\partial ^2 z}{\partial \eta^2}  \right) v = 0,
\end{equation}
where $z \equiv a^2 (\partial \phi/\partial \eta )/(\partial a/\partial \eta )$.
We can relate the curvature perturbation to $v$ at linear order in perturbation theory using $\zeta = v/z = H \delta \phi / (\partial \phi / \partial t) = \delta x /  y$. 
Assuming initial adiabatic vacuum, each mode $k$ is fixed in the asymptotic past at $\eta_{\rm in}  \ll -1/k $ as 
\begin{equation}
    v(\eta ) = 
    e^{-i k (\eta - \eta_{\rm in})}/ \sqrt{2 k}, 
\end{equation}
which translates into boundary conditions for Eq.~\eqref{eq:MS equation} of the form
\begin{equation}
    v_{\rm in} = \frac{1}{\sqrt{2 k}}
,
\qquad 
    \frac{ \partial v_{\rm in}}{\partial \eta} =
    -i \sqrt{\frac{k}{2}}.
\end{equation}
One can improve the stability of the numerical computation by defining a rescaled variable $\widetilde {\delta \phi} \equiv a_{\rm in} e^{i k (\eta - \eta_{\rm in})} \delta \phi \sqrt{2 k}$ and the corresponding dimensionless quantity $\widetilde {\delta x} \equiv \widetilde{\delta \phi}/\mpl$. The prefactor absorbs the sub-Hubble oscillations, simplifying the time evolution of the sub-Hubble phase. 
Next, we introduce the dimensionless momentum 
$\kappa \equiv k \mpl / V_0^{1/2}$.
Finally, moving to the number of $e$-folds as the time variable, the MS equation becomes 
\begin{equation}
\label{eq:MSequationadim}
\frac{{\rm d}^2
\widetilde{\delta x}}
{\d N^2}
+ 
\left( 3 - \frac{1}{2} y^2 - \frac{2i \kappa}{e^N h} \right) 
\frac{{\rm d}
\widetilde{\delta x}}
{\d N}
+ \left(
\frac{U_{,xx} + 2U_{,x} y}{h^2} 
+ 3y^2 - \frac{1}{2} y^4 
- \frac{2i\kappa}{e^N h}\right)
\widetilde{\delta x} = 0,
\end{equation}
to be solved with initial conditions $\widetilde{\delta\phi}_{\rm in} = 1$, and 
$\widetilde{\delta\phi}'_{\text{in}} = -1$, 
while the curvature power spectrum is extracted using
\begin{equation}\label{eq:Pzeta_from adim}
    {\cal P}_\zeta (k)
    = 
    \frac{1}{4\pi^2} 
    \frac{V_0}{\mpl^4}
    \left( \frac{\kappa}{e^{N_{\text{in}}}} \right)^2 
    \frac{|\widetilde{\delta x}|^2}{y^2}.
\end{equation}

\begin{figure}[t]
\centering
\includegraphics[width=0.7\textwidth]{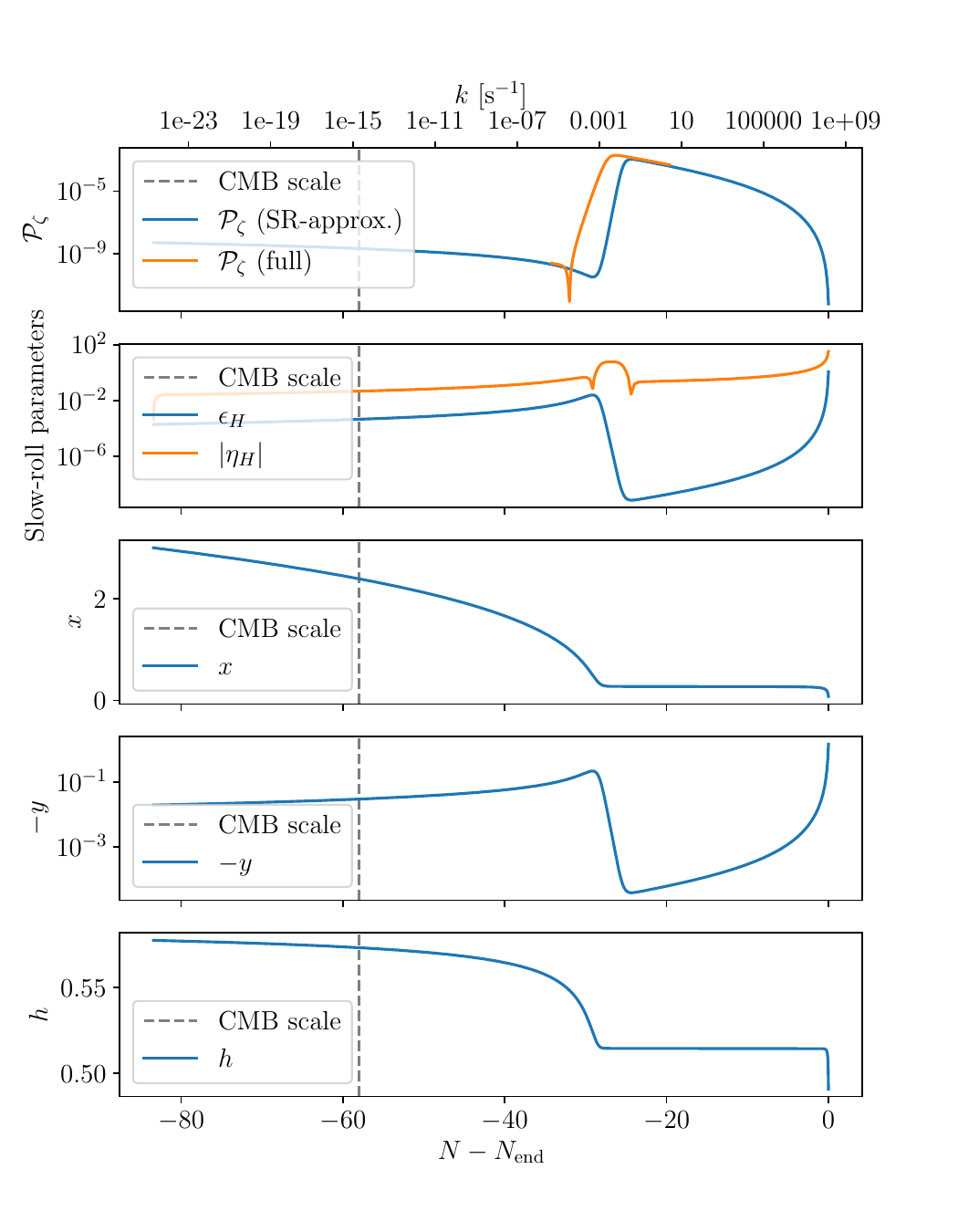}
\caption{From top to bottom:  
curvature power spectrum ${\cal P}_\zeta$; 
Hubble SR parameters $\epsilon_H$ and $\eta_H$; 
dimensionless inflaton field $x$;  
dimensionless inflaton field velocity $y$; 
rescaled Hubble parameter $h$.
In the second panel, $\eta_H$ is initially small and negative and it transitions to positive values close to $\eta_H\sim 3$ during the USR phase. 
All quantities are plotted as a function of the number of $e$-folds to the end of inflation. 
The vertical dashed lines indicate the epoch when the CMB pivot scale $k_{\rm ref} = 0.05/{\rm Mpc}$ crossed the Hubble sphere.
    }
\label{fig:USRpzeta}
\end{figure}

In Fig.~\ref{fig:USRpzeta} we show the background evolution of $x(N)$, $y(N)$ and $h(N)$ obtained in the benchmark scenario \eqref{eq:benchUSRvals}.
For convenience, we define $N_{\rm end}$ as the number of $e$-folds at the end of inflation.
In the same plot (top panels), we report the evolution of the Hubble SR parameters and the curvature power spectrum. 
For simplicity, we identify the momentum with the corresponding time of Hubble crossing $k = a(N)H(N)$. 
On top, we also indicate the momentum in units of $1/s$.
The CMB reference scale crosses the Hubble sphere $N-N_{\rm end}=-58$ $e$-folds before the end of inflation, which is indicated with a dashed vertical line. 
In the top panel, the curvature power spectrum is shown both using the SR approximation \eqref{eq:PzetaSR} (blue line),
which however fails to reproduce the spectrum around the peak where the USR phase takes place. We show with an orange line the full spectrum computed solving Eq.~\eqref{eq:MSequationadim}.
At $k = 0.05/{\rm Mpc}$ where the SR expressions~\eqref{eq:PzetaSR}--~\eqref{eq:r_slow} are valid, we find
$
{\cal P}_\zeta =  2.12 \times 10^{-9}
$,
$n_s =  0.952$, 
and 
$r =  0.00726$. These numbers are compatible 
with the latest observational bounds~\cite{Akrami:2018odb,BICEP:2021xfz}
\begin{equation} \label{eq:CMB_observables}
    {\cal P}_\zeta = (2.10 \pm 0.03)\times 10^{-9} , \quad
    n_s = 0.9649 \pm 0.0042, \quad
    r < 0.036 
\end{equation}
at the scale $k_{\rm ref} \equiv 0.05 / {\rm Mpc}$, reported here at $68 \%$ CL for $A_s$ and $n_s$, and at $95 \%$ for $r$.
The low value of $n_s$ is a rather common feature of models featuring an USR phase not sufficiently far from the region of the potential controlling the CMB scales~\cite{Kannike:2017bxn, Drees:2019xpp, Ballesteros:2017fsr, Madge:2023dxc}, as it is the case if one considers enhancements in the LISA band. 
This benchmark USR scenario leads to a power spectrum within the LISA band which can be fitted with a BPL template with parameters \eqref{eq:BPL_bench}.

Let us mention here that it was recently suggested that USR dynamics, 
in the extremal case of large spectral enhancement with $\Pz \sim {\cal O}(10^{-2})$ leading to PBH formation, may violate perturbativity and induce loop corrections that could also affect much longer modes associated with the scales observed through the CMB \cite{Kristiano:2022maq}.
While the existence and magnitude of this effect for soft modes is still under debate \cite{Inomata:2022yte,Cheng:2021lif,Riotto:2023hoz,Kristiano:2023scm,Riotto:2023gpm,Kristiano:2024vst,Franciolini:2023lgy, Davies:2023hhn,Firouzjahi:2023ahg, Iacconi:2023ggt,Motohashi:2023syh,Tasinato:2023ioq, Tasinato:2023ukp,Firouzjahi:2023aum,Fumagalli:2023hpa, Cheng:2023ikq,Tada:2023rgp,Firouzjahi:2023bkt,Braglia:2024zsl, Kawaguchi:2024lsw,Ballesteros:2024zdp,Inomata:2024lud,Fumagalli:2024jzz,Kawaguchi:2024rsv,Green:2024fsz},
recent analyses show that for realistic transitions into and out of USR perturbativity is retained (see e.g.~for analytical \cite{Riotto:2023gpm,Franciolini:2023lgy,Ballesteros:2024zdp} and lattice results \cite{Caravano:2024moy}). 
In this work, we restrict to adopting linear perturbation theory to compute the $\Pz$ at the LISA scales and leave further refinements including loop corrections to the spectrum of curvature perturbations for future work.

\paragraph{Non-Gaussianities.} As a final note, let us comment on the 
NGs
% non-Gaussianities (NGs)   ALREADY DEFINED
expected in this benchmark scenario. 
In this class of USR models, the peak in $\mathcal{P}_{\zeta}$ is generated during a brief phase of USR inflation, which transitions into constant-roll (i.e.~constant $\eta_H$) inflation afterwards~\cite{Atal:2018neu, Biagetti:2018pjj, Karam:2022nym}. 
In these scenarios, NGs are controlled by the details of the transition between USR and the subsequent phase, which is related to the UV tilt of the spectrum~\cite{Atal:2019cdz, Pi:2022ysn,Tomberg:2023kli,Franciolini:2023pbf,Ballesteros:2024pwn}. 
We can expect the non-linear curvature perturbation to take the form
\begin{align}
\zeta = -\frac{2}{\beta}\log\left(1-\frac{\beta}{2}\zeta_{\rm G}\right)
= \zeta_{\rm G} + \frac{3}{5} f_{\rm NL} \zeta_{\rm G}^2 + \dots ,
\end{align}
with  $f_{\rm NL} = 5 \beta/12 \simeq 0.092$ in our benchmark scenario \eqref{eq:BPL_bench}. Note, however, that this only takes into account the local generation of non-Gaussianity on super-Hubble scales by considering Gaussian inflaton fluctuations, while it has recently been shown using simulations that nonlinear interactions can also generate a large amount of non-Gaussianity intrinsic to the inflaton \cite{Caravano:2024moy}. Finally, let us mention that such small NG for perturbations of the typical amplitude considered here would lead to negligible contributions to the SIGW spectrum (see more discussion in Sec.~\ref{sec:NG-SGWB}), even beyond standard perturbation theory \cite{Iovino:2024sgs}. 

\section{Computation of the scalar-induced GW background}
\label{sec:computation}
Primordial scalar perturbations are frozen  %at the end of inflation and 
during their super-Hubble evolution; only when re-entering the Hubble radius after inflation ends do they start evolving in time. Moreover, even if scalar, vector and tensor modes are independent at the first order in perturbation theory, they couple when going to higher orders in fluctuations. For example, scalar modes source tensor modes and thus produce GWs.\footnote{The production of scalar modes from primordial tensor modes is discussed in \cite{Bari:2021xvf, Bari:2022grh} Moreover, very recently, SIGWs sourced by scalar-tensor perturbations have also been analyzed \cite{Bari:2023rcw, Picard:2023sbz}.}.

In the metric defined in Eq.~\eqref{eq:metric newtonian}, we ignore tensor perturbations generated at first order $h_{ij}^{(1)}$ and consider scalar perturbations that act as a source of second order tensor modes at  $h_{ij}^{(2)}$. The exact evolution of $h_{ij}^{(2)}$ can be obtained from the spatial part of the Einstein equations after applying the projection tensor $\mathcal T_{ij}^{lm}$, which selects the transverse-traceless component.
In the absence of anisotropic stress, at second order (note that we drop the superscript indicating the order in perturbation theory from now on) one obtains \cite{Acquaviva:2002ud, Mollerach:2003nq, Ananda:2006af, Baumann:2007zm}
\begin{equation}
h_{ij}''(\eta,{\bm x})
+2\mathcal H h_{ij}'(\eta,{\bm x})
-\nabla^2 h_{ij}(\eta,{\bm x})
=-4 \mathcal T_{ij}{}^{lm}\mathcal S_{lm}(\eta,{\bm x}),
\label{eq:eom GW}
\end{equation}
where $'$ is the derivative with respect to conformal time $\eta$, $\mathcal H=a'/a$ denotes the conformal Hubble parameter, and $\mathcal S_{ij}$ is the source term 
\begin{equation}
\mathcal S_{ij} (\eta,{\bm x}) = 4\Phi\partial_i\partial_j\Phi+2\partial_i\Phi\partial_j\Phi-\frac{4}{3(1+w)}\partial_i\left(\frac{\Phi'}{\mathcal H}+\Phi\right)\partial_j\left(\frac{\Phi'}{\mathcal H}+\Phi\right)\,.
\label{eq:source S}
\end{equation}
In the last equation $w$ is the equation-of-state parameter of the Universe, and the scalar perturbation $\Phi(\eta, {\bm x})$ can be related to the gauge-invariant comoving curvature perturbation $\zeta$. Solving Eq.~\eqref{eq:eom GW} in Fourier space, as we review in Sec.~\ref{sec:SIGW derivation}, we obtain the (time-averaged) dimensionless power spectrum $ \calP_h (\eta_f,k)$ of GWs at a time $\eta_f$ after the end of their production.
The fractional energy density of GWs per logarithmic interval in frequency is given by
\begin{equation}
\Omega_{\rm GW}(\eta_f, k)\equiv 
  \frac{\rho_{\rm GW}(\eta_f, k)}{\rho_{\rm c}(\eta_f)}
  = \frac{1}{24} \left (\frac{k}{\cal H(\eta)}\right )^2 \overline{\calP_h(\eta_f, k)} \,,
\label{eq:OmegaGW} 
\end{equation}
with
\begin{equation}
\rho_{\rm GW}(\eta) = \int \d \ln k \, \rho_{\rm GW}(\eta, k) \,,
\end{equation}
the bar denoting an oscillation average and $\rho_{\rm c}$ is the critical energy density.
GWs redshift as relativistic species and the current abundance of the GWB can be obtained by  accounting for entropy injections in the standard thermal history of the Universe:
\begin{equation}
\Omega_{\rm GW}(k)h^2 =
    \Omega_{r,0}h^2
    \left ( \frac{g_*(\eta_f)}{g_*^0} \right) \left( \frac{g_{*\textsc{s}}^0}{g_{*,\textsc{s}}(\eta_f)}\right)^{4/3} \Omega_{\rm GW}(\eta_f, k)\,.
\label{eq:OmegaGW2}
\end{equation}
Here $g_*$ ($g_{*,\textsc{s}}$) are the relativistic degrees of freedom in energy (entropy), $\Omega_{r,0}h^2  = 4.2 \cdot 10^{-5}$ is the current radiation density if the neutrino were massless \cite{Planck:2018vyg}. 
Assuming standard model degrees of freedom, one finds that \cite{Borsanyi:2016ksw}
\begin{equation}
    c_g(\eta_f)\equiv  \left ( \frac{g_*(\eta_f)}{g_*^0} \right) \left( \frac{g_{*\textsc{s}}^0}{g_{*,\textsc{s}}(\eta_f)}\right)^{4/3} \simeq 0.39,
\end{equation}
for $\eta_f$ of relevance for LISA.

\subsection{Source of GWs at second order in the scalar perturbations}
\label{sec:SIGW derivation}
The solution to \cref{eq:eom GW} can be easily found in Fourier space, where the equation for the GW amplitude $h$ for each polarization state $s$ becomes
\begin{equation}
    h''_{s}(\boldsymbol{k}, \eta) + 2\mathcal{H}h'_{s}(\boldsymbol{k}, \eta) + k^2h_{s}(\boldsymbol{k}, \eta) = 4\mathcal{S}_{s}(\boldsymbol{k}, \eta)
    \label{eq:eom_Fourier}
\end{equation}
and where $\mathcal{S}_{s}(\boldsymbol{k}, \eta)$ encloses the Fourier transform of the (projected) source given by
\begin{equation}
    \mathcal{S}_{s}(\boldsymbol{k}, \eta) 
    = \int\frac{d^3\boldsymbol{p}}{(2\pi)^{3}}Q_{s}(\boldsymbol{k},\boldsymbol{p})f(|\boldsymbol{k}-\boldsymbol{p}|,p,\eta)\zeta(\boldsymbol{p})\zeta(\boldsymbol{k}-\boldsymbol{p})\,.
    \label{eq:ftsource}
\end{equation}
In the latter equation, we introduced 
\begin{align}
    f(|\boldsymbol{k}-\boldsymbol{p}|,p,\eta) = & \frac{3(1+w)}{(5+3w)^2}\left[2(5+3w)T(|\boldsymbol{k}-\boldsymbol{p}|\eta)T(p\eta)
    + \frac{4}{\mathcal{H}^2}T'(|\boldsymbol{k}-\boldsymbol{p}|\eta)T'(p\eta) \right. \nonumber\\
    & \qquad \left. +\frac{4}{\mathcal{H}}\left(T(|\boldsymbol{k}-\boldsymbol{p}|\eta)T'(p\eta)+T'(|\boldsymbol{k}-\boldsymbol{p}|\eta)T(p\eta)\right)\right]
\end{align}
and 
we introduced the curvature perturbation transfer function $T(k\eta)$, and the spherical coordinates $(p,\theta,\phi)$ for the internal momentum $\vp$ and aligned the axes $(\hat x, \hat y, \hat z)$ with $(e_+(\vk), e_{\times}(\vk),\vk)$ (with $e_+, e_{\times}$ being the polarisation tensors for the GW) so that
\begin{equation}
    Q_{s}(\boldsymbol{k},\boldsymbol{p}) = e_{s}^{ij}(\hat{\boldsymbol{k}})p_ip_j = \frac{p^2}{\sqrt{2}}\sin^2\theta\times
    \begin{cases}
    \cos2\phi, \hspace{0.2cm} s = + \\
    \sin2\phi, \hspace{0.2cm} s = \times
    \end{cases}\,.
\end{equation}
A solution to the Fourier transform of the inhomogeneous equation of motion for $h_{ij}(\eta, k)$, Eq.~\eqref{eq:eom GW}, can be obtained using the Green's function  $G_{\vk}(\eta,\bar{\eta})$, that solves the homogeneous equation
\begin{equation}
\label{Green function equation}
G^{\prime\prime}_{\vk}(\eta,\bar{\eta})  + \left( k^{2} -\frac{a^{\prime\prime}}{a}\right)G_{\vk}(\eta,\bar{\eta}) = \delta\left(\eta-\bar{\eta}\right),
\end{equation}
with the boundary conditions $\lim_{\eta\to \bar{\eta}}G_{\vk}(\eta,\bar{\eta}) = 0$ and $ \lim_{\eta\to \bar{\eta}}G^{\prime}_{\vk}(\eta,\bar{\eta})=1$. 
The Green's function depends on $k=|\vk|$ by isotropy and can be expressed analytically in terms of Bessel functions as
\begin{equation}\label{Green function analytical}
k\,G_{\vk}(\eta,\bar{\eta}) = \sqrt{x\bar{x}}\left[y_\mathrm{\nu}(x)j_\mathrm{\nu}(\bar{x})  - j_\mathrm{\nu}(x)y_\mathrm{\nu}(\bar{x})\right],
\end{equation}
where $x=k\eta, \bar x=k\bar \eta$,  and $\nu = {3(1-w)}/[{2(1+3w)}]$ and $j_\nu$ and $y_\nu$ are respectively the spherical Bessel function of the first and second kind.
 For example, during RD, $k G_k(\eta,\bar\eta)=\sin(x-\bar x) \Theta (\eta-\bar\eta)$, where $\Theta$ is the Heaviside function.

The amplitude of the tensor modes can then be written as
\begin{align}
    h_{s}(\boldsymbol{k}, \eta) = &{4}\int_{\eta_i}^{\eta}d\Bar{\eta}\, G_{\boldsymbol{k}}(\eta,\Bar{\eta})\frac{a(\Bar{\eta})}{a(\eta)}\mathcal{S}_{s}(\boldsymbol{k},\Bar{\eta})\nonumber\\
    = &4\int_{\eta_i}^{\eta}d\Bar{\eta} \, G_{\boldsymbol{k}}(\eta,\Bar{\eta})\frac{a(\Bar{\eta})}{a(\eta)}\int\frac{d^3\boldsymbol{p}}{(2\pi)^{3}}Q_{s}(\boldsymbol{k},\boldsymbol{p})f(|\boldsymbol{k}-\boldsymbol{p}|,p,\Bar{\eta})\zeta(\boldsymbol{p})\zeta(\boldsymbol{k}-\boldsymbol{p})
    \,,
    \label{solution-h}
\end{align}
with $s$ indicating the polarisation and $\eta_{i}$ the initial emission time. The GW two-point function, needed to obtain the energy density of GWs as a function of the scalar power spectrum, is defined in terms of the GW power spectrum
\begin{equation}
\label{eq: def P_h}
\left\langle h^r(\eta,\vk_1) h^s(\eta,\vk_2)\right\rangle \equiv (2\pi)^3 \delta(\vk_1+\vk_2)\, \delta^{rs}\,\frac{2\pi^2}{k_1^3} \Ph(k_1) \,.
\end{equation}
After substituting Eq.~\eqref{Green function analytical} into Eq.~\eqref{solution-h}, the final expression for the second-order induced tensor power spectrum reads~\cite{Espinosa:2018eve,Kohri:2018awv} 
\begin{multline}
    \langle h^{s_{1}}(\eta,\boldsymbol{k}_1)h^{s_{2}}(\eta,\boldsymbol{k}_2)\rangle \hspace{0.1cm}
    = \hspace{0.1cm}16\int\frac{d^3\boldsymbol{p}_1}{(2\pi)^{3}}\frac{d^3\boldsymbol{p}_2}{(2\pi)^{3}}Q_{s_{1}}(\boldsymbol{k}_1,\boldsymbol{p}_1)Q_{s_{2}}(\boldsymbol{k}_2,\boldsymbol{p}_2)\\ 
    \times I(|\boldsymbol{k}_1-\boldsymbol{p}_1|,p_1,\eta_1) I(|\boldsymbol{k}_2-\boldsymbol{p}_2|,p_2,\eta_2)\langle \zeta_{\boldsymbol{p}_1}\zeta_{\boldsymbol{k}_1-\boldsymbol{p}_1}\zeta_{\boldsymbol{p}_2}\zeta_{\boldsymbol{k}_2-\boldsymbol{p}_2}\rangle\,.
    \label{general}
\end{multline}
In the Gaussian case, where only the disconnected part of the trispectrum survives (hence the ``d'' in the following equation, we discuss in Sec.~\ref{sec:NG-SGWB} the impact of primordial NG), one obtains
\begin{equation}
\overline{\mathcal{P}_{h,\text{d}} (\eta, k)} =  4 
\int_0^\infty \text{d}v \int_{|1-v|}^{1+v}\text{d} u \left [ \frac{4v^2-(1+v^2-u^2)^2}{4uv} \right ]^2 \overline{I^2 (u,v,k, \eta)} \mathcal{P}_\zeta ( k v ) \mathcal{P}_\zeta ( k u ),
\label{P_h_uv}
\end{equation}
where we introduced the dimensionless variables
\begin{equation}
v\equiv \frac pk , \qquad u\equiv \frac{|\vk-\vp|}{k}\,.
\label{def-u-v}
\end{equation}
The overline stands for an oscillation average, and the kernel function $I(u,v,\eta)$ is defined in terms of Green's function as
\begin{equation}
    I(|\boldsymbol{k}-\boldsymbol{p}|,p,\eta) = \int_{\eta_i}^{\eta}d\Bar{\eta}\,G_{\boldsymbol{k}}(\eta,\Bar{\eta})\frac{a(\Bar{\eta})}{a(\eta)}f(|\boldsymbol{k}-\boldsymbol{p}|,p,\Bar{\eta})\,.
    \label{eq:kernelfunc}
\end{equation}
Since the integration domain of \eqref{P_h_uv} is rectangular, for computational purposes, it is convenient to rotate the coordinates into
\begin{equation}\label{eq:t-s}
t\equiv u+v-1 , \qquad s\equiv u - v\,
\end{equation}
and the SIGW spectrum $\overline{\mathcal{P}_h (\eta, k)}$ can be then rewritten as  
\begin{multline}
\overline{\mathcal{P}_h (\eta, k)} = 4 
\int_0^\infty \text{d}t \int_{0}^{1}\text{d} s \left [ \frac{t(2+t)(1-s^2)}{
(1-s+t)(1+s+t)
} \right ]^2 \overline{I^2 (t,s,k,\eta)} \\
\qquad\times \mathcal{P}_\zeta \left(k\, 
\frac{t+s+1}{2} \right) \mathcal{P}_\zeta \left(k\, 
\frac{t-s+1}{2} \right) , 
\label{eq:P_h_ts}
\end{multline}
where the integration in $s$ is restricted to positive values due to the integrand being an even function of $s$. 
The integration kernel $I(t,s,\eta)$ contains information about the time evolution of the source during emission, as well as the propagation of the emitted GWs after emission, and thus depends on the thermal history when the relevant modes re-entered the Hubble radius. Let us consider different assumptions on the thermal history in the following sections.

The generation of tensor modes at second-order in perturbation theory raises concerns about the potential gauge dependence of results commonly calculated in the Newtonian gauge. Unlike first-order tensor modes, which are gauge invariant, second-order tensor modes are gauge dependent \cite{Malik:2008im}.
During the phase when the source is still active, the result is expected to remain gauge-dependent, as one cannot directly identify the tensor mode with the freely propagating GW. However, when the source becomes inactive after the GWs are produced, it effectively decouples from the GWs.
This happens for example during the radiation-dominated era of the Universe and in the other cases considered in this draft.
Therefore, in the late-time limit well inside the cosmological horizon, tensor mode behaves as linear metric perturbations and the initial gauge dependence no longer affects the final result \citep{DeLuca:2019ufz,Inomata:2019yww,Yuan:2019fwv,Domenech:2020xin}, ensuring that the spectra computed in this work are unaffected by this issue.

\subsection{Radiation-dominated era}
\label{sec_RDE}

If the emission takes place in a RD universe, the kernel function in the deep sub-horizon regime $k\eta \to \infty$ takes the form\,\cite{Espinosa:2018eve,Kohri:2018awv,Garcia-Saenz:2022tzu} 
\begin{align}
&\overline{I_{\text{RD}}^2(t, s, k\eta \to \infty)} =\frac{1}{2 (k \eta)^2}
I_A^2(u,v)\left[I_B^2(u,v)+I_C^2(u,v)\right] 
\label{I_RD_osc_ave_ts}
\end{align}
where
\begin{equation}
\begin{aligned}
I_A(u,v)&= \frac{3(u^2+v^2-3)}{4u^3v^3} \,,\\
I_B(u,v)&= -4uv+(u^2+v^2-3)\log\left|\frac{3-(u+v)^2}{3-(u-v)^2}\right| \,,\\
I_C(u,v)&= \pi(u^2+v^2-3)\Theta(u+v-\sqrt{3}) \,,
\label{IABC-functions}
\end{aligned}
\end{equation}
$u$ and $v$ have been introduced above and again 
$\Theta(x)$ is the Heaviside function. Notice that the unphysical divergence obtained in the limit 
$|\vk-\vp|+p=\sqrt{3}k$, which is also retained in the spectrum produced by monochromatic scalar perturbations \cite{Ananda:2006af}.
The factor of $\sqrt{3}$ originates from the (inverse) sound speed in RD appearing in the transfer function, and in the limit $|\vk-\vp|+p=\sqrt{3}k$ the contributions from some of the source terms add up constructively and build up logarithmically over time \cite{Espinosa:2018eve}. Assuming RD up to $\eta\to \infty$ leads to this logarithmic divergence, which is regularized in the integral for $\Ph$ if $\Pz(k)$ is smooth enough (e.g.~has a nonzero width), or if the emission time is finite.
We do not introduce this regulator here, as it is numerically irrelevant for spectra with finite width \cite{LISACosmologyWorkingGroup:2024hsc}.

\subsection{Transition from an early matter-dominated to the radiation-dominated era}
\label{sec_tran}

We further consider the alternative thermal history wherein an early matter-dominated (eMD) era may precede the RD era. 
We follow the prescription of \cite{Inomata:2019ivs}, assuming a sudden transition at a conformal time $\eta=\eta_{\mathrm{R}}$, where the subscript $R$ indicates the reheating time \cite{Kohri:2018awv}. In this scenario, the dominant contribution comes from GWs induced during the RD era by the scalar perturbations that have entered the horizon during an eMD era. 

Interestingly, in this case, one observes a resonantly enhanced production of GWs. In particular, during the transition, the time derivative of the Bardeen potential $\Phi$, which is the source of the SIGW signal, goes very quickly from $\Phi^\prime = 0$ (since in an eMD era $\Phi$ = constant) to $\Phi^\prime\neq 0$ (see~\cite{Inomata:2019yww,Domenech:2020ssp} for more details), leading to an enhanced secondary tensor mode production sourced mainly by the $\mathcal{H}^2\Phi^{\prime 2}$ term in Eq.~\eqref{eq:source S}. In a more physical scenario, the transition happens more gradually \cite{Inomata:2019zqy,Kumar:2024hsi,Pearce:2023kxp}.

For the kernel function $I(t,s,k,\eta)$, one finds two dominant contributions at the onset of the late RD era, i.e.~at $\eta = \eta_\mathrm{R}$, when most of the GWs are expected to be produced~\cite{Inomata:2019yww,Inomata:2020lmk}. The first contribution to $I(t,s,k,\eta)$ is given by $k_{\max} / k \sim 1$, at $t_0 =\sqrt{3}-1$ ~\cite{Inomata:2019ivs}
\begin{equation}\label{eq:irdapp}
\overline{I^2_{\rm lRD,{\rm res}}}(t_0,s,k,\eta_\mathrm{R}) \simeq Y \frac{9\left(-5 + s^2+2t+t^2\right)^4x_{\rm R}^8}{2^{17} 5^4(1-s+t)^2(1+s+t)^2}{\rm Ci}^2\left(y\right),
%81920000
\end{equation}
where $x_{\rm R} = k\eta_\mathrm{R} $. The variable $y$ is defined as $y \equiv \frac{|t+1-c_s^{-1}|x_{\rm R}}{2c_s^{-1}} =  \frac{|t+1-\sqrt{3}|x_{\rm R}}{2\sqrt{3}} $, and $Y$ is a fudge factor to absorb the uncertainty in the integration boundary, set here to be $2.3$ as in \cite{Inomata:2019ivs}.
At $t_0=\sqrt{3}-1$, the logarithmic singularity of the function $\mathrm{Ci}$ is reached, giving rise to the peak in the spectrum.

The second contribution to $I(t,s,k,\eta)$ comes from the wave-numbers satisfying $k_{\max} / k \gg 1$, hence the integrations are dominated by the large $t$ region $u \sim v \sim t \gg 1$.
Therefore, the dependence on $s$ is lost. 
Setting $s=0$, the kernel function reads
\begin{equation}
\label{eq:irdapp2}
\overline{I^2_{\rm lRD,{\rm LV}}}( t \gg 1, s, k, \eta_\mathrm{R})\simeq \frac{9t^4x_{\rm R}^8}{%81920000
2^{17}5^4}\Big[4{\rm Ci}^2(x_{\rm R}/2)+\big(\pi - 2{\rm Si}(x_{\rm R}/2)\big)^2\Big]\,.
\end{equation}%
The integration region is
\begin{equation}
0 \leq s \leq 1 \quad \text{and} \quad 0 \leq t \leq-s+2 \frac{k_{\max}}{k}-1 \qquad \text{for} \quad k \leq k_{\max},
\end{equation}
and the result obtained from this integration region is then doubled to account for $s \rightarrow-s$.
The two contributions are computed separately and added to give 
\begin{equation}
\Omega_{\mathrm{GW}}^{\rm eMDRD} 
\simeq 
\Omega_{\mathrm{GW}}^{(\mathrm{LV})}
+
\Omega_{\mathrm{GW}}^{(\mathrm{res})}.    
\end{equation}

\subsection{Computation of SIGW with the binned spectrum approach}\label{sec:OmegaGWijk}

In this section, we discuss in more detail the procedure sketched in 
Sec.~\ref{sec:binnedPzeta}  for computing the SIGW in terms of a template-free approach to the curvature power spectrum. We express the power spectrum as a sum over momentum bins, as in Eq.~\eqref{eq:PzetaBinning}. The specific profile for the power spectrum ${\cal P}_\zeta$ is then associated with a vector of coefficients $A_i$. By plugging Eq.~\eqref{eq:PzetaBinning} into Eqs.~\eqref{eq:OmegaGW} and \eqref{eq:P_h_ts}, we recast the SIGW density into the sum
\begin{equation}
\label{eq:OmegaMatrix}
    \Omega_{\rm GW} \left(k\right)
    =
    \sum_{i,j}^{N -1} \Omega_{\rm GW}^{\left(i,j\right)} \left(k\right)
    A_i A_j 
\end{equation}
performed over the momentum bins. The kernel for this sum is the matrix
\begin{align}
&\Omega_{\rm GW}^{(i,j)} (k)
=
\frac{1}{12}
\left( 
\frac{k}{a H}
\right)^2
\int_0^\infty dt \,
\int_0^1 d s
\,\left [ \frac{t(2+t)(s^2-1)}{\left(1-s+t\right)\left(1+s+t\right)} \right ]^2 \overline{I^2 \left(t,s,k,\eta\right)}
\nonumber
\\
&\times
\Theta \left(k\, v \left(s, t \right)-p_i \right)\,\Theta \left(p_{i+1}-k \,v \left(s, t\right)\right)\,\Theta \left(k\, u \left(s, t\right)-p_j\right)\Theta \left(p_{j+1}-k \,u \left(s, t \right) \right)
\,,
\label{eq:matroij}
\end{align}
where $p_i, p_j$ are the boundaries of the momentum bins 
entering in Eq.~\eqref{eq:PzetaBinning}.

Importantly, we stress that the matrix 
$\Omega_{\rm GW}^{(i,j)}$ of Eq.~\eqref{eq:OmegaMatrix} 
 is {\it independent} of the specific scalar spectrum considered, and the information on ${\cal P}_{\zeta}$ 
is stored only in
the coefficients $A_i$
appearing quadratically in Eq.~\eqref{eq:OmegaMatrix}. This implies that the computation
of Eq.~\eqref{eq:matroij} depends only on the kernel function, and its entries can be computed once for all for any given cosmology: for example,  we can use one of the kernels discussed in Sec.~\ref{sec_RDE} or Sec.~\ref{sec_tran}. 
The nested integrals appearing in Eq.~\eqref{eq:matroij} should then be performed a single time for each kernel. 
Once $\Omega_{\rm GW}^{(i,j)}$ is determined, it can be used to swiftly compute the resulting 
$\Omega_{\rm GW}$ for {\it any}  ${\cal P}_\zeta$, by  means of the contractions in Eq.~\eqref{eq:OmegaMatrix}. In fact, with this method, we  reduce  the problem of 
computing $\Omega_{\rm GW}$ to perform the 
simple sum of  Eq.~\eqref{eq:OmegaMatrix}.

This approach
is useful in scenarios where the underlying shape of the curvature spectrum is not accurately known, for example, due to the presence of peaks or breaks, whose position depends on the underlying physics we wish to probe. In fact, the method allows us to scan over different sets of  $A_i$ components, swiftly computing the SIGW frequency profile, which can then be compared with data. Other approaches for reconstructing the properties of the underlying ${\cal P}_\zeta$ from  SIGW data can be found for example in \cite{Yi:2023tdk,Kuhnel:2024zpv,Zeng:2024ovg}.

We tabulate the matrix \eqref{eq:OmegaMatrix} assuming a varying number of bins 
$N$ in the range of relevance for LISA, which is $k\in [1.26 \times 10^{-4},6.28]/{\rm s}$ both for the internal $(i,j)$ indices, as well as external momentum $k$. This range is chosen to match the one used by the \texttt{SGWBinner} code adopted to perform the LISA forecasts. See the discussion in Sec.~\ref{sec:tempbinnerpipe}.

\subsection{Non-Gaussian imprints on the SIGW spectrum}
\label{sec:NG-SGWB}

From the solution of the SIGW, Eq. \eqref{solution-h}, one can relate the tensor power spectrum to the four-point correlation function of the curvature perturbation, see Eq.~\eqref{general}. As anticipated above, the latter can be decomposed into disconnected and connected contributions, where the connected part vanishes when primordial fluctuations are drawn
from a Gaussian distribution. The disconnected contribution gives rise to Eq. \eqref{P_h_uv} that can be solely expressed in terms of the scalar power spectrum ${\cal P}_\zeta(k)$. However, the connected contribution depends on the primordial trispectrum
 $\langle \zeta_{\bm{k}_1} \zeta_{\bm{k}_2} \zeta_{\bm{k}_3} \zeta_{\bm{k}_4} \rangle_{\text{c}}'= T_{\zeta}(\bm{k}_1,\bm{k}_2,\bm{k}_3,\bm{k}_4)$, whose corresponding tensor power spectrum reads \cite{Garcia-Saenz:2022tzu}\footnote{In \cite{Garcia-Saenz:2022tzu}, which uses different conventions, their Eq.~(2.6) should be divided by $4$, as well as subsequent results. This has been corrected in \cite{Garcia-Saenz:2023zue} in a study of SIGWs including parity violation.}
\begin{align}
    \overline{\mathcal{P}_{h, \text{c}}} 
    &=\frac{1}{4 \pi}\int_0^{\infty}\d v_1 \int_{|1-v_1|}^{1+v_1} \d u_1\int_0^{\infty}\d v_2 \int_{|1-v_2|}^{1+v_2} \d u_2\int_0^{2\pi}\d\psi\; 
   \nonumber  \\
    &\times \frac{\cos(2\psi)}{(u_1v_1u_2v_2)^{5/4}}
    [4v_1^2-(1+v_1^2-u_1^2)^2]
    [4v_2^2-(1+v_2^2-u_2^2)^2] 
    \nonumber \\
    &\times 
\overline{I(u_1,v_1,k,\eta)I(u_2,v_2,k,\eta)}\mathcal{T}_\zeta(u_1,v_1,u_2,v_2,\psi) \,,
\label{eq:connected-contribution}
\end{align}
and we use the variables $u_i$ and $v_i$ defined above. The dimensionless trispectrum function $\mathcal{T}_\zeta$ is defined as
\begin{equation}
\mathcal{T}_\zeta(\bm{k}_1,\bm{k}_2,\bm{k}_3,\bm{k}_4)=\frac{(k_1k_2k_3k_4)^{9/4}}{(2\pi)^6}T_\zeta(\bm{k}_1,\bm{k}_2,\bm{k}_3,\bm{k}_4)\,,
\end{equation}
and is evaluated at $\bm{k}_1=\bm{p}_1, \bm{k}_2=\bm{k}-\bm{p}_1$, $\bm{k}_3=-\bm{p}_2$, $\bm{k}_4=-\bm{k}+\bm{p}_2$, with $\psi=\phi_1-\phi_2$ the difference between the azimuthal angles of $\bm{p}_1$ and  $\bm{p}_2$ with respect to $\bm{k}$. The integration kernel for emission during radiation domination is given by
\begin{align}
      \overline{I(u_1,v_1,k, \eta)I(u_2,v_2,k, \eta)}&=
      \frac{1}{2(k \eta)^2} I_A(u_1,v_1)I_A(u_2,v_2)
      \nonumber \\
      & \times \left[I_B(u_1,v_1)I_B(u_2,v_2)+I_C(u_1,v_1)I_C(u_2,v_2)\right]   \,,
      \label{eq:kernel}
\end{align}
in terms of $I_{A,B,C}$ defined in Eq.~\eqref{IABC-functions} and with $x=k\eta$. 
As for the disconnected contribution, it is numerically convenient to change the integration variables from $(u_i,v_i)$ to $(t_i,s_i)$, with 
\begin{equation}
    \int_0^{\infty}\d v_i \int_{|1-v_i|}^{1+v_i} \d u_i (\ldots)=\frac12 \int_0^\infty \d t_i \int_{-1}^{1} \d s_i (\ldots)\,,
\end{equation}
where we did not assume symmetry between positive and negative $s$, to retain full generality in this case.

On general grounds,  the properties of the trispectrum and the symmetries of the kernel Eq.~\eqref{eq:kernel} enable one to split the connected contribution \eqref{eq:connected-contribution} into three inequivalent channels \cite{Garcia-Saenz:2022tzu}. Hence, any trispectrum function of the unordered set $\{\bm{k}_1,\bm{k}_2,\bm{k}_3,\bm{k}_4\}$, can be written as
\begin{equation}
    T_\zeta\left[\{\bm{k}_1,\bm{k}_2,\bm{k}_3,\bm{k}_4\}\right]=\tilde{T}_\zeta\left[(\bm{k}_1,\bm{k}_2,\bm{k}_3,\bm{k}_4)\right] + \,\, \text{23 perm.} \,,
\end{equation}
where the individual contributions $\tilde{T}_\zeta$ are not in general invariant under permutations of their arguments. Moreover, it can be conveniently written as
\begin{equation}
\label{eq: def channels}
    T_\zeta = \left( T_{\s} + T_{\tc} +T_{\uc} \right)  + \,\, \text{7 perm.}  \,, \,\,\text{with}  \begin{cases}
      T_{\s} = \tilde{T}_\zeta\left[(\bm{k}_1,\bm{k}_2;\bm{k}_3,\bm{k}_4)\right] \\
      T_{\tc} = \tilde{T}_\zeta\left[(\bm{k}_1,\bm{k}_3;\bm{k}_2,\bm{k}_4)\right] \\
      T_{\uc} = \tilde{T}_\zeta\left[(\bm{k}_1,\bm{k}_4;\bm{k}_2,\bm{k}_3)\right] 
    \end{cases}
    \,,
\end{equation}
where the channels $\s,\tc$ and $\uc$ correspond to the three unordered pairs $\{\{\bm{k}_1,\bm{k}_2\},\{\bm{k}_3,\bm{k}_4\}\}$, $\{\{\bm{k}_1,\bm{k}_3\},\{\bm{k}_2,\bm{k}_4\}\}$ and $\{\{\bm{k}_1,\bm{k}_4\},\{\bm{k}_2,\bm{k}_3\}\}$ together with their ``exchanged momenta'' $\s=|\bm{k}_1+\bm{k}_2|$, $\tc=|\bm{k}_1+\bm{k}_3|$ and $\uc=|\bm{k}_1+\bm{k}_4|$ respectively,  and where the seven permutations in \eqref{eq: def channels} preserves the exchanged momentum of each channel. 
Furthermore one can show that the trispectrum-induced GW spectrum \eqref{eq:connected-contribution}
can be written in terms of only three fundamental contributions corresponding to the seeds $T_\s$, $T_\tc$ and $T_\uc$:
\begin{equation}
\label{P-3-channels}
\overline{\mathcal{P}_{h, \text{c}}}=8 (\overline{\mathcal{P}_{h, \text{c}}^{\s}}+\overline{\mathcal{P}_{h, \text{c}}^{\tc}}+\overline{\mathcal{P}_{h, \text{c}}^{\uc}})\,,    
\end{equation}
with $\overline{\mathcal{P}_{h, \text{c}}^{\s}}$ simply corresponding to $\overline{\mathcal{P}_{h, \text{c}}}$ with $T_\zeta$ replaced by $T_{\s}$, etc.

Computing from first principles the trispectrum generated in models relevant for GW astronomy is a difficult task, as these scenarios often involve a strong breaking of scale invariance as well as enhanced fluctuations that can jeopardize perturbative computations, see e.g.~\cite{Fumagalli:2020nvq,Inomata:2021zel,Inomata:2021tpx,Fumagalli:2021mpc,Inomata:2022yte,Unal:2023srk,Iacconi:2023slv,Caravano:2024tlp,Caravano:2024moy}. In the following, we assume that the trispectrum is of the local $\tauNL$ type:
\begin{align}
\label{eq:tauNL-shape}
    T_\zeta^\mathrm{loc} &= \tauNL \left[\left(P_{\zeta}(\s) P_{\zeta}(k_{1}) P_{\zeta}(k_{3}) 
    + \,\, \text{3 perm.} \right)  + \,\,  (\s \leftrightarrow \tc)  +  (\s \leftrightarrow \uc)  \right]\,.
\end{align}
We give more details on the computation of the spectrum in the presence of local NGs in App.~\ref{App:NG_tec}.

Local NGs, typical of multi-field models, generically arise from the non-linear evolution of cosmological fluctuations on super-Hubble scales (see e.g.~\cite{Wands:2010af} for a review). Besides the $\tauNL$ type, which emerges microscopically from the exchange of scalar particles through cubic interactions, the local trispectrum also acquires in general a $g_\mathrm{NL}$ component, coming from contact quartic interactions. However, its momentum dependence is such that it does not contribute to the GW spectrum, and hence we can disregard it for our purpose. We stress that our choice, Eq.  \eqref{eq:tauNL-shape}, is a first methodological step motivated by simplicity, in particular because the momentum dependence of the trispectrum is fully characterized by the one of the power spectrum $P_{\zeta}(k)$, which is not the case in general (see \cite{Garcia-Saenz:2022tzu} for a study of the impact of various trispectrum shapes on the GW spectrum). 
Let us also highlight a conceptual aspect. Several works in the literature consider a local ansatz in which the real-space curvature perturbation $\zeta(\bm{x})$ is expanded in powers of a Gaussian variable $\zeta_{\textrm{G}}(\bm{x})$ as
\begin{align}
\zeta = \zeta_{\rm G} + \frac{3}{5} f_{\rm NL} \zeta_{\rm G}^2+\ldots \,,
\label{eq:zeta-local-anzatz}
\end{align} 
and compare the corresponding GW spectrum with the one obtained by keeping only the first term, fully characterized by the power spectrum $P_{\zeta_\textrm{G}}$, see e.g.~\cite{Cai:2018dig,Unal:2018yaa,Yuan:2020iwf,Atal:2021jyo,Adshead:2021hnm,Ragavendra:2021qdu,Abe:2022xur,Yuan:2023ofl,Tasinato:2023ioq,Perna:2024ehx,Zeng:2024ovg,Ruiz:2024weh,Inui:2024fgk}. At leading order, such an expansion does lead to the trispectrum \eqref{eq:tauNL-shape} with $P_{\zeta}$ replaced by 
$P_{\zeta_\textrm{G}}$, and $\tauNL=\left(6 f_\mathrm{NL}/5\right)^2$. However, the nonlinear terms in Eq. \eqref{eq:zeta-local-anzatz} also imply that the curvature power spectrum does not coincide with $P_{\zeta_\textrm{G}}$. Instead, keeping only the quadratic term shown in Eq.~\eqref{eq:zeta-local-anzatz} for definiteness, one finds the power spectrum
\begin{align}
P_{\zeta}(k)=P_{\zeta_\textrm{G}}(k)+\frac12 \left( \frac{6}{5} f_{\rm NL}\right)^2 \int \frac{\mathrm{d}^3\bm{p}}{(2\pi)^3} P_{\zeta_\textrm{G}}(p)P_{\zeta_\textrm{G}}(|\bm{k}-\bm{p}|) \,.
\label{P-PG}
\end{align}
 Hence, as described in \cite{Garcia-Saenz:2022tzu}, in this approach, one considers on similar grounds the impact of primordial NG on the SIGW, through the trispectrum, and the difference between a putative $P_{\zeta_\textrm{G}}$ to which we have no access, and the power spectrum of $\zeta$ which is anyway the only observable quantity. Again, we emphasize that the effects of non-linearities on the SIGW spectrum may not always be fully captured by the local ansatz, Eq. \eqref{eq:zeta-local-anzatz}. As a result, the predicted amplitude of the resulting SGWB could differ significantly \cite{Iovino:2024sgs}, potentially being suppressed or enhanced by several orders of magnitude.
 
 In our analysis, whose results are shown in Sec.~\ref{sec:NG}, we find it conceptually clearer to take as a benchmark the disconnected prediction from the purely Gaussian theory \eqref{P_h_uv} with a given power spectrum ${\cal P}_\zeta(k)$, which will take to be of the log-normal form \eqref{eq:PLN}, and to compare it with the addition of the non-Gaussian, connected, contribution, Eq. \eqref{eq:connected-contribution} with trispectrum \eqref{eq:tauNL-shape}. Note that for the latter, the $\s$-channel contribution vanishes as the corresponding ${\cal T}_\zeta$ in Eq.~\eqref{eq:connected-contribution} does not depend on the azimuthal angle $\psi$. We are thus left with the two $\tc$ and $\uc$ contributions in Eq. \eqref{P-3-channels}. Overall, we stress that the parameter to be constrained from observations is $\tauNL$, which measures the non-Gaussian contribution to the SIGW spectrum coming from the trispectrum of curvature perturbations. On scales relevant to LISA, there is \textit{a priori} no constraint on $\tauNL$ except that it is positive in known concrete realizations of inflation. Its size is also \textit{a priori} arbitrary, although from a theoretical perspective, perturbative control during inflation typically implies $\tauNL \calP_\zeta<1$.

\section{Mock signal reconstructions with the \texttt{SGWBinner} and \SIGWAY\  codes}
\label{sec:tempbinnerpipe}

This section outlines the analysis method employed in this work.  
Before presenting the LISA data model adopted in our analysis (Sec.~\ref{sec:data}) and functionalities of the code (Sec.~\ref{sec:data_analysis}), let us briefly illustrate the measurement of GWs with LISA.

The observatory will consist of three satellites ($\alpha = 1,2,3$) that orbit at the vertices of an approximately equilateral triangle with sides about 2.5 million kilometers long. Each satellite contains two Test Masses (TMs), whose positions are constantly monitored, and two lasers emitting toward the other satellites. 
By monitoring the fractional Doppler frequency shifts of photons traveling along the arms between satellites, LISA measures the relative displacements of the TMs.
The path connecting two satellites is typically dubbed ``link'' and the single link measurement can be denoted as $\eta_{\alpha\beta}(t)$, where the laser emitted from the satellite $\beta$ at time $t - L_{\alpha \beta}/c$ %and 
is recorded at time $t$ in the satellite $\alpha$.
These measurements are, however, dominated by laser frequency noise, which is expected to be several orders of magnitude greater than the required sensitivity~\cite{LISA:2017pwj}.
To suppress this noise contribution, LISA will employ a post-processing technique called Time-Delay Interferometry (TDI)~\cite{Armstrong_1999,Prince:2002hp,Shaddock:2003bc,Shaddock:2003dj,Tinto:2003vj,Vallisneri:2005ji,Muratore:2020mdf,Tinto:2020fcc,Muratore:2021uqj}. 
In practice, TDI can be understood as the operation of $3\times6$ matrix on the six single link measurements $\eta_{\alpha\beta}(t)$~\cite{Baghi:2023qnq,Hartwig:2023pft} that returns the three TDI channels where the laser frequency noise is strongly suppressed.

As in the previous studies using the \texttt{SGWBinner} code~\cite{Caprini:2019pxz,Flauger:2020qyi,Caprini:2024hue,Blanco-Pillado:2024aca,LISACosmologyWorkingGroup:2024hsc}, in this work we assume for simplicity i) equal and static arm lengths and ii) equality of noise at each link. While, in reality, these hypotheses will not be perfectly satisfied\footnote{
With realistic orbits, LISA will not be perfectly equilateral and arm-lengths vary at the percent level~\cite{Martens:2021phh} (see also Appendix A of~\cite{Mentasti:2023uyi}).}, it has been shown that the signal reconstruction is almost unaffected by unequal (but static) arm length and unequal noise amplitudes~\cite{Hartwig:2023pft, Kume:2024sbu}.
Under the equal and static arm length assumption, the so-called first-generation TDI variables suffice to achieve laser noise cancellation.\footnote{To account for non-static arm lengths and the associated Doppler shifts, the second-generation TDI variables~\cite{Vallisneri:2005ji,Muratore:2020mdf,Muratore:2021uqj,Hartwig:2021mzw} would be required.} In the $\left\{\rm X,Y,Z\right\}$ basis, they are expressed as
\begin{equation}
    {\rm X} \equiv (1 - D_{13}D_{31})(\eta_{12} + D_{12}\eta_{21}) + (D_{12}D_{21} - 1)(\eta_{13} + D_{13}\eta_{31})\,,
\end{equation}
with ${\rm Y}$ and ${\rm Z}$ being cyclic permutations of ${\rm X}$.
Here $D_{\alpha\beta}$ is the delay operator acting on any time-dependent function $x(t)$ as $D_{\alpha\beta} \, x(t) = x(t - L_{\alpha\beta})$ and we take $L_{\alpha\beta}=L=2.5 \times 10^9\,{\rm m}$. 
For SGWB signal searches, it is convenient to combine the XYZ variables to obtain the so-called AET basis~\cite{Hogan:2001jn,Adams:2010vc}, defined as
\begin{equation}
\label{eq:XYZ_to_AET}
    \rm{A} \equiv \frac{{\rm Z-X}}{\sqrt{2}}, \qquad
    \rm{E} \equiv \frac{{\rm X - 2Y + Z}}{\sqrt{6}}, \qquad
    \rm{T} \equiv \frac{{\rm X + Y + Z}}{\sqrt{3}} \; , 
\end{equation}
which, in the limit of equal arms and equal noises, can be shown to have vanishing cross-correlations and simplify the likelihood computation. 
Moreover, due to its symmetric structure, the T channel strongly suppresses GW signals at small frequencies compared to instrumental noise. For this reason, the T channel can be treated as a quasi-null channel that is mostly sensitive to instrumental noise.\footnote{The T channel does not remain a null channel in general with unequal and flexing arms and differing noise levels in the different spacecraft, although other quasi-null channels are available \cite{Muratore:2022nbh}.}

\subsection{Data streams from LISA TDI channels}\label{sec:data}
We represent the three time-domain data streams as $d_i(t)$, where $i$ runs over the channels of the TDI basis. These quantities are real-valued functions defined on the interval $\left[ -\tau/2, \tau/2\right]$ with $\tau$ being the duration of a data segment.
The Fourier transforms of these data streams are then given by
\begin{equation}
\tilde{d}_i\left(f\right) =  \int_{-\tau/2}^{
\tau/2} {\rm d} t  \;\textrm{e}^{ 2\pi i f t} d_i \left(t\right)\,.
\end{equation}
Our central assumption is that all transients including loud deterministic signals and glitches in the noise are subtracted from the time stream through some appropriate methods within the LISA global fit scheme~\cite{Cornish:2005qw,Vallisneri:2008ye,MockLISADataChallengeTaskForce:2009wir,Littenberg:2023xpl, Strub:2024kbe, Katz:2024oqg}.\footnote{See Ref.~\cite{Alvey:2023npw} for the application of simulation-based inference to the SGWB search performed by LISA in the presence of transient signals, which goes beyond the framework of purely stochastic analysis.}
That is, as adopted in previous studies~\cite{Caprini:2019pxz,Flauger:2020qyi,Caprini:2024hue,Blanco-Pillado:2024aca,LISACosmologyWorkingGroup:2024hsc,Kume:2024sbu}, 
the data considered in our analysis 
only contain the stochastic contributions to the noise, $\tilde{n}_i^{\nu}$, and the stochastic signal $\tilde{s}_i^{\sigma}$ due to the unresolved binary signals and, possibly, the SGWB:
\begin{equation}
    \tilde{d}_i(f) = \sum_{\nu}\tilde{n}_i^{\nu}(f) + \sum_{\sigma}\tilde{s}_i^{\sigma}(f)\;,
\end{equation}
where $\nu$ and $\sigma$ run over different noise and signal components, respectively.
In the following, we will assume that all these components are stationary and obey Gaussian statistics with zero mean and variance given by
\begin{equation}
\langle \tilde{n}_i^{\nu}(f) \tilde{n}_j^{\nu*}(f') \rangle = \frac{1}{2} \delta \left(f-f'\right) P^{\nu}_{N,ij}(f) \;, \quad\quad 
\langle \tilde{s}_i^{\sigma}(f) \tilde{s}_j^{\sigma*}(f') \rangle = \frac{1}{2} \delta \left(f-f'\right) P^{\sigma}_{S,ij}(f) \;,
\end{equation}
where we define the one-side power-spectral density (PSD) (for $i = j$) and cross-spectral density (CSD) (for $i \neq j$) of noise and signal components as $P^{\nu}_{N,ij}(f)$ and $P^{\sigma}_{S,ij}(f)$, respectively. 
Assuming all these components to be uncorrelated with one another, we obtain
\begin{equation}
\begin{aligned}
\label{eq:singlesidedsps}
\langle \tilde{d}_i(f) \tilde{d}_j^*(f') \rangle &= \frac{1}{2} \delta \left(f-f'\right) \left[ \sum_{\nu}P^{\nu}_{N,ij}(f) +\sum_{\sigma}P^{\sigma}_{S,ij}(f)\right] \\
&\equiv \frac{1}{2} \delta \left(f-f'\right)
\left[ P_{N,ij}(f) + P_{S,ij}(f)\right] \; ,
\end{aligned}
\end{equation}
where $P_{N,ij}(f)$ and $P_{S,ij}(f)$ are the total noise and signal PSDs and CSDs.
By denoting the response functions for isotropic SGWB signals as $\mathcal{R}_{ij}(f)$ (see Refs.~\cite{Robson:2018ifk,Flauger:2020qyi} for expressions for this quantity), the SGWB (in either strain $S^{\sigma}_h(f)$ or abundance $\Omega_{\rm GW}^{\sigma}(f)$) projected onto the data PSDs and CSDs can be expressed as
\begin{equation}
		\label{eq:signal_psd}
		P_{S,ij}(f) =  \mathcal{R}_{ij} (f) \sum_{\sigma}S_h^{\sigma} (f) \; = \mathcal{R}_{ij} (f) \frac{3H_0^2}{4\pi^2f^3}\sum_{\sigma}\Omega_{\rm GW}^{\sigma}(f)\;,
\end{equation}
where $H_0$ is the present Hubble constant and $h$ is the normalized one as $H_0/h \simeq 3.24 \times 10^{-18}$ 1/s.
Note again that under the assumptions stated above in this section, one finds that $R_{ij}(f)$ is diagonal in the AET basis. 

It is common practice to quantify the predicted primordial SGWB signal in terms of $h^2 \Omega_{\rm GW}(f)$; therefore, for later convenience, we define
\begin{equation}
\label{eq:omega_sens}
		P_{N,ij}^{\Omega}(f) = \frac{4 \pi^2 f^3}{3 H_0^2} %\mathcal{R}_{ij}^{-1} (f)
  P_{N,ij}(f) \;.
\end{equation}

In the following, we provide more detailed descriptions of the noise PSDs in the AET basis and of the astrophysical foregrounds which are included in $h^2\Omega_{\rm GW}^{\sigma}(f)$.

\subsubsection{Instrumental noise}
\label{sec:noise_and_signal_model}

Our current knowledge of the LISA noise is based on the LISA Pathfinder~\cite{PhysRevLett.116.231101} and laboratory tests.
As a first approximation, the stochastic component of the noise in each TDI channel can be grouped into two effective components: ``Optical Metrology System" (OMS) noise and TM noise.
The former accounts for noise in the readout frequency, such as laser shot noise, while the latter models the noise sources causing accelerations of the TMs, e.g., by environmental disturbances.
Introducing the transfer functions for these two noise sources $\mathcal{T}^{\nu}_{ij,\alpha\beta}(f)$ (for details, see e.g.~Refs.~\cite{Flauger:2020qyi,Hartwig:2021mzw,QuangNam:2022gjz,Hartwig:2023pft}), which project those contributions onto the TDI channels, the total noise PSDs and CSDs can be expressed as 
\begin{equation}
		\label{eq:two_signal_correlators}
		P_{N,ij}(f) = \sum_\nu P_{N,ij}^{\nu}(f)
  = 
  \sum_{\alpha\beta}
  \left[ \mathcal{T}_{ij,\alpha\beta}^{\rm TM} (f) S_{\alpha\beta}^{\rm TM} (f) 
  + \mathcal{T}_{ij,\alpha\beta}^{\rm OMS} (f) S_{\alpha\beta}^{\rm OMS} (f)
  \right].
\end{equation}

As customary in the literature, we assume stationary, Gaussian, and uncorrelated noises at each link with identical spectral shapes given by
\begin{align}
\label{eq:TM_noise_def}
S^\text{TM}_{\alpha\beta}(f) & = 7.7\times 10^{-46} \times A_{\alpha\beta}^2 \left( \frac{f_c}{f} \right)^2 \;  \left[1 + \left(\frac{0.4 \textrm{mHz}}{f}\right)^2\right]\left[1 + \left(\frac{f}{8  \textrm{mHz}}\right)^4\right] \;  \times \mathrm{s} \;,  \\ 
\label{eq:OMS_noise_def}
S^\text{OMS}_{\alpha\beta}(f) & = 1.6 \times 10^{-43} \times P_{\alpha\beta}^2 \left( \frac{f}{f_c} \right)^2 \; \left[1 + \left(\frac{2\textrm{mHz}}{f}\right)^4 \right] 
\;  \times \mathrm{s}  \;,
\end{align}
where $A_{\alpha\beta}$ and $P_{\alpha\beta}$ represent the amplitudes of the TM and OMS noises in the different links. Moreover, we have introduced $f_c \equiv (2\pi L/c)^{-1} \simeq 19$mHz representing the characteristic frequency of the detector.
As mentioned above, we assume the noise amplitudes for all links to be identical, i.e.~$A_{\alpha\beta} = A_{\rm noise}$ and $P_{\alpha\beta} = P_{\rm noise}$, and, following the ESA mission specifications~\cite{Colpi:2024xhw}, with fiducial values $A_{\rm noise} = 3$ and $P_{\rm noise} = 15$. 
In this case, the noise spectra reduce to $S^\text{TM}_{\alpha\beta}(f) = S^\text{TM}(f, A)$ and $S^\text{OMS}_{\alpha\beta}(f) = S^\text{OMS}(f, P)$. With $\mathcal{T}^{\nu}_{ij,\alpha\beta}(f)$ in the equal arm length limit, the PSDs in the AET basis read 
\begin{equation}
\begin{aligned}
    P_{N,{\rm AA}}(f)  &= P_{N,{\rm EE}}(f) \\
    &=  8 \sin^2 x \left\{ 4 \left[1 +\cos x + \cos^2 x\right] S^\text{TM}(f,A_{\rm noise}) \;  +  \left[ 2 +\cos x \right]S^\text{OMS}(f,P_{\rm noise})  \right\}  \; ,
\end{aligned}\label{eq:psdAA}
\end{equation}
and
\begin{equation}
\begin{aligned}
    P_{N,{\rm TT}}(f) =  16 \sin^2 x \left\{ 2 \left[ 1 - \cos x \right]^2  S^\text{TM}(f,A_{\rm noise}) + \left[ 1 - \cos x \right] S^\text{OMS}(f,P_{\rm noise}) \right\} \; ,
\end{aligned}
\label{eq:psdTT}
\end{equation}
where we have defined $x \equiv f/f_c$.
Here the CSDs vanish, i.e.~$P_{N,ij}(f) = 0$ ($i\ne j$), so that the noise covariance matrix is diagonal.

\subsubsection{Astrophysical foregrounds}\label{sec:foreground}
Numerous weak and unresolvable signals from astrophysical sources will superimpose incoherently generating astrophysical SGWB~\cite{10.1046/j.1365-8711.2001.04217.x, Farmer:2003pa, Regimbau:2011rp,LISA:2022yao,Pozzoli:2023kxy,Babak:2023lro,Staelens:2023xjn,Toubiana:2024qxc}. 
There are at least two guaranteed components in the LISA band. 
Below a few millihertz, the dominant contribution will come from Compact Galactic Binaries (CGBs) mostly composed of Double White Dwarfs (DWDs)~\cite{Evans:1987qa,Bender:1997hs}. 
At higher frequencies, another contribution is expected from all the extragalactic compact objects including Stellar Origin Binary Black Holes (SOBBHs) and binary neutron stars (BNS)~\cite{LIGOScientific:2019vic}.
In the remainder of this section, we provide the templates for these foreground components implemented in the \texttt{SGWBinner} code that was recently used in Refs.~\cite{Caprini:2024hue,Blanco-Pillado:2024aca,LISACosmologyWorkingGroup:2024hsc,Kume:2024sbu}.

\noindent
\paragraph{Galactic foreground.}
This component represents the contribution from the unresolved sub-threshold mergers of CGBs that remain after the removal of loud signals from the population of CGBs in the galactic disk~\cite{Nissanke:2012eh}.
Due to the angular dependence of the response functions and LISA yearly orbit, this component exhibits an annual modulation. While, in principle, this characteristic can help distinguish the galactic component from other stationary contributions, e.g., by accounting for variations in each segment~\cite{Adams:2013qma,Mentasti:2023uyi,Hindmarsh:2024ttn}, we average over anisotropies, which leads to suboptimal (but conservative\footnote{Keeping track of anisotropic nature of the signal, requires the analysis to be time-frequency. To consistently work this out, one would also need to keep track of the non-stationary nature of the noise (see e.g.~\cite{Baghi:2021tfd,Robson:2018jly,Seoane:2021kkk,Alvey:2024uoc,Rosati:2024lcs}) and the presence of gaps in the data. }) foreground extraction.
Similarly, because this foreground is formed by the superposition of many unresolvable sources, it is expected to have Gaussian statistics. Recently, Refs. \cite{Buscicchio:2024wwm, Karnesis:2024pxh, Rosati:2024lcs} have called into question whether the populations entering the foreground are sufficient for the central limit theorem to apply at all frequencies, and imply a Gaussian description of the foreground may be biased. The non-stationarity of the foreground also in principle induces some non-Gaussianity when time-averaging.

Nevertheless, we use the empirical model from Ref.~\cite{Karnesis:2021tsh}, which describes the sky-averaged and Gaussian contribution by
\begin{equation}\label{eq:gal}
    h^2\Omega^{\textrm{Gal}}_{\rm GW}(f)=\frac{1}{2}\left(\frac{f}{ 1\,\textrm{Hz}}\right)^{2/3}
    e^{-(f/f_1)^\alpha}
    \left[1+\tanh{\frac{f_{\textrm{knee}}-f}{f_2}}\right]h^2\Omega_{\textrm{Gal}}\;,
    \end{equation}
    where the value of $f_1$ and $f_{\textrm{knee}}$ depends on the total observation time $T_{\rm obs}$ as
    \begin{eqnarray}
    \log_{10} (f_1/{\rm Hz}) &=& a_1 \log_{10}(T_{\rm obs}/{\rm year}) + b_1\,,\nonumber\\
    \log_{10} (f_{\textrm{knee}}/{\rm Hz}) &=& a_k \log_{10}(T_{\rm obs}/{\rm year}) + b_k\,.
\end{eqnarray}
The exponential factor $e^{-(f/f_1)^{\alpha}}$ accounts for the loss of stochasticity at higher frequency \cite{Karnesis:2021tsh}, while the last $\tanh$ term models the expected complete subtraction of CGBs signal at frequencies $f > f_{\rm knee}$.
In order to keep the notation compact, we define $\log_{10}(h^2\Omega_{\rm Gal}) \equiv \alpha_{\rm Gal}$.
From Ref.~\cite{Karnesis:2021tsh}, we set the fiducial values $a_1 = -0.15$, $b_1 = -2.72$, $a_k = -0.37$, $b_k = -2.49$, $\alpha = 1.56$, $f_2 = 6.7\times10^{-4}$Hz and $\alpha_{\rm Gal} = -7.84$.

\noindent
\paragraph{Extragalactic foreground.}
The extragalactic foreground, arising from the incoherent superposition of all extragalactic compact object mergers, includes potential contributions from SOBBHs, BNSs, EMRIs, and DWDs in their inspiral phase.
In this work, we focus on only the SOBBH+BNS contribution, leaving any potential EMRI and DWD contribution to future work.
Recent studies suggest that extreme mass-ratio inspirals can largely contribute to the foreground but only in somewhat extreme population synthesis scenarios~\cite{Pozzoli:2023kxy}. Extragalactic DWDs may also be more abundant than previously estimated, with a relevant impact on the extragalactic foreground~\cite{Staelens:2023xjn, Hofman:2024xar} which ongoing analyses are verifying~\cite{LambertsTOappear}. In the lack of a firmer understanding,  we assume these contributions to be below the foregrounds of galactic binaries and extragalactic SOBBHs and BNSs.

We now focus on what we will assume to be the dominant contribution, the SOBBH and BNS foreground.
The vast majority of these signals cannot be individually resolved by LISA~\cite{Seto:2022xmh, Lehoucq:2023zlt, Ruiz-Rocha:2024xjt} and, for the most part individual detections are possible for the few multi-band sources~\cite{Sesana:2016ljz} (see~\cite{Buscicchio:2024asl}, for a more accurate study of such sources). The best estimates for the populations of these objects are based on observations from ground-based detectors~\cite{KAGRA:2021duu,KAGRA:2021kbb}.
Due to the relatively uniform distribution of the sources and the limited angular resolution of LISA, this component can be well modeled as an isotropic SGWB signal with the power-law shape
\begin{equation}\label{eq:ext}
    h^2 \Omega^{\rm Ext}_{\rm GW}(f) 
    = h^2\Omega_{\rm Ext} \left( \frac{f}{1 {\rm mHz}}\right)^{2/3} \; ,
\end{equation}
where $h^2 \Omega_{\rm Ext}$ is the amplitude at 1\,mHz. Recent observations by LIGO-Virgo-KAGRA collaboration estimate the magnitude of SGWB signal from SOBBHs and BNS as~\cite{KAGRA:2021duu}
\begin{equation}
    \Omega_{\rm Ext} =  7.2^{+3.3}_{-2.3} \times 10^{-10}  
    \text{ at } 
    f = 25
    \,{\rm Hz}\;.
\end{equation}
In order to keep the notation compact, we define $\log_{10}(h^2\Omega_{\rm Ext}) \equiv \alpha_{\rm Ext}$. Extrapolating this amplitude to the LISA band~\cite{Babak:2023lro}, yields the fiducial value $\alpha_{\rm Ext} = -12.38$.

\subsection{Analysis of the simulated data}\label{sec:data_analysis}
In this section, we summarise the data analysis scheme implemented in the \texttt{SGWBinner} code (see Refs.~\cite{Caprini:2019pxz,Flauger:2020qyi} for more details). Let us start with the generation of simulated data. 
Given the effective observation time $T_{\rm obs}$ and the number of segments $N_d$ (which define the duration of each segment $\tau = T_{\rm obs} / N_d$), the code generates the data $\tilde{d}^s_i(f_k)$ ($s = 1, ..., N_d$) segment-by-segment in the frequency-domain. 
For each frequency bin $f_k$ (spanning $[3\times10^{-5}, 0.5]$ Hz with spacing $\Delta f = 1/\tau$), $N_d$ Gaussian realizations of the signal, noise, and foregrounds are generated with zero mean and variances defined by their respective PSDs. These data are then averaged over segments to define $\bar{D}_{ij}^k \equiv \sum_{s=1}^{N_d} d_i^s(f_k)d_j^{s*}(f_k)/N_d$, which gives an estimate of the total power at all frequencies. The next step consists of coarse-graining the data using inverse variance weighting. 
This results in a coarser set of frequency bins $f_{ij}^k$ and a data set $D_{ij}^k$ with weights $n_{ij}^{k}$, retaining similar statistical properties of the original dataset. Similarly to Refs.~\cite{Flauger:2020qyi,Caprini:2024hue,Blanco-Pillado:2024aca,LISACosmologyWorkingGroup:2024hsc,Kume:2024sbu}, we set $\tau = 11.4$ days ($\Delta f = 10^{-6}$ Hz), $N_d = 126$, and $T_{\rm obs} = 4$ years in our analysis.

The likelihood employed in the code reads~\cite{Flauger:2020qyi}
\begin{equation}
\ln  \mathcal{L}(D|{\bm \theta}) = \frac{1}{3} \ln \mathcal{L}_{\rm G}(D|{\bm \theta})+ \frac{2}{3} \ln  \mathcal{L}_{\rm LN}(D|{\bm \theta}) \; ,\label{eq:full_likelihood}
\end{equation}
with
\begin{equation}
    \label{eq:gaussian_likelihood}
	\ln \mathcal{L}_{\rm G}(D|{\bm \theta}) = - \frac{ N_d }{2} \sum_{i \in \{ {\rm AET} \}}\sum_{k} n_{ii}^{k} \left[ \frac{ \mathcal{D}_{ii}^{th}(f_{ii}^{k}, {\bm \theta}) -\mathcal{D}_{ii}^{k}}{ \mathcal{D}_{ii}^{th}(f_{ii}^{k}, {\bm \theta})} \right]^2 \;,
\end{equation}
\begin{equation}
\label{eq:lognormal_likelihood}
\ln \mathcal{L}_{\rm LN}(D|{\bm \theta}) = - \frac{N_d}{2} \sum_{i \in \{ {\rm AET} \}} \sum_{k} n_{ii}^{k}  
\ln ^2 \left[ \frac{ \mathcal{D}_{ii}^{th}(f_{ii}^{k}, {\bm \theta})  }{ \mathcal{D}_{ii}^{k} } \right]  \; , 
\end{equation}
where the index $k$ runs over the coarse-grained data points and $\mathcal{D}_{ii}^{th}(f, {\bm \theta})$ denotes the theoretical predictions for the data, depending on some parameters ${\bm \theta}$. The model can be further expressed as $ \mathcal{D}_{ii}^{th}(f, {\bm \theta}) \equiv  \mathcal{R}_{ii} \, 
h^2\Omega_{\rm GW}(f, {\bm \theta}_{\rm cosmo}, {\bm \theta}_{\rm fg}) +  P_{N,ii}^{\Omega}(f, {\bm \theta}_{\rm n})$, with ${\bm \theta}_{\rm cosmo}, {\bm \theta}_{\rm fg}$ and ${\bm \theta}_{\rm n}$ denoting the signal, foreground, and noise parameters, respectively. Notice that the diagonality of the AET basis has been exploited, and no cross terms appear in the likelihood. Given some priors $\pi({\bm \theta})$ for the parameters, the posterior distribution reads
\begin{equation}
	p({\bm \theta}|D) \equiv \frac{\pi({\bm \theta})  \mathcal{L}(D | {\bm \theta})} {Z(D)}\; ,\label{eq:full_post}
\end{equation}
where $Z(D)$ is the model evidence defined as 
\begin{equation}
    Z(D) \equiv 
    \int {\mathrm d}{\bm \theta}\, \pi({\bm \theta})  \mathcal{L}(D | {\bm \theta})\;.\label{eq:evidence}
\end{equation}

To compare the validity of two different models $ M_i({\bm \theta}_i) $, each characterized by a set of parameters $ {\bm \theta}_i $, we can use the evidence $Z_i$ as a measure of the quality of the models given the data. The \textit{Bayes factor} for two models $i, j$ is defined as $ B_{ij} \equiv Z_i / Z_j $. This Bayes factor can then be compared to the Jeffreys' scale~\cite{Kass:1995loi} to determine which model is favored by the data. 

As key functionalities, the \texttt{SGWBinner} code offers {\it i)} model-agnostic signal reconstruction and {\it ii)} template-based signal reconstruction. The former fits the signal in each frequency bin using a power-law template, i.e.~the signal parameters are 
\begin{equation}
{\bm \theta}_{\rm cosmo} = \left\{ \alpha_1, n_{\rm T, 1}, \dots , \alpha_n, n_{{\rm T}, n} \right\} \;,
\end{equation}
with $n$ denoting the number of bins. The number and width of the bins are dynamically adjusted as described in Refs.~\cite{Caprini:2019pxz,Flauger:2020qyi}. In practice, this method enables a preliminary identification of the spectral shape of the signal, which can guide the choice of the template for the template-based analysis. For the latter, the vector of parameters of the cosmological component ${\bm \theta}_{\rm cosmo}$ corresponds to the template parameters. 

In this work, we assume the fiducial noise and foreground parameters to be
\begin{equation}
  {\bm \theta}_{\rm  n}= \{A_{\rm noise}, P_{\rm noise}\}, 
  \quad 
  {\bm \theta}_{\rm fg} 
  = 
  \{ 
\alpha_{\rm Gal}
   , 
\alpha_{\rm Ext}
    \},  
\end{equation}
 while we assume that the other foreground parameters are known\footnote{The effect of loosening this assumption on the signal reconstruction has been discussed in Ref.~\cite{Kume:2024sbu}.}, and ${\bm \theta}_{\rm cosmo}$ is model dependent. Moreover, when fitting the simulated data, we use the same noise model applied to generate the data.\footnote{We note that any differences between the instrumental noise and the model could introduce bias. This issue will have to be closely monitored in future upgrades of the code.}
Both for the noise and foreground amplitudes, we assume Gaussian priors centered on their fiducial values.
For the former, we set the standard deviation to be $20\%$ of the fiducial mean value. For $\alpha_{\rm Gal}$ and $\alpha_{\rm Ext}$, we set the standard deviation to be 0.21 and 0.17, respectively.
To sample the parameter space the code relies on the \texttt{Cobaya}~\cite{Torrado:2020dgo} inference framework.
To facilitate template-based analysis specifically for SIGW signals, we develop the dedicated \SIGWAY\  code. As detailed in App.~\ref{app:SIGWBinner}, the \SIGWAY\  code implements the parameterization of curvature perturbations discussed in Sec.~\ref{sec:modellingPzeta} and performs the numerical computation of SIGW signals.

Finally, the code also supports Fisher analysis. 
In practice, the Fisher Information Matrix (FIM) can be computed by the continuous integral over the frequency range, expressed as
\begin{equation}
    \label{eq:FIM_final}
    F_{ab} \equiv T_{\rm obs} \sum_{i \in \{ {\rm AET} \} }\int_{f_{\mathrm{min}}}^{f_{\mathrm{max}}} 
    \textrm{d}f \;
    \left.
    \frac{\partial \ln \mathcal{D}_{ii}^{th}}{\partial \theta^a} \frac{\partial \ln \mathcal{D}_{ii}^{th}}{\partial \theta^b} \right\vert_{{\bm \theta} = {\bm \theta}_{\rm fid}} \,   , 
\end{equation}
where $f_{\mathrm{min}}$ and $f_{\mathrm{max}}$ represent the detector's minimal and maximum measured frequencies, assumed to be $f_{\mathrm{min}} = 3 \times 10^{-5}$\,Hz and $f_{\mathrm{max}} = 0.5$\,Hz~\cite{Colpi:2024xhw}. 
If non-trivial (log-)priors are included in the analysis, the code consistently adds their derivatives to Eq.~\eqref{eq:FIM_final} to obtain the full FIM. 
The relative uncertainty on the reconstruction parameters can then be estimated from the covariance  $\mathrm{cov}_{a b} = F_{ab}^{-1}$.
Given its computational efficiency, we also employed the FIM approach to assess the prospect of signal reconstruction with some level of accuracy.
Note that in the case of SIGW signals, the FIM can be efficiently computed using the automatic differentiation feature of the JAX library~\cite{jax2018github} by applying the ${\bm \theta}$ derivative in Eq.~\eqref{eq:FIM_final} directly to $\mathcal{P}_{\zeta}$, before it is integrated to yield $\mathcal{P}_{h}$. 
We stress that the FIM formalism only works under the assumption that the likelihood is well approximated by a Gaussian distribution in the model parameters around the best fit (and that, when dealing with real data, the true values of the parameters lie within the region where the FIM is evaluated). 

Finally, to complement the visualization of relative uncertainties in the parameter space, we will plot the signal-to-noise ratio (SNR) defined as
\begin{equation}
\label{eq:SNR_def}
\textrm{SNR} \equiv \sqrt{T_{\rm obs} \;\sum_{i \in \{ {\rm AET} \} } \int_{f_{\rm min}}^{f_{\rm max}} \left( \frac{P^{\sigma}_{S,ii}}{P_{N,ii}} \right)^2 \; \textrm{d}f }  \; , 
\end{equation}
which scales linearly with the signal amplitude.

\section{Results}
\label{sec:results}

In this section, we summarise our main results presenting different analyses based on the \SIGWAY\  code outlined in Sec.~\ref{sec:data_analysis} (see App. \ref{app:SIGWBinner} for more details). 
We adopt the three approaches discussed in Sec.~\ref{sec:modellingPzeta}, namely 
{\it i)} binned spectrum agnostic approach; 
{\it ii)} template-based approach; 
{\it iii)} first principle USR model of inflation -- 
all limited to the leading order SIGW and assuming an RD universe. 
We then consider specific examples including non-standard early universe evolution and non-Gaussianities. 

We report results for  both $\Omega_{\rm GW}$ and ${\cal P}_\zeta$. In the former case, we include the noise curves as well as the foregrounds as discussed in Sec.~\ref{sec:tempbinnerpipe}.
Since the SIGW backgrounds we consider are emitted at very high redshift, when scales currently associated with mHz re-enter the Hubble sphere, they contribute to the energy budget in the early Universe and can affect cosmological observables as any other relativistic free-streaming component beyond the standard model. 
In particular, the SIGW contributes to the effective number of neutrino species as $N_{\rm eff}\equiv 3.044+\Delta N^{\rm GW}_{\text{eff}}$, with $\Delta N^{\rm GW}_{\text{eff}}=\rho_{\rm GW}/\rho_{\nu,1}$ and $\rho_{\nu,1}$ is the energy density of a single neutrino species. 
Specifically, the total (integrated) GW abundance is 
$\Omega_{\rm GW} h^2\simeq 1.6\cdot 10^{-6}\left(\Delta N^{\rm GW}_{\text{eff}}/0.28\right)$~\cite{Caprini:2018mtu}.  
Measurements of the CMB \cite{Planck:2018vyg} and Baryon Acoustic Oscillations (BAO) constrain $\Delta N_{\rm eff}\leq 0.28$ at $95\%$ C.L.
We report this bound for reference as shaded gray regions in the $\Omega_{\rm GW} h^2$ plots.

Strong primordial density perturbations can lead to the copious formation of PBHs with masses of the order of the horizon mass
$
    M_H = 1.3 \times 10^{-15} M_{\odot} 
\left[ (k/\kappa_{\rm rm})/\mathrm{s}^{-1} \right]^{-2}\,,
$
where we kept track of the additional prefactor 
$\kappa_{\rm rm} \equiv k r_m \sim {\cal O}(3)$ that relates the perturbation scale to the characteristic perturbation size $r_m$
at Hubble crossing ~\cite{Germani:2018jgr,Musco:2018rwt,Escriva:2019phb,Young:2019osy}.
Thus, the overproduction of dark matter in the form of PBHs in the asteroid mass range implies a bound ${\cal P}_\zeta \leq \mathcal{O}(10^{-2})$~\cite{Saito:2008jc,Bugaev:2009kq,Bugaev:2010bb,Byrnes:2018txb,Gow:2020bzo} on the scalar curvature perturbations and thus also on the strength of the SIGW in the mHz frequency band. The abundance and the mass distribution of PBHs depend on the shape of the curvature power spectrum, non-Gaussianities, and on the equation of state of the universe during their formation~\cite{LISACosmologyWorkingGroup:2023njw}, so does the implied upper bound on SIGWs.

\subsection{Binned spectrum method}
\label{sec_resBM}

\begin{figure}[t]
    \centering
    \includegraphics[width=0.49\textwidth]{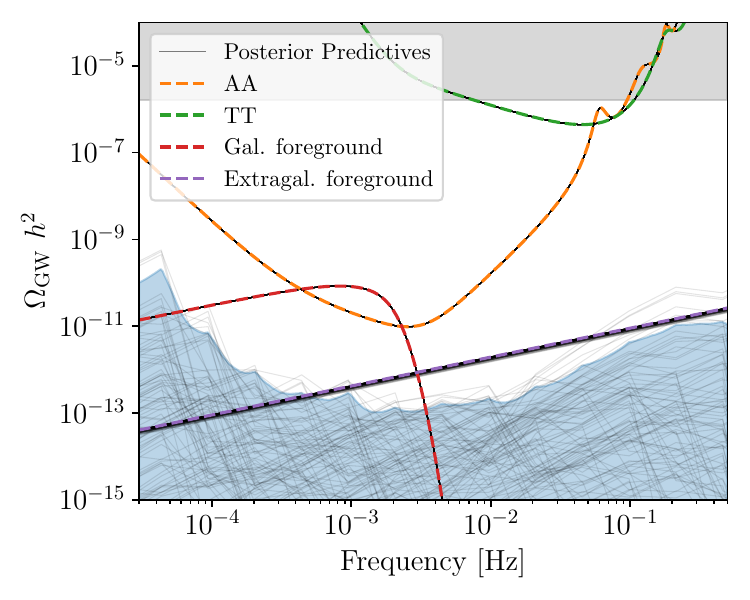}
    \includegraphics[width=0.49\textwidth]{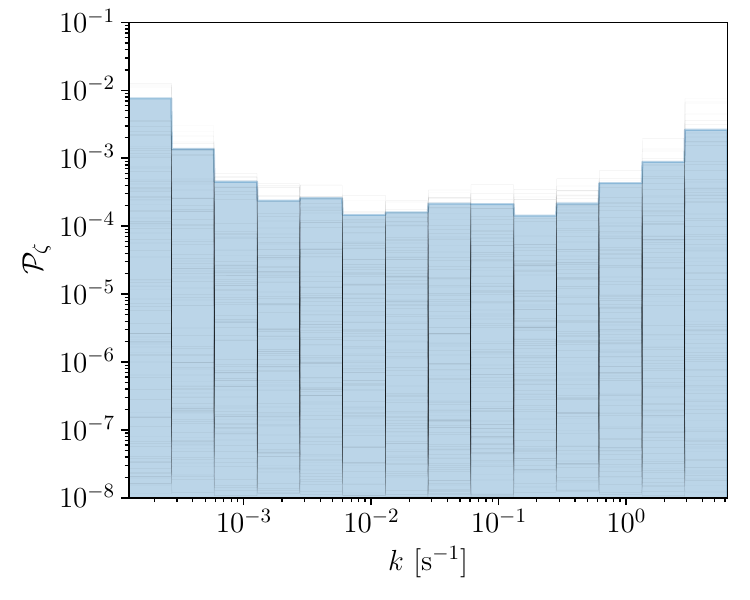}
    \caption{
    Posterior predictive distribution for both $\Omega_{\rm GW}h^2$ (left panel) and $\Pz$ (right panel).
    We represent the binned reconstruction of Sec.~\ref{sec_resBM} in the case without an injected signal. 
    The posterior saturates the lower bound of the prior for the amplitudes $A_i$, due to the absence of a resolvable signal. 
    We therefore can only set upper bounds on both $\Omega_{\rm GW} h^2$ and $\mathcal{P}_\zeta$. 
    The light blue line shows the $95\%$ credible intervals, while the pale black lines individual realisations of a signal sampled from the parameters' posterior distribution.
    The upper bound from $\Delta N_{\rm eff}$ is shown with a gray shading.}
    \label{fig:binned_no_injection_15}
\end{figure}

In Fig.~\ref{fig:binned_no_injection_15} we report the constraints obtained with the binned method (see Sec. \ref{sec:binnedPzeta} and \ref{sec:OmegaGWijk}) when injecting no SIGW signal.
This analysis forecasts the model-independent upper bounds on both the SIGW energy density spectrum (left panel) and the primordial curvature power spectrum (right panel) in case of no SGWB detection at LISA: this only relies on observational data, without assuming a specific signal model.
For this analysis, we assume that the spectrum is divided into $N =15$ bins.
The free parameters in this model are
\begin{equation}
    {\bm \theta}_{\rm cosmo} =\{ A_1, \cdots, A_{15}\}.
\end{equation}
In the left panel of Fig.~\ref{fig:binned_no_injection_15} we indicate the posterior predictive distribution for $\Omega_\text{GW}$ with the shaded light blue region, denoting the 95\% credible interval (CI). 
The upper bound effectively reflects the LISA sensitivity, which falls around two orders of magnitude below the noise components in the AA channel because of the long observation time  $T_{\rm obs} = 4 {\rm yr}$ (see  Sec.~\ref{sec:data_analysis}).
In the right panel, we then show how the LISA sensitivity translates into the ${\cal P}_\zeta$ parameter space.
The figure displays the upper bounds on $\mathcal{P}_\zeta$ across the range of momenta $k$ considered, which is $k\in [1.26 \times 10^{-4},6.28]/{\rm s}$. 
The posterior saturates at the lower edge of the prior on the amplitude parameters $A_i>10^{-8}$ reflecting the absence of detectable power beyond the noise level.
The posterior predictive bands illustrate that the method can constrain $\mathcal{P}_\zeta$ across several orders of magnitude in $k$.
In the most sensitive range, this bound reaches ${\cal P}_\zeta \lesssim 2 \times 10^{-4}$. 

This sensitivity is sufficient for probing a wide range of viable scenarios for asteroid mass PBHs. 
In particular, a non-detection of a SIGW by LISA would close the asteroid mass window for PBH dark matter formed from the collapse of moderately non-Gaussian curvature fluctuations, including models of PBHs from first-order phase transitions~\cite{Lewicki:2023ioy, Cai:2024nln, Lewicki:2024ghw}. 
However, as with $\mu$-distortion bounds on heavy PBHs~\cite{Nakama:2017xvq, Unal:2020mts, Byrnes:2024vjt, Iovino:2024tyg}, extremely strong non-Gaussianities could enhance PBH production and potentially allow evading these constraints. 
Such extreme scenarios and their theoretical consistency should be studied case by case.

\begin{figure}[t]
    \centering
    \includegraphics[width=0.49\textwidth]{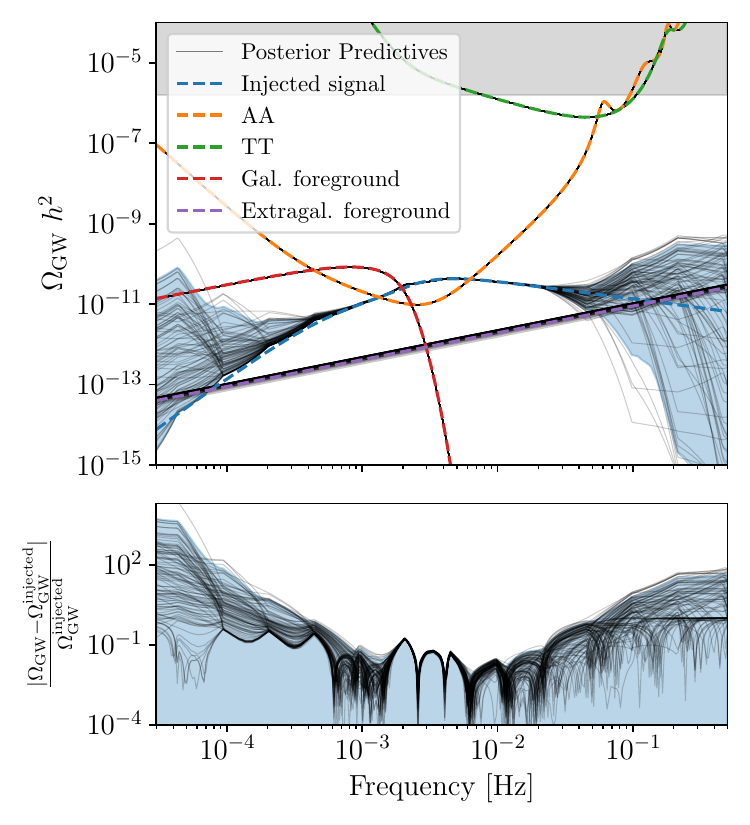}
    \includegraphics[width=0.49\textwidth]{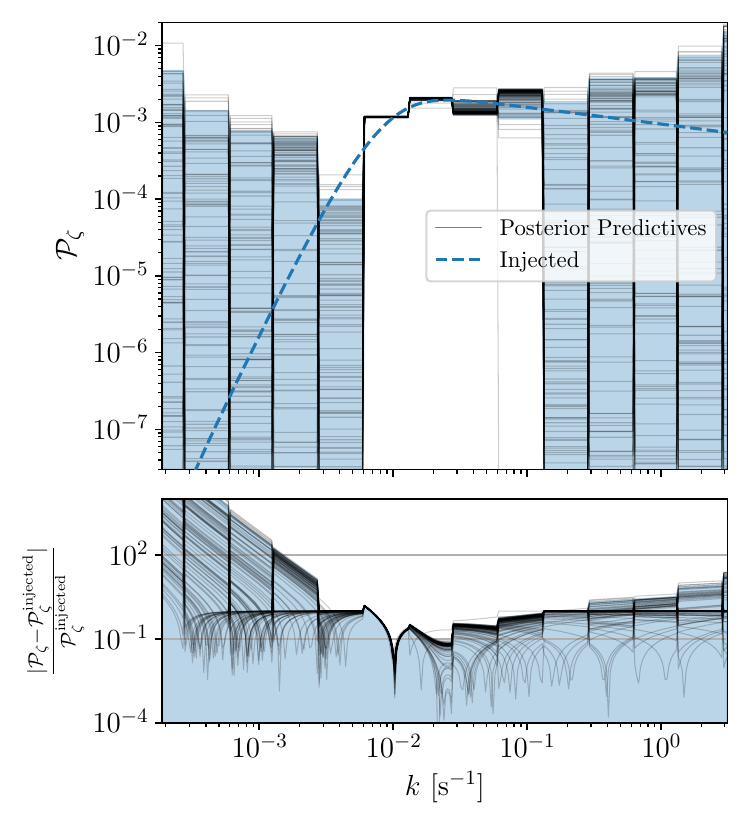}
    \caption{
    Same as Fig.~\ref{fig:binned_no_injection_15}, but simulating the observation of a signal obtained in the benchmark USR model scenario. The quantity 
    $\mathcal{P}_\zeta$ is reconstructed with the model-independent binning method with 15 bins. The blue band in the upper panels shows the $90\%$ (symmetric) credible interval, while the blue band in the bottom shows the $95\%$ upper bound on the residuals.}
    \label{fig:binned_injection_15}
\end{figure}

In Fig.~\ref{fig:binned_injection_15} we report the constraints obtained with the binned method when injecting the benchmark SIGW signal derived from the single field USR model of Sec \ref{sec:benchmodelUSR}. 
The curvature power spectrum has a BPL shape (see Eq.~\eqref{eq:BPL}), with a peak at around ${\cal P}_\zeta \sim 2 \cdot 10^{-3}$.
Again, we use a template with $N =15$ bins.
In the left panel of Fig.~\ref{fig:binned_injection_15}, we show the injected SIGW signal (blue dashed line), along with the posterior predictive distribution (light blue band).
Although the low number of bins reduces the frequency resolution of our model compared to the one achieved by LISA, the SIGW spectrum is well reconstructed, reaching a precision of the order of a few percent around the peak. At the edges of the observable range of frequencies, the blue bands widen up indicating a poor constraining power on the tail regions. 
The right panel of  Fig.~\ref{fig:binned_injection_15} indicates the SIGW bounds translate into four bins being well constrained in the range $k\sim [10^{-2}, 10^{-1}]/$s with around ${\cal O}(10)\%$ precision, while the other ones being subject to an upper bound of similar amplitude as in Fig.~\ref{fig:binned_no_injection_15}.

One could in principle enhance the frequency resolution by using a template with a larger number of bins, at the cost of drastically increasing the computational cost of the Bayesian MCMC inference. In App.~\ref{App:largeNbinning}
we discuss these issues in more detail.

\subsection{Template based method}\label{sec:template_method}

In this section, we present a forecast on reconstructing the SIGW signal using a template-based method, addressing different scenarios discussed in Sec.~\ref{sec:analyticaltemplates}. 

\subsubsection{Smooth spectra}
\label{sec:results_template_smooth}

\noindent
\paragraph{Lognormal scalar spectrum.}
The first signal injection we  consider is a log-normal shape of $\mathcal{P}_\zeta$ as defined in Eq.~\eqref{eq:PLN} with 
the benchmark values defined in Eq.~\ref{eq:PLN injected}, which we report here
$\log_{10} A_s=-2.5$, 
$\log_{10}\Delta=\log_{10}(0.5)$,
$\log_{10} \left(k_*/\mathrm{s}^{-1}\right)=-2$.
This produces a loud signal with the typical double peak being fully within the LISA band.
This choice gives us a concrete measure of the precision that is achievable in measuring the SIGW background in this scenario for a reasonably loud signal.
The free parameters for this approach are
\begin{equation}
    {\bm \theta}_{\rm cosmo} =\{\log_{10} A_s,\log_{10}\Delta,
    \log_{10} \left(k_*/\mathrm{s}^{-1}\right) 
    \}.
\end{equation}
\begin{figure}[h!]\centering
\includegraphics[width=0.57\textwidth]{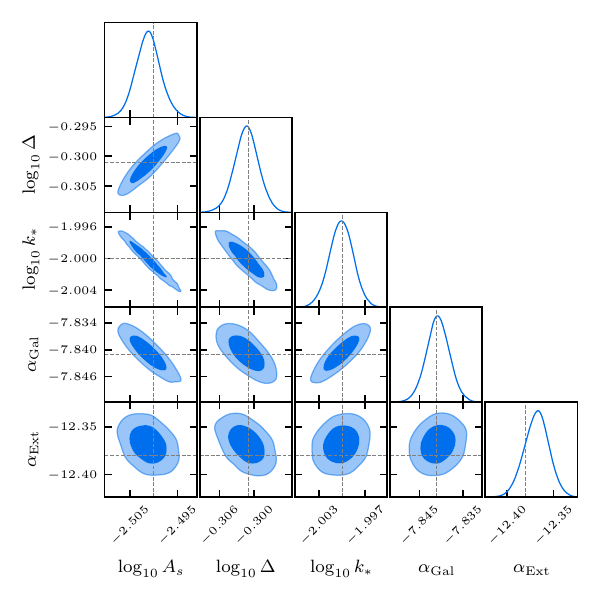}
\caption{
Corner plot with the posterior distribution for an injected LN spectrum $\Pz$, as defined in Eqs.~\eqref{eq:PLN} and \eqref{eq:PLN injected}. 
$k_*$ is expressed in $1/\mathrm{s}$ units.}
\label{fig:ln_corner}
\end{figure}
\begin{figure}[h!]
    \centering
    \includegraphics[width=0.49\linewidth]{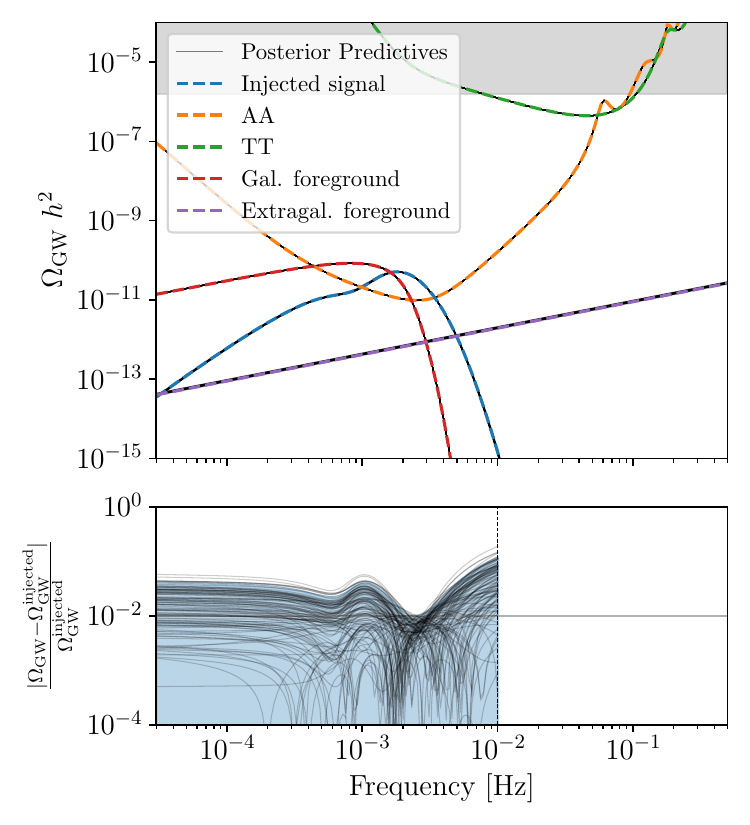}
    \includegraphics[width=0.49\linewidth]{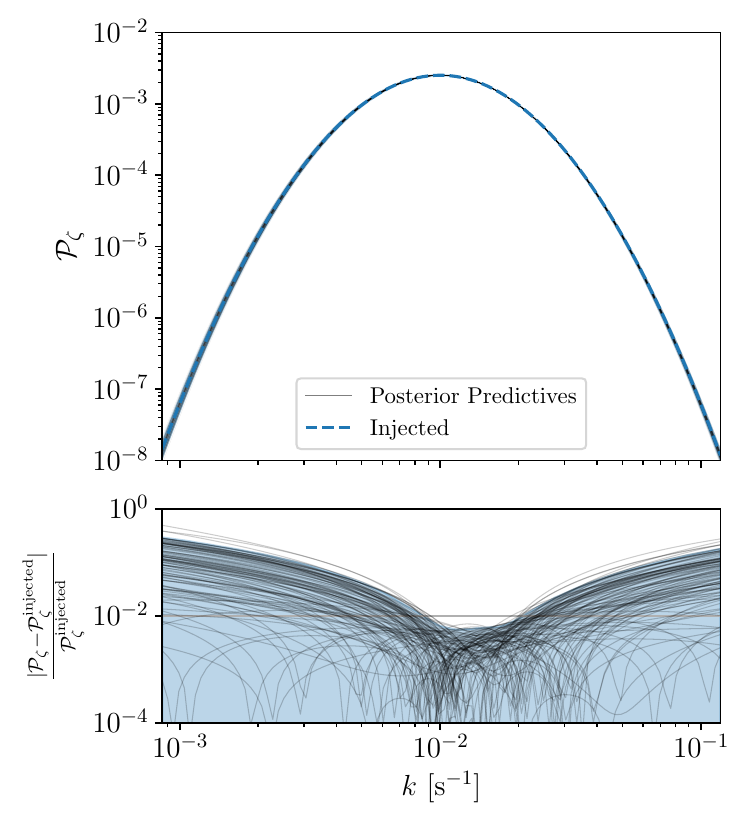}
    \caption{
    Same as Fig.~\ref{fig:binned_no_injection_15}, for an injected log-normal spectrum, as defined in Eqs.~\eqref{eq:PLN} and \eqref{eq:PLN injected}.
    In the left panel, we arbitrarily cut the posterior predictive where the signal falls below $\Omega_{\rm GW} \lesssim 10^{-15}$.}
    \label{fig:ln_posterior_predictives}
\end{figure}

Fig.~\ref{fig:ln_corner} shows the posterior distribution for each parameter of the LN template, alongside the ones describing the galactic/extragalactic foregrounds. 
We omit in these plots the posterior distributions for the noise parameters, as they are weakly correlated with the others in all cases. 
The injected values are indicated with a dashed gray line. 
As we can see, due to the relatively high SNR of the injected signal, the parameters of this template are very accurately reconstructed, with $\Pz$ being reconstructed with a relative error of a few percent at its peak.
The correlation between $A_s$ and $\Delta$ originates from the definition of $\Pz$. As customary in the literature~\cite{Pi:2020otn}, the amplitude at the peak is $A_s/\Delta$, while $A_s$ is the integrated power spectrum $\int_{-\infty}^\infty \Pz \, \d \log k = A_s$. Therefore, $A_s$ enters the power spectrum only through the ratio $A_s/\Delta$, thus the positive correlation between the two parameters. Defining $A_s$ to be the peak amplitude would avoid this degeneracy.
The correlation of $A_s$, $k_*$, and $\alpha_\mathrm{Gal}$ is instead specific of our choice of fiducial parameters. As can be seen from the posterior predictive distribution in the left panel in Fig.~\ref{fig:ln_posterior_predictives}, the injected signal is close enough to the galactic foreground that a slight increase in $A_s$ with a decrease in $k_*$ can be compensated by a small decrease in the background amplitude $\alpha_\mathrm{Gal}$.
We expect that these correlations fade away with a larger injected $k_*$, when the signal and the galactic background are more distinct.

Fig.~\ref{fig:ln_posterior_predictives} shows the posterior predictive distribution for the SIGW (left panel) and curvature power spectrum (right panel).
We see that the signal reconstruction achieves better than percent uncertainties on $\Omega_{\rm GW}$ and $\Pz$ around the peak. 
The uncertainty on the low-frequency tail of $\Omega_{\rm GW}$ saturates at around a few percent, due to the universal behavior of the causality tail \cite{Hook:2020phx} sufficiently deep in the IR. 
See e.g.~Appendix B of \cite{LISACosmologyWorkingGroup:2024hsc} for a discussion of logarithmic corrections to the IR tail of the SIGW spectrum.
The uncertainty on the tails of $\Pz$ remains low even at very small values, because of the rigid assumption about the LN template.

\begin{figure}[h]
    \centering
    \includegraphics[width=0.49\textwidth]{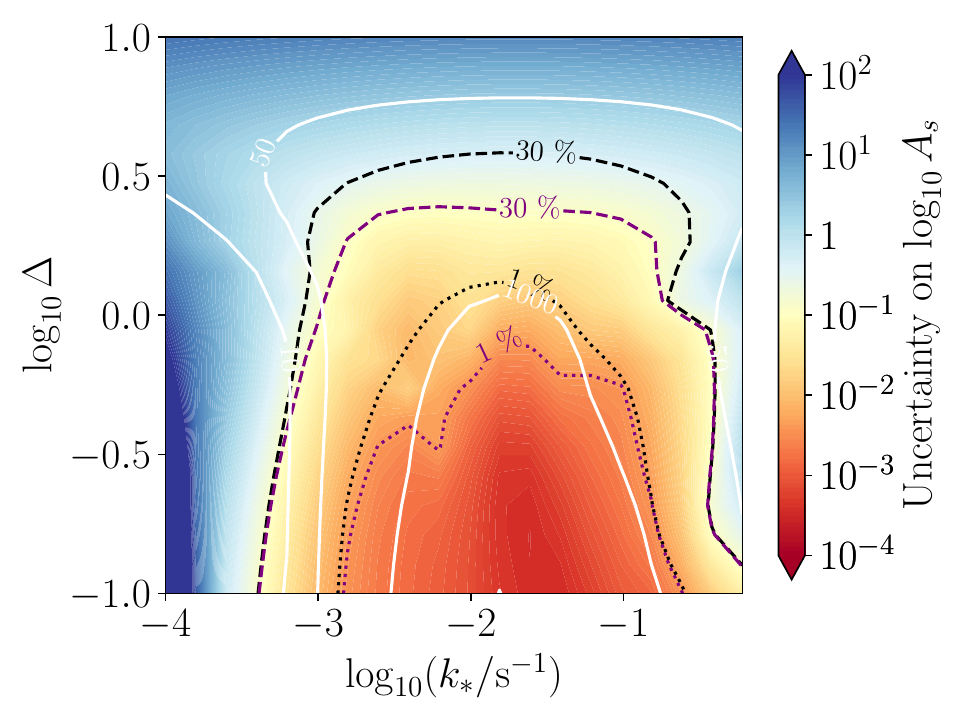}
    \includegraphics[width=0.49\textwidth]{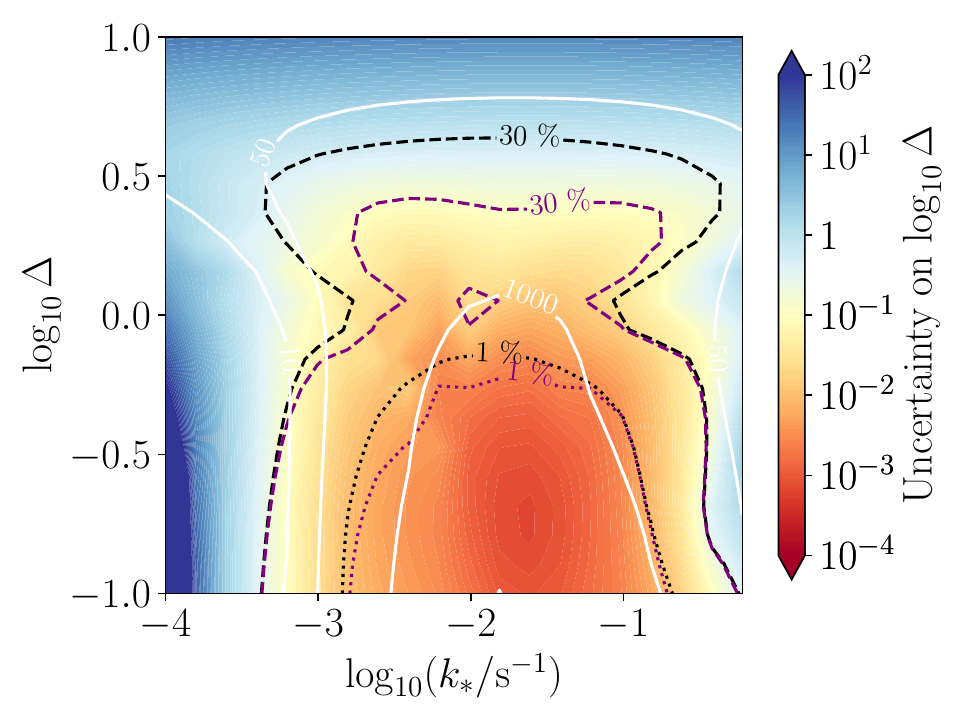}
    \includegraphics[width=0.49\textwidth]{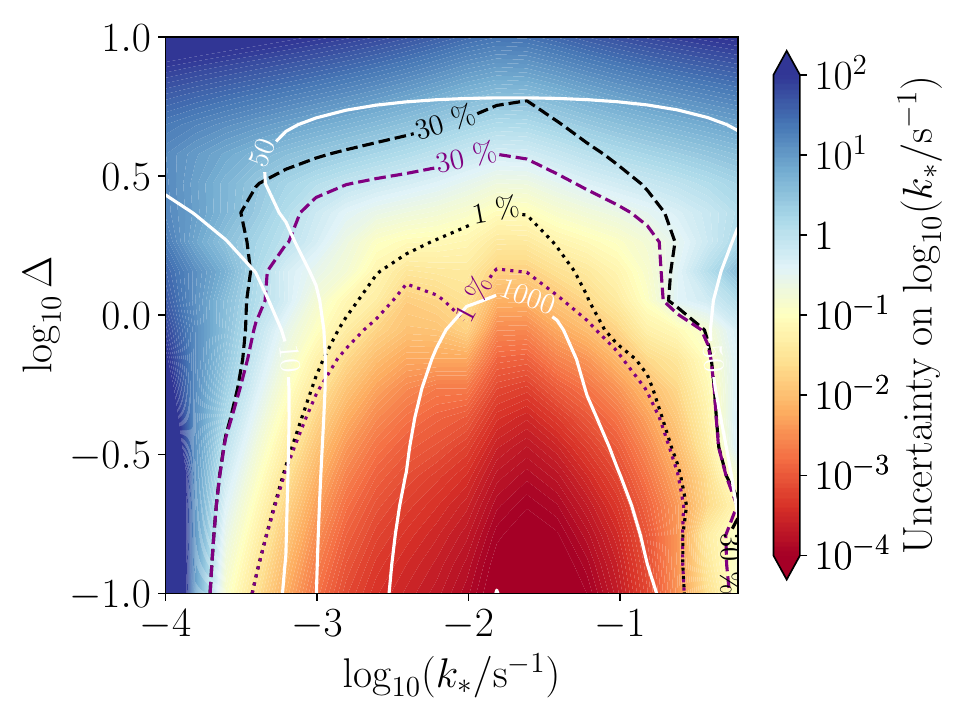}
    \caption{
    From left to right:
    Relative uncertainties of each of the parameters of the LN template, computed with an FIM forecast injecting an LN $\Pz$ with fixed amplitude $\log_{10} A_s = -2.5$ and varying $k_*$ and $\Delta$. 
    Black (purple) contours show uncertainties without (with) astrophysical foregrounds. The white line indicates SNR values. }
    \label{fig:LN_scan_Fisher}
\end{figure}
In Fig.~\ref{fig:LN_scan_Fisher}, we scan the parameter space in $k_*$ and $\Delta$ estimating the relative uncertainties on all LN parameters using the FIM method. 
We fix the amplitude of $\Pz$ to $\log_{10} A_s=-2.5$.
The SIGW amplitude scales like $\Omega_{\rm GW} \sim A_s^2$, and in the high SNR limit we expect the uncertainties on the parameter to scale inversely $\sim 1/A_s^2$.
The results highlight the great sensitivity that is achievable on a SIGW background if the peak lies around the peak sensitivity of LISA, $k_*\sim 10^{-3}-10^{-1}\,\mathrm{s}^{-1}$. 
In that range, the width $\Delta$ for an LN scalar spectrum can be measured with an accuracy of order 10\% or better if $\Delta\lesssim \mathcal O(1)$.
Notice from Fig.~\ref{fig:LN_scan_Fisher} that the purple contours, marking the sensitivity on the primordial SIGW background accounting for astrophysical foregrounds, degrade when $k_*$ coincides with the expected peak of the white-dwarfs (WD) galactic foreground, and the two GW backgrounds are less distinguishable \cite{Racco:2022bwj}.

\begin{figure}[h!]
    \centering
    \includegraphics[width=0.7\linewidth]{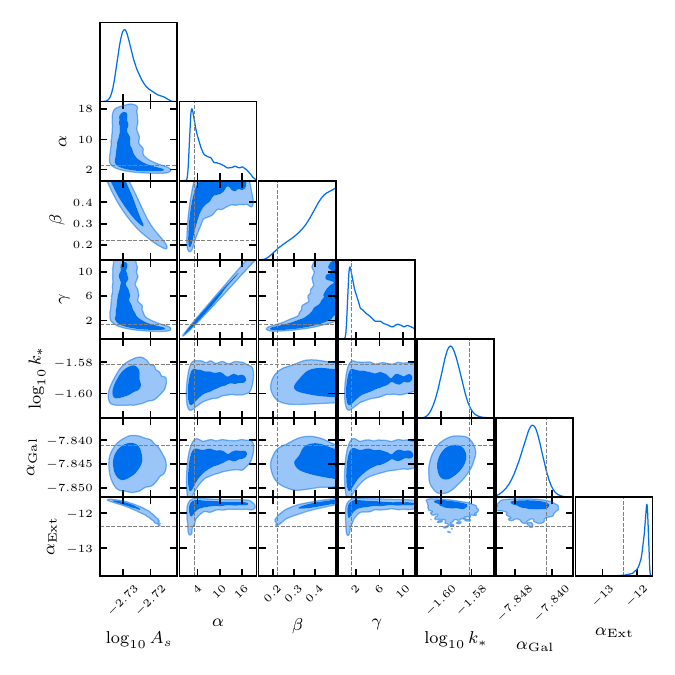}
    \caption{
Same as Fig.~\ref{fig:ln_corner}, but for a 
 recovered BPL curvature power spectrum assuming an injected
 signal motivated by the USR benchmark model. 
$k_*$ is expressed in $1/\mathrm{s}$ units.}
    \label{fig:usr_bpl_corner}
\end{figure}
\begin{figure}[h!]
    \centering
    \includegraphics[width=0.49\linewidth]{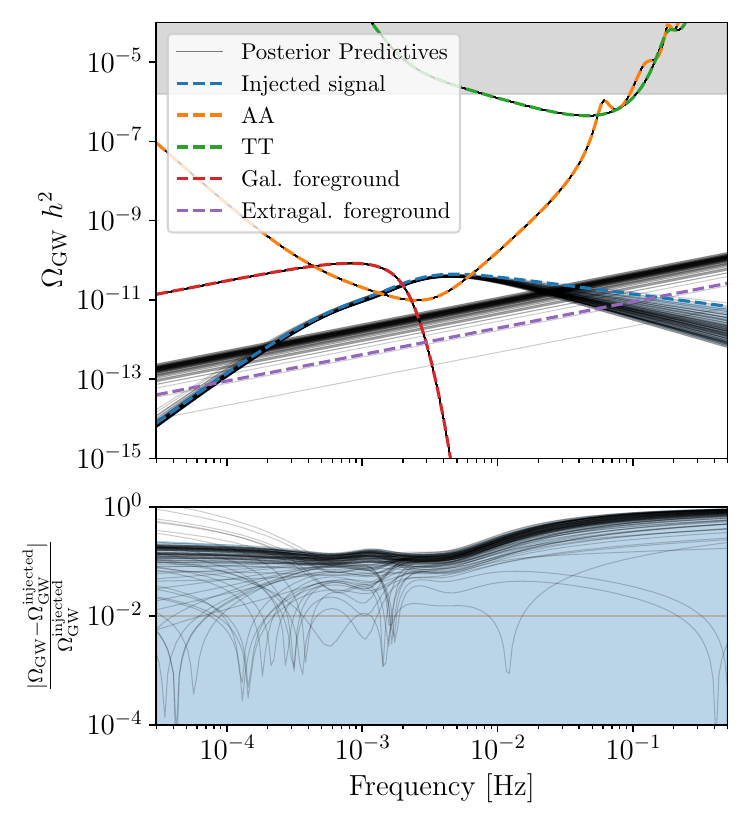}
    \includegraphics[width=0.49\linewidth]{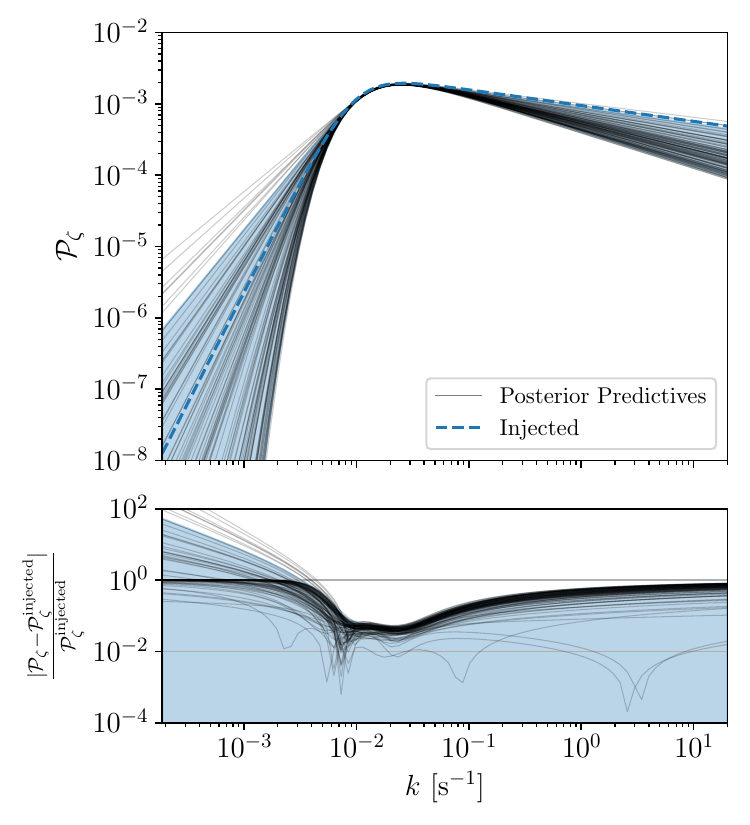}
    \caption{
Same as Fig.~\ref{fig:binned_no_injection_15}, but for an injected signal from a BPL $\Pz$, recovered using the BPL template. The injected parameters are motivated by the USR benchmark model.  
}
\label{fig:usr_bpl_posterior_predictives}
\end{figure}

\noindent
\paragraph{Broken power law.} 

The second injected signal is a BPL that is derived from the USR model discussed in Sec.~\ref{sec:benchmodelUSR} and~\ref{sec:USR_pzeta_computation}, with input parameters as in Eqs.~\eqref{eq:BPL} and \eqref{eq:BPL_bench}: $\log_{10}A_s = -2.71$, $\log_{10} (k_*/ \mathrm{s}^{-1}) = -1.58$, $\alpha = 3.11$, $\beta = 0.221$, $\gamma = 1.25$.
We reconstruct the signal using a BPL template.
The results of the reconstruction of the signal using the USR model will be discussed below.
The free parameters for this approach are
\begin{equation}
    {\bm \theta}_{\rm cosmo} =\{
    \log_{10} A_s,
    \log_{10} \left(k_*/\mathrm{s}^{-1}\right),
    \alpha,
    \beta,
    \gamma\}.
\end{equation}

In Fig.~\ref{fig:usr_bpl_corner} we show a corner plot of the reconstructed parameters of the broken power-law template, while Fig.~\ref{fig:usr_bpl_posterior_predictives} displays the posterior predictive distribution for the SIGW (left panel) and curvature power spectrum (right panel). 
While the amplitude of the $\Pz$ peak $A_s$ and the peak position $k_*$ are well reconstructed, the infrared (IR) spectral index $\alpha$ and the smoothing coefficient $\gamma$ are poorly constrained, and $\alpha$ and $\gamma$ appear to be very degenerate as the IR tail of the signal is hidden by the galactic foreground. 
This can be seen in the corner plot of Fig.~\ref{fig:usr_bpl_corner} as well as from the right panel of Fig.~\ref{fig:usr_bpl_posterior_predictives}, where the slope for $k<k_*$ has a large uncertainty.
In order to reconstruct $\alpha$ and $\beta$ with some precision, one would need to be sensitive to the tails of the signal outside the peak region. 
This would be only possible for a much larger signal, which nevertheless would have to compete with stringent bounds from PBH overproduction.
With the relatively low SNR injected here, changes in the tilt can be traded for a smoother turnover around the peak, and vice versa. 
Furthermore, there is a residual correlation between $\beta$ and $\gamma$, although it is less pronounced.
Biases in $A_s$, $\beta$, and $\alpha_{\rm Ext}$ are also evident due to their degeneracy in the high-frequency tail. 
Specifically, there is a tendency to reconstruct higher foreground values compared to the UV part of the SIGW spectrum. 
Importantly, we have verified that this bias is not an artifact introduced by the additional degeneracies induced by $\gamma$, which is correlated with both tilts and amplitude. This conclusion is supported by tests we ran with $\gamma$ fixed to its injected value.

Notably, this bias does not appear when the same signal is reconstructed using the USR model. The USR model is inherently less flexible, and its UV tilt is better constrained, thereby mitigating the impact of degeneracies.

\begin{figure}[h]
    \centering
    \includegraphics[width=0.49\textwidth]{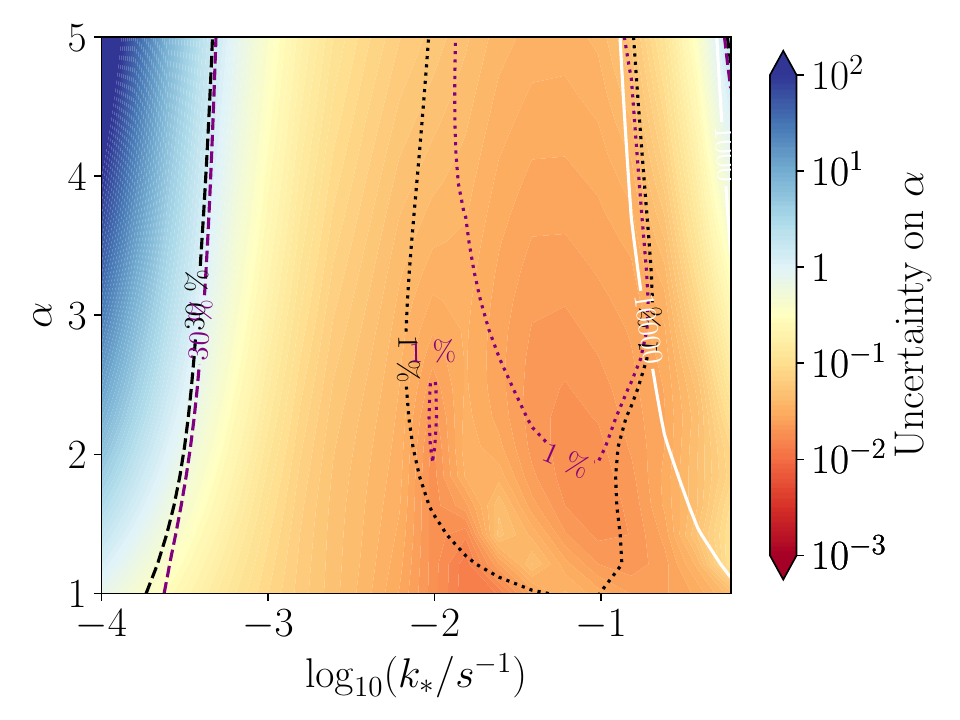}
    \includegraphics[width=0.49\textwidth]{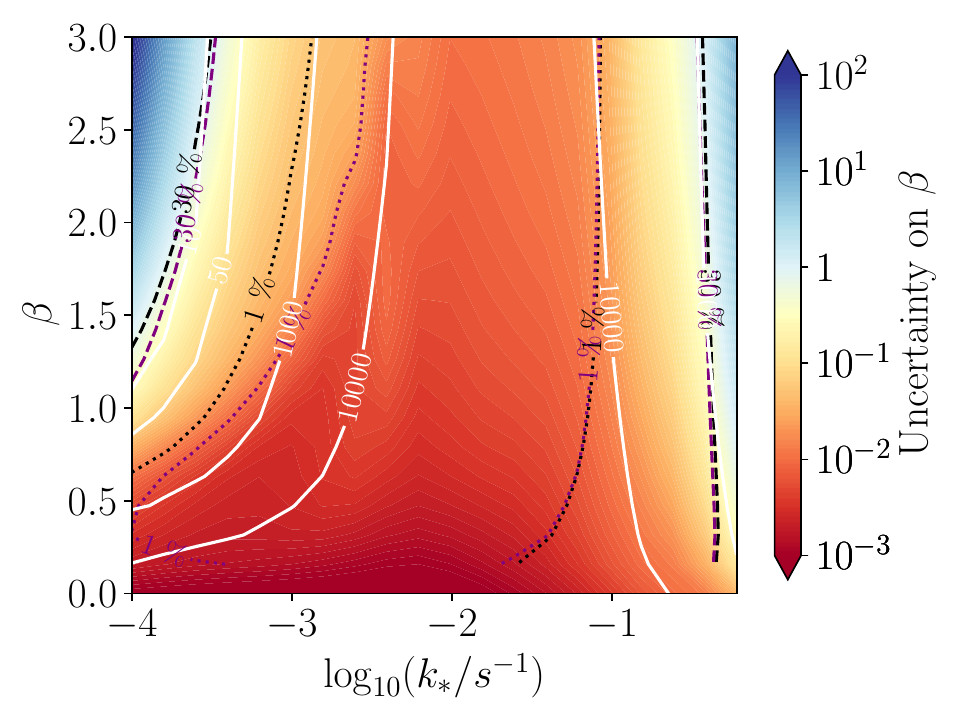}
    \caption{
    Uncertainty on $\alpha$ (left panel) and $\beta$ (right panel) computed with the FIM approach for the broken power law ${\cal P}_\zeta$. The remaining parameters injected are those of the benchmark scenario in Eq.~\eqref{eq:BPL_bench}. 
    We fix $\gamma$ to the injection to remove the degeneracies with the two tilts in the small SNR regions of the parameter space.}
    \label{fig:FIM_BPL}
\end{figure}

In Fig.~\ref{fig:FIM_BPL} we show FIM estimates of uncertainties on the tilt parameters depending on the injected BPL shape and $k_*$.
We stress that both tilts and $\gamma$ can be independently resolved only if one observes the tail of the signal with a sufficiently high SNR.  
If one only observed the SIGW peak, a shallower (steeper) tilt can be traded off for a smoother (faster) transition.
We see this happening in Figs.~\ref{fig:usr_bpl_corner} and \ref{fig:usr_bpl_posterior_predictives}, where large degeneracies exist between $\gamma$ and the other parameters. As a consequence, these posterior can hardly be approximated with the FIM.
Furthermore, in part of the parameter space explored in Fig.~\ref{fig:FIM_BPL}, we would obtain ill-conditioned FIM because of this degeneracy. In order to avoid this, only in this case we remove $\gamma$ from the parameters of the model and fix it to the injected value. This should be considered a slightly less flexible model for ${\cal P}_\zeta$, and an optimal choice should be made when performing a real GW data inference.

The left panel of Fig.~\ref{fig:FIM_BPL} shows the absolute uncertainty achievable on $\alpha$ in the $(k_*, \alpha)$ plane, and the right panel shows the same for $\beta$.  We checked that these uncertainties do not depend on the injected value for the other tilt parameter.
The slope of the IR tail of the GW spectrum is only mildly dependent on $\alpha$ when $\alpha \gtrsim 1.5$, as discussed before, so most of the sensitivity comes from the signal in the frequency range around the peak. For this reason, the uncertainty on $\alpha$ reduces to $0.1$ or better only if the SIGW is well within the LISA range, and on the right of the galactic WD foreground ($10^{-2}<k_* /s^{-1}<10^{-1}$).
Still, it is very interesting to notice that steep values of the tilt, higher than $\alpha \simeq 1.5$, which are fully covered by the causality tail $\Omega_{\rm GW}\sim f^3$ \cite{Caprini:2009fx,Cai:2019cdl,Allahverdi:2020bys,Hook:2020phx} in the SIGW spectrum (up to log-corrections), can be well constrained, as information on the tilt is still retained in the shape of the double peak feature of the signal close to the dominant peak, provided no large degeneracies with other parameters controlling the shape of the peak are present.  
The sensitivity to $\beta$ (right panel of Fig.~\ref{fig:FIM_BPL}) is instead much better, as it determines the UV slope $\Omega_{\rm GW}(f)\sim f^{-2\beta}$. Therefore, $\beta$ cannot be measured with an uncertainty smaller than $0.1$ only if the SIGW lies outside LISA's peak range ($k_*>10^{-1}\, \mathrm{s}^{-1}$) or if $\beta\gtrsim{-2}$, where the SIGW background falls too quickly in the UV.

\subsubsection{Spectra with oscillations}

\noindent
\paragraph{Turns in multi-field inflation.}

As a benchmark example of a primordial feature in the power spectrum, we analyze a signal arising from turns in multi-field space as introduced in Sec.~\ref{sec:temposc}.
The free parameters for this analysis are
\begin{equation}
    {\bm \theta}_{\rm cosmo} =\{
    \log_{10} A_s,
    \log_{10} \left(k_*/\mathrm{s}^{-1}\right),
    \delta,
    \eta_\perp,
    F\}.
\end{equation}

\begin{figure}[h]
    \centering
    \includegraphics[width=0.62\linewidth]{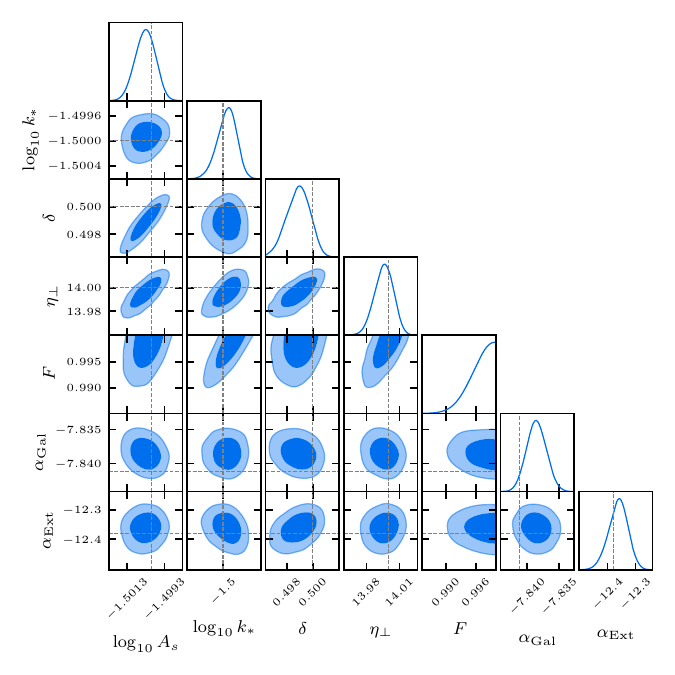}
    \caption{
Same as Fig.~\ref{fig:ln_corner}, but for a simulated signal motivated by the multi-field scenario with sharp turns from \eqref{eq:Pzeta-strong-sharp-turn}. 
$k_*$ is expressed in $1/\mathrm{s}$ units.
 }    \label{fig:oscillations_multi_field_corner}
\end{figure}
\begin{figure}[h!]
    \centering
    \includegraphics[width=0.49\linewidth]{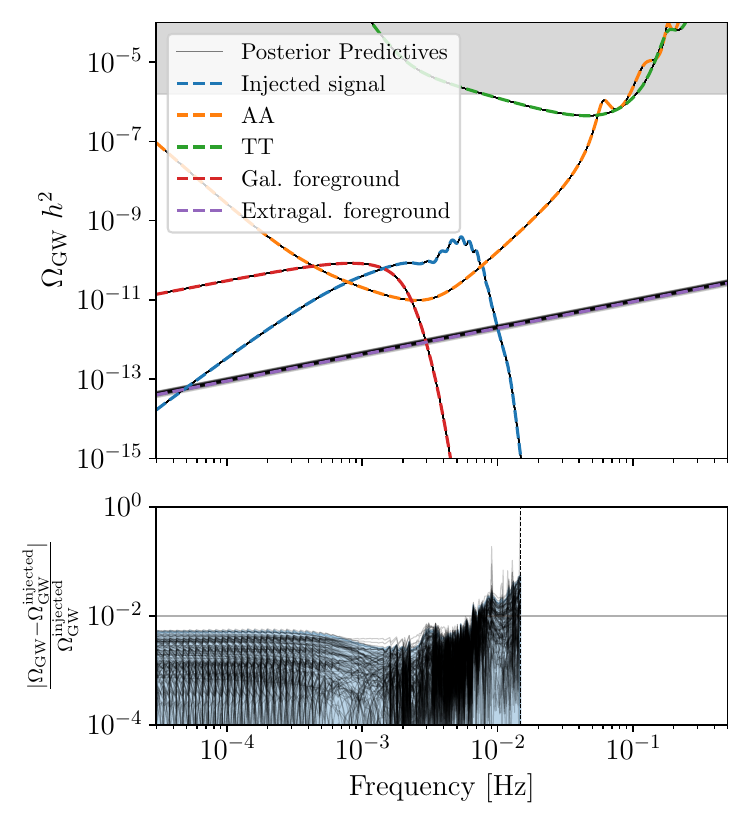}
    \includegraphics[width=0.49\linewidth]{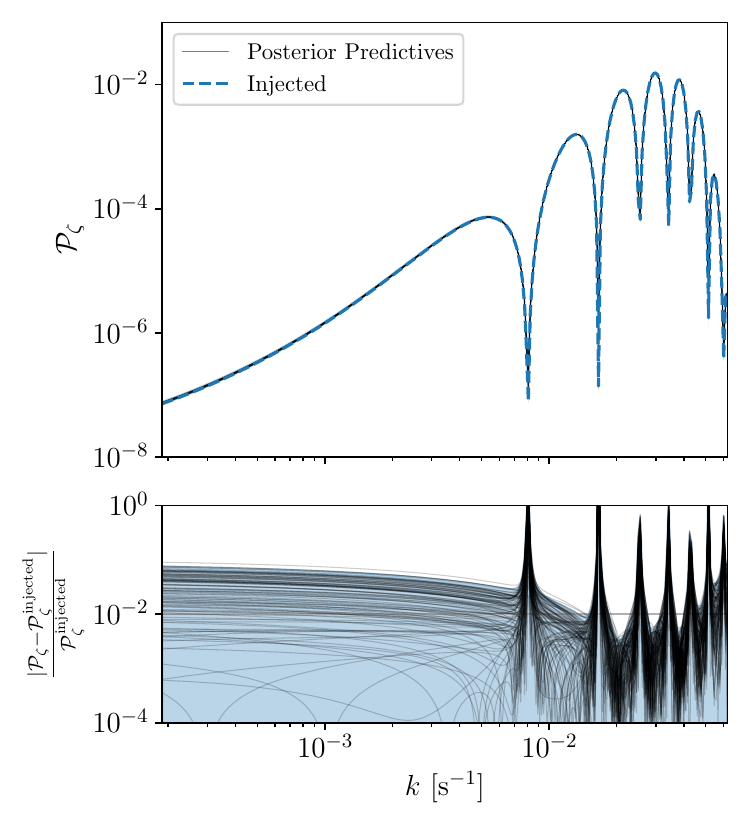}
    \caption{
    Same as Fig.~\ref{fig:binned_no_injection_15}, for the signal generated through a multi-field scenario with sharp turns from Eq.~\eqref{eq:Pzeta-strong-sharp-turn}.}
\label{fig:oscillations_multi_field_posterior_predictives}
\end{figure}

In Fig.~\ref{fig:oscillations_multi_field_corner} we show the posterior distributions for key parameters governing the sharp-turn scenario in multi-field inflation, along with the foreground parameters. The injection assumes the benchmark scenario where the parameters controlling the template \eqref{eq:Pzeta-strong-sharp-turn} are fixed as in Eq.~\eqref{eq:benchES}.
We see that in this case, the parameter $F$ is constrained to be close to unity with better than percent precision, showing the high sensitivity to the template oscillations. In this case, the signal amplitude and central scale $k_*$ are weakly correlated, while the former is still positively correlated to both $\delta$ and $\eta_\perp$ which control the enhancement factor. 
Due to the ideal location of the SIGW peak, we also observe weak correlations between the foreground parameters and the signal parameters.

Figure \ref{fig:oscillations_multi_field_posterior_predictives} shows the posterior predictive distribution for $\Omega_{\rm GW}$ and $\Pz$. In this example, the main peak of the SGWB lies within the LISA sensitivity band and above both astrophysical foregrounds. As a result, both the shape and amplitude of the peak in $\Omega_{\rm GW}$, along with the $O(20\%)$ modulations, are reconstructed at the percent level. Since the frequency of these modulations is linked to the oscillations in $\mathcal{P}_{\zeta}$ through the assumed thermal history at horizon re-entry, the oscillations are also reconstructed with high accuracy--see the right bottom plot. In this fortunate case, it would be possible to pinpoint the duration and strength of the field-space turn, as well as the inflationary time scale of the phenomenon. The latter is related to the oscillation frequency, as it is customary from the sharp feature phenomenon, while the former two can be disentangled by combining the peak amplitude, its location, and the frequency of the modulations.

\begin{figure}[t]
    \centering
    \includegraphics[width=0.5\linewidth]{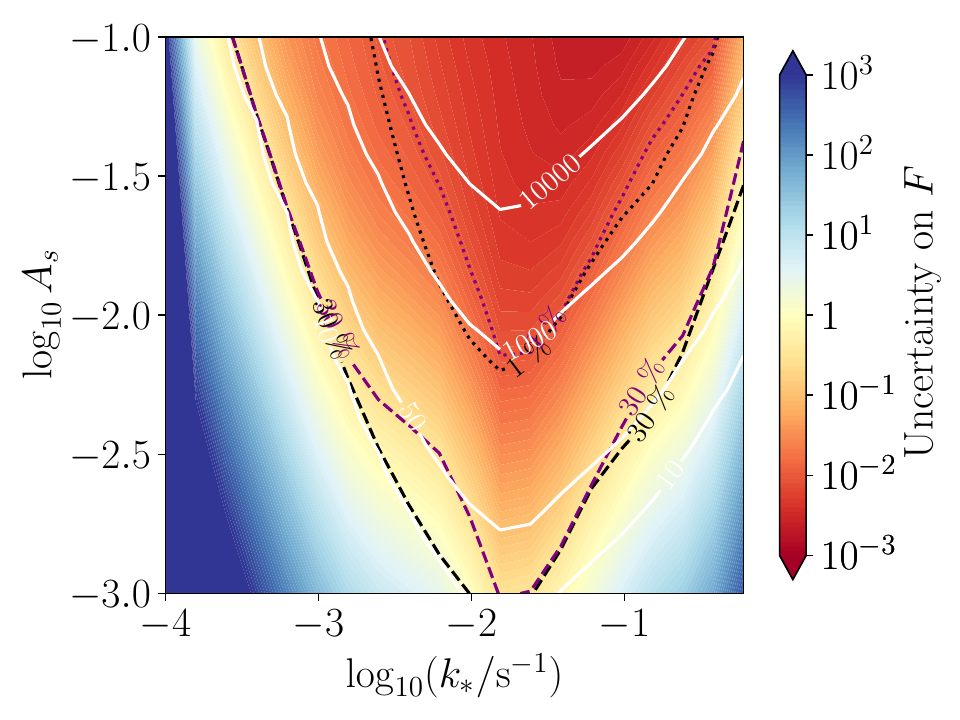}
    \caption{Fisher analysis for the oscillation template from multi-field inflation with turns.
    We vary $k_*$ and $A_s$ while keeping the remaining parameters fixed to the benchmark values \eqref{eq:benchES}. } 
    \label{fig:oscillations_multi_field_fisher}
\end{figure}

Finally, the results of a Fisher analysis, highlighting the uncertainty in the parameter $ F $ associated with the oscillatory behavior, are presented in Fig. \ref{fig:oscillations_multi_field_fisher}. There, we vary the power spectrum amplitude $ A_s $ and the position of the main peak, while keeping the other parameters fixed to the benchmark values discussed just above. This simplification is useful for illustrative purposes, as the parameters in the current model are not independent. Notably, when the signal is centered near the LISA sweet spot at $ \log_{10}( k_*/s^{-1} )\simeq -2 $, the oscillations can be accurately detected even with a moderate enhancement such as $\log_{10}A_s \simeq -3 $.

\noindent
\paragraph{Rapid transitions between SR and USR phases.}
\label{sec:rapid_transitions}

The other injected spectrum with oscillatory features is characteristic of single field models with fast transitions from an SR to a USR phase described by Eqs.~\eqref{eq:USR_osc},~\eqref{eq:Pzeta_sharpfeature}.
The free parameters for this analysis are
\begin{equation}\label{eq:theta_rapid_trans}
    {\bm \theta}_{\rm cosmo} =\{
    \log_{10} A_s,
    \log_{10} \left(k_*/\mathrm{s}^{-1}\right),
    \nu_{\rm I},
    \nu_{\rm II},
    F\}.
\end{equation}
We consider the benchmark scenario with the input parameters listed Eq.~\eqref{eq:bench_sharpfeature}: $\log_{10}A_s = -2.58$, $\log_{10} (k_*/\mathrm{s}^{-1}) = -2.02$, $\nu_{I} = 1.95$, $\nu_{II} = 1.61$, $\gamma = 1.67$, $F = 1$.
We perform the MCMC Bayesian inference modelling of the
signal using the template~\eqref{eq:Pzeta_sharpfeature}, which allows us to turn on the oscillations smoothly by varying the parameter $F$ from 0 to 1. 
The value $F=0$ corresponds to a featureless BPL similar to the one considered above. Furthermore, we fix $\gamma$ as the strong degeneracy between $\alpha$ and $\gamma$ (see \cref{fig:usr_bpl_corner}) makes sampling challenging and since our main concern lies in constraining $F$.

\begin{figure}[h!]
    \centering
    \includegraphics[width=0.6\linewidth]{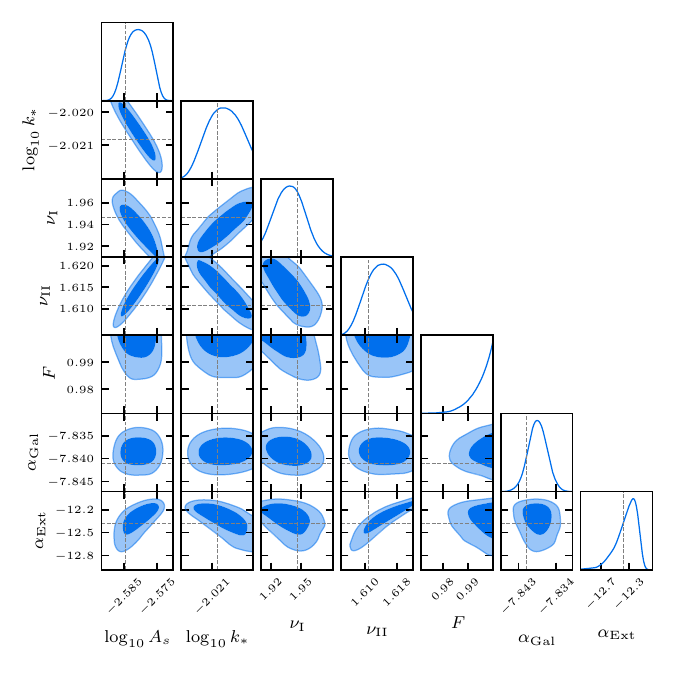}
    \caption{
Same as Fig.~\ref{fig:ln_corner}, but for a simulated signal from Eq.~\eqref{eq:Pzeta_sharpfeature}. 
$k_*$ is expressed in $1/s$ units. Note the injected value $F=1$ is not visible sitting at the edge of the plot. 
}
\label{fig:oscillations_sharpfeature_corner}
\end{figure}

\begin{figure}[h!]
    \centering
    \includegraphics[width=0.49\linewidth]{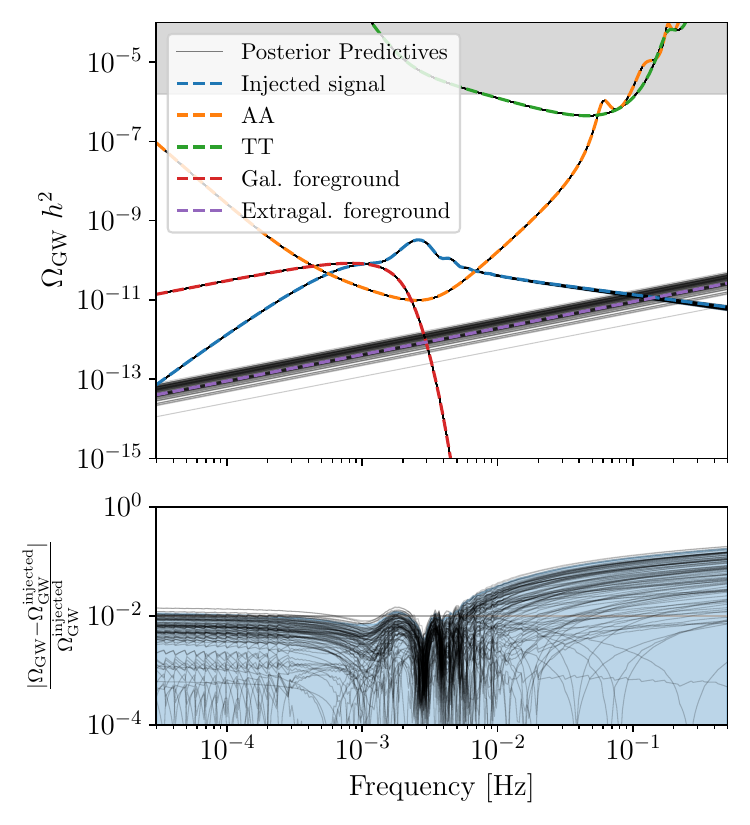}
    \includegraphics[width=0.49\linewidth]{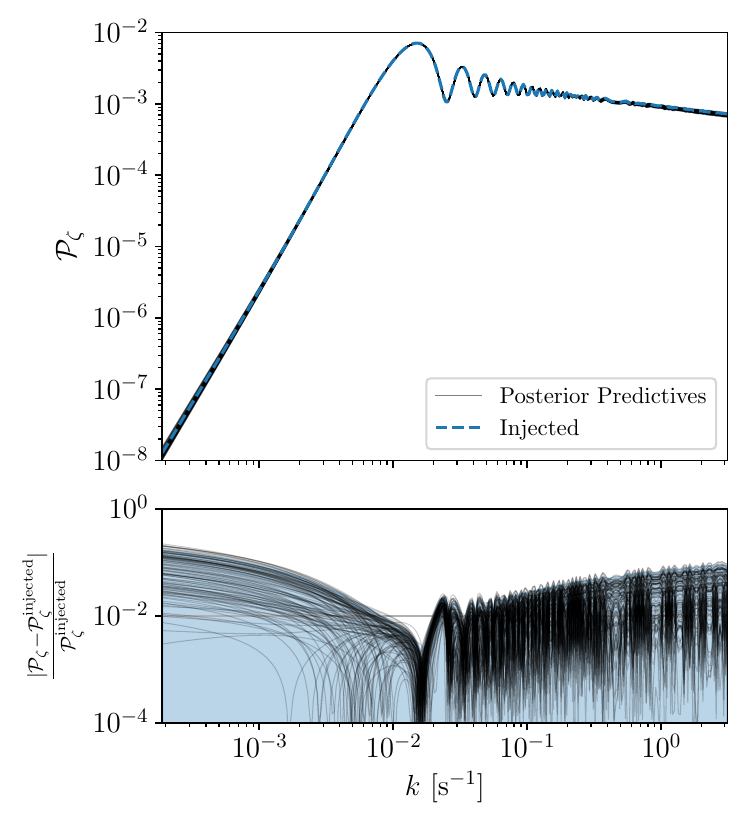}
    \caption{
    Same as Fig.~\ref{fig:binned_no_injection_15}, for the signal generated through a fast transition from SR to USR from \eqref{eq:Pzeta_sharpfeature}. Both $\mathcal{P}_\zeta$ and $\Omega_{\rm GW}$ are reconstructed very well.}\label{fig:oscillations_sharpfeature_posterior_predictives}
\end{figure}

In Fig.~\ref{fig:oscillations_sharpfeature_corner} we show the posterior distribution for each parameter of the signal and foregrounds.
First of all, we see the parameter $F$ controlling the relevance of the oscillations over the smooth BPL is very tightly constrained around unity. This tells us the presence of oscillations can be resolved with high accuracy for such a high-SNR signal. 
The sensitivity to oscillations is mainly driven by the dominant peak, as we will discuss in the following. 
We also find tight correlations between the parameters, which are non-trivially connected in the signal template \eqref{eq:USR_osc}. In particular, we observe a strong correlation between the BPL tilts and the amplitude, due to the large impact of the former on the overall amplitude of the dominant peak. 
The negative correlation in the $(A_s,k_*)$ plane is probably induced by the way the dominant peak, contributing to most of the SNR, can be adjusted, as one could lower the characteristic scale by enhancing the amplitude.

In Fig.~\ref{fig:oscillations_sharpfeature_posterior_predictives} we show the posterior predictive distribution for both $\Omega_{\rm GW}$ and $\Pz$.
The presence of a dominant peak at scales around $k_*$ in the right plot leads to a distinctive large enhancement of the SIGW signal around peak frequencies seen in the left plot.
Additional oscillations in the SIGW spectrum can be observed at larger frequencies, although the second-order emission soon washes out further oscillations in the UV tail. 
The residuals of $\Omega_{\rm GW}$ show that the IR tail of the signal is reconstructed at around the percent level, with a flat behavior due to the causality tail dominating the IR.
On the other hand, the relative deviation grows larger than ${\cal O}(10)$\% percent in the UV part of the plot, due to the finite precision at reconstructing $\nu_{\rm II}$.
The best accuracy is obtained around the peak, as expected.
Correspondingly, in the right plot, $\Pz$ is reconstructed with better than percent accuracy around the peak, while the reconstruction degrades in both tails. 
The oscillations are reconstructed for a few cycles in $k$, while they are lost in the UV as seen in the bottom panel showing the relative deviation from the injected signal. 
The envelope of the out-of-phase oscillations behaves following the underlying power-law tail. 
Note that the tails are reconstructed better than in the pure BPL scenario, as in this template \eqref{eq:Pzeta_sharpfeature}, the shape of the dominant peak also brings information on the parameter $\nu_{\rm I,II}$ controlling the tails. 

\begin{figure}[t!]
    \centering
    \includegraphics[width=0.5\linewidth]{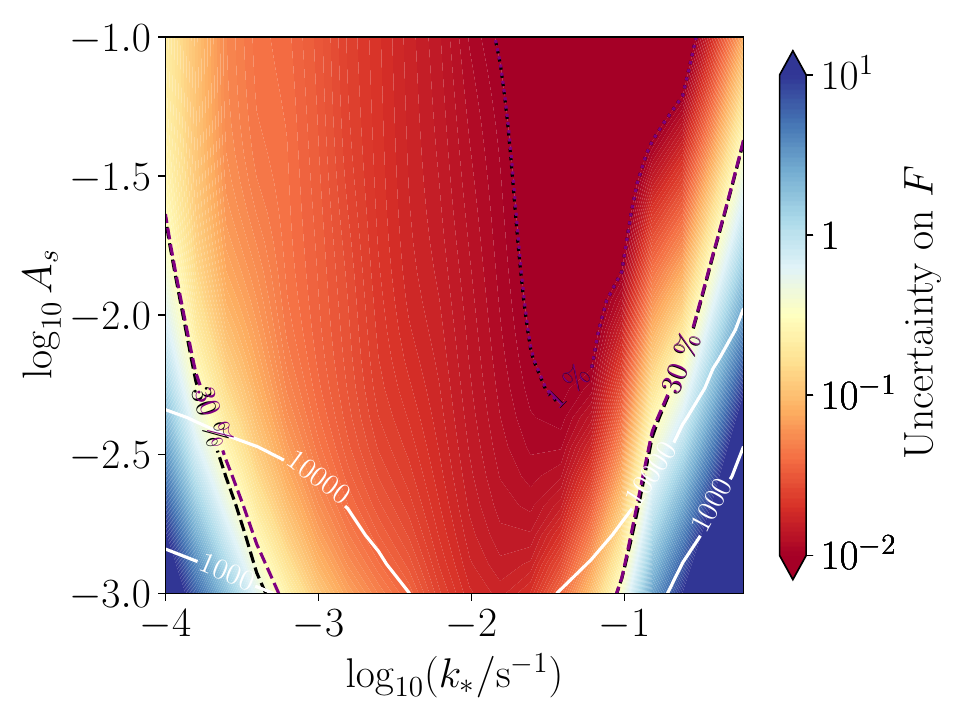}
    \caption{
    Absolute uncertainty on $F$ estimated using the FIM for the oscillation template from a sharp transition between SR and USR.
    We vary $\log_{10} k_*$ and $\log_{10} A_s$ while keeping the remaining parameters fixed to the benchmark values \eqref{eq:bench_sharpfeature}. } 
    \label{fig:oscillations_sharpfeature_fisher}
\end{figure}

We perform a Fisher analysis focusing on the uncertainty on $F$, by varying $\log_{10}A_s$ and $\log_{10} k_*$, see Fig.~\ref{fig:oscillations_sharpfeature_fisher}. 
The oscillations are very well recovered, to $\mathcal{O}(10^{-2})$ uncertainty, if the peak of the signal falls within the LISA band.
This is because the oscillations in $\mathcal{P}_\zeta$ translate into oscillations mainly around the peak in the SGWB, as visible in Fig.~\ref{fig:oscillations_sharpfeature_posterior_predictives}.

\subsection{Single field USR inference}\label{sec:USR_inference}
\begin{figure}[t]
    \centering
    \includegraphics[width=0.62\linewidth]{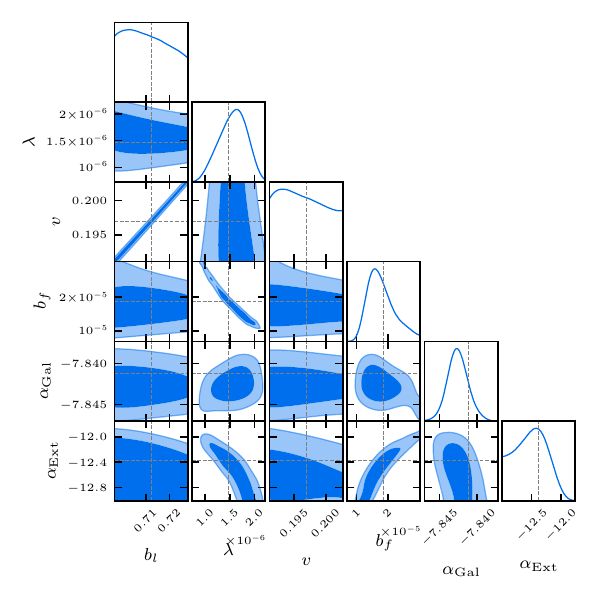}
    \caption{
    Same as Fig.~\ref{fig:ln_corner}, but for the USR reconstruction of the benchmark scenario.}
    %The parameter $v$ is expressed in $\mpl$ units.
    \label{fig:usr_mcmc_corner}
\end{figure}

In Fig.~\ref{fig:usr_mcmc_corner} we show the reconstruction capability of the USR model parameters of  Sec.~\ref{sec:benchmodelUSR}, obtained by running a MCMC Bayesian inference using the USR inflationary model with
free parameters
\begin{equation}
    {\bm \theta}_{\rm cosmo} =\{
    \lambda, v, b_l, b_f
    \}.
\end{equation}
controlling the inflaton potential. We assume that the CMB scale crosses the Hubble sphere $N = 58$ $e$-folds before the end of inflation. We therefore avoid modelling the reheating era, and postopone its inclusion for future work (see e.g. \cite{Allegrini:2024ooy}). 
The input values determining the injected signal were introduced in Eqs.~\eqref{eq:Germani-potential} and \eqref{eq:benchUSRvals}, but we report them here for convenience: 
$\lambda = 1.47312\times  10^{-6}$, 
$v = 0.19688$, 
$b_l  = 0.71223$, 
$b_f = 1.87\times  10^{-5}$.

We understand the results as follows. The height of the peak in $\Pz$ is proportional to $\lambda$ and it is also sensitive to the tuning of $b_f$.\footnote{With other potential parameters fixed, we did find the approximate behavior $\Pz \propto (1-b_f/b_{f,*})^{-n}$ in the parameter region supporting peaked $\Pz$. Here, $b_{f,*}$ and $n > 2$ are parameters that depend on the remaining parameters of the potential. Such scaling is observed in other models~\cite{Karam:2023haj}.} This results in the negative correlation between $\lambda$ and $b_f$. 
The self-coupling $\lambda$ can be constrained, even though with only $\mathcal{O}(1)$ precision, because it controls the slope of the potential and therefore the SR parameter $\eta_H$ \textit{before} the inflection point, which determines the growth of $\Pz$ before the peak as discussed below Eq.~(\ref{eq:BPL}). 
The parameters $b_l$ and $v$ appear to be strongly correlated, meaning that the linear term in $b_l/v$ in Eq.~\eqref{eq:Germani-potential} gives the dominant dependence on $b_l$ in the potential.
The galactic background is well reconstructed due to its large magnitude, while the extragalactic one is completely hidden by the USR signal.

\begin{figure}[t]
    \centering
    \includegraphics[width=0.49\linewidth]{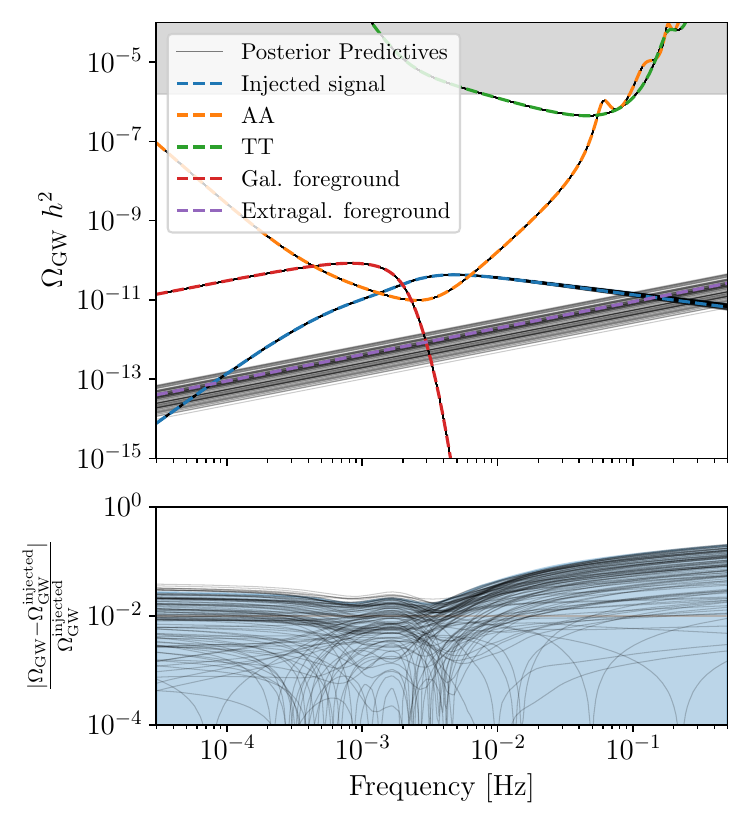}    \includegraphics[width=0.49\linewidth]{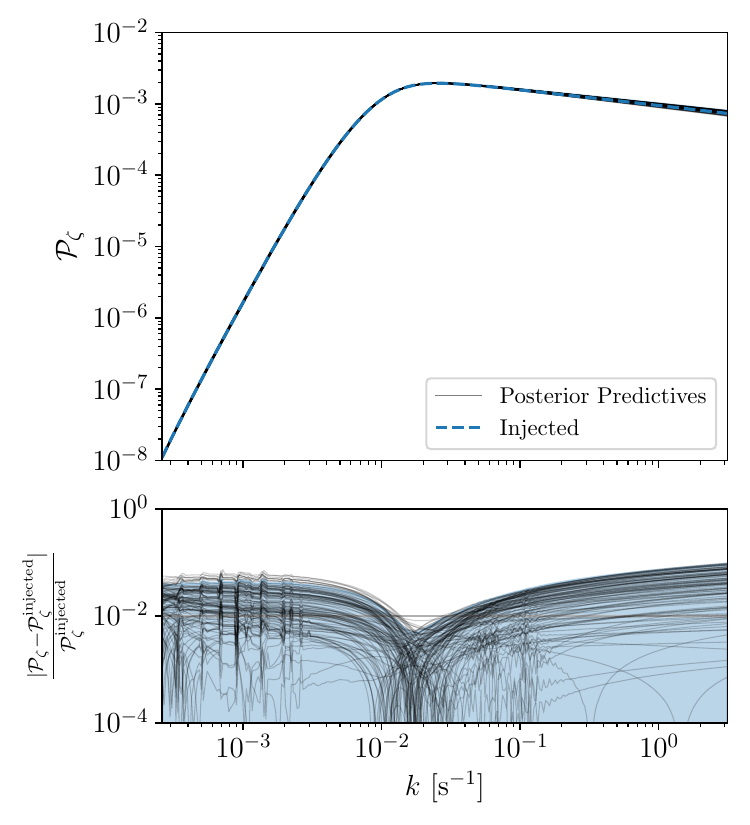}
    \caption{
    Same as Fig.~\ref{fig:binned_no_injection_15}, but for the USR reconstruction of the benchmark scenario. }
    \label{fig:usr_posterior_predictives}
\end{figure}

Figure~\ref{fig:usr_posterior_predictives} shows the posterior predictives in $\Omega_{\rm GW} h^2$ and in $\mathcal{P}_\zeta$. 
It is interesting to compare the right panel of this figure with that of Fig.~\ref{fig:usr_bpl_posterior_predictives}, which is obtained with the same injected signal but a different template for the reconstruction. In the present case, the spectrum of scalar perturbations $\Pz$ is reconstructed with excellent precision, even if LISA is sensitive only to the peak. 
This comes from the fact that the spectrum for the USR model has a universal slope $\sim k^4$ in the IR,
 whereas the IR slope is a free parameter for the BPL model.

\begin{figure}[t]
    \centering
    \includegraphics[width=0.49\linewidth]{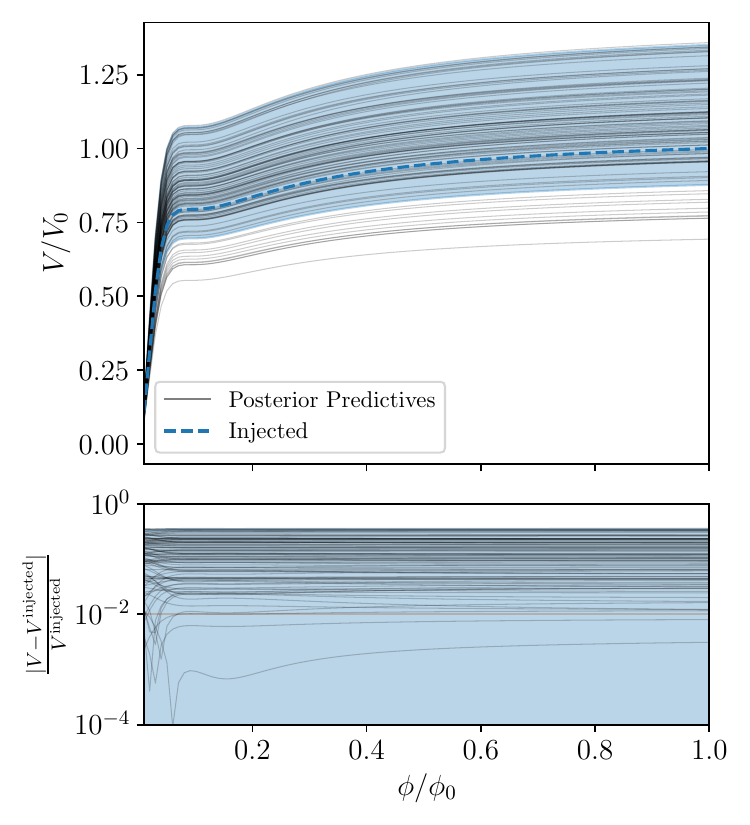}
    \caption{Posterior predictive distribution of the potential $V(\phi)$ for the USR reconstruction of the benchmark scenario.}
    \label{fig:usr_posterior_predictives_V}
\end{figure}

The relatively large uncertainty on the overall potential amplitude $V(\phi)$ in Fig.~\ref{fig:usr_posterior_predictives_V} is due to the degeneracies between the overall 
scale $V_0 \sim \lambda v^4$
and the parameter $b_f$ controlling the enhancement.
As we are only constraining the enhanced part of the spectrum, there is a tight correlation between 
$\lambda$ and $b_f$.
Adding information from CMB data in the inference would reduce this uncertainty by adding an independent constraint on $V_0$.

\begin{figure}[t]
    \centering
    \includegraphics[width=0.49\textwidth]{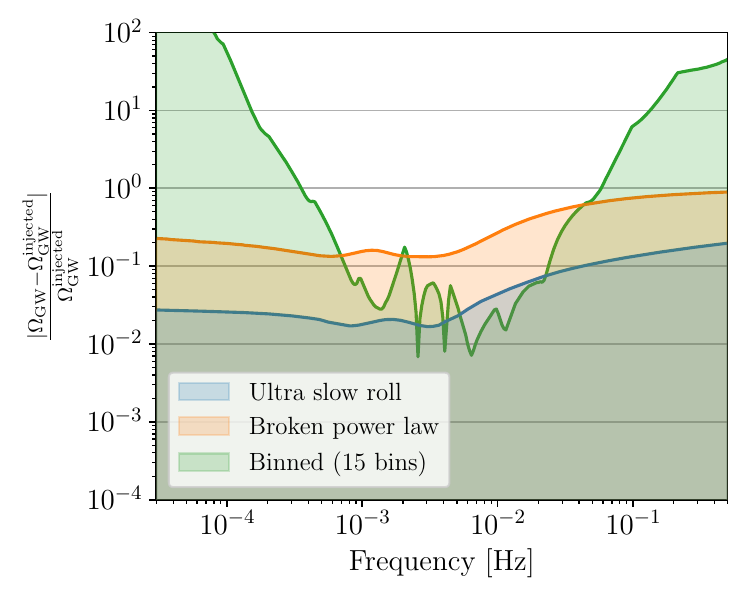}
    \includegraphics[width=0.49\textwidth]{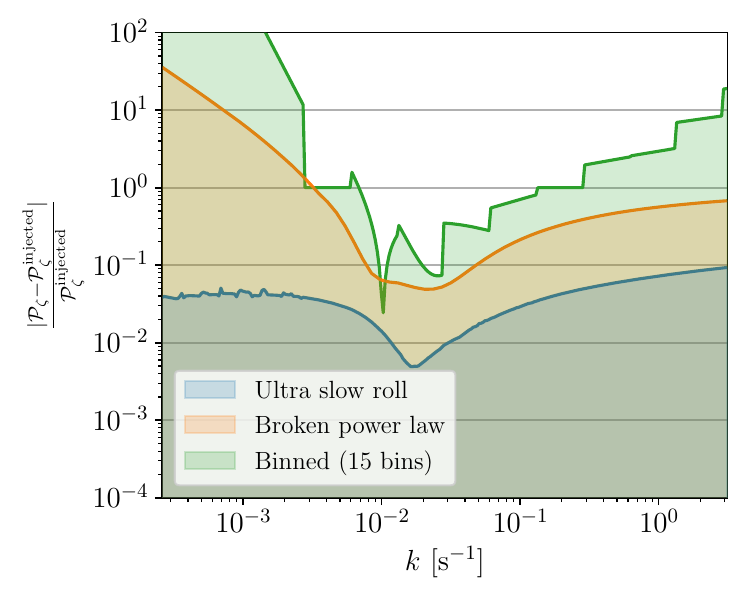}
    \caption{Comparison of the residuals between the three different methods for recovering the signal injected assuming the benchmark USR model.}
    \label{fig:residuals_comparison}
\end{figure}

\paragraph{Comparison between different methods.}
We can compare the performance of different methods when fitting the same injected signal, which is taken to be the USR benchmark scenario. 
In Fig.~\ref{fig:residuals_comparison} we show the upper bound at 95\% C.L. on the relative difference between the posterior predictive distribution and the injected signal for the binned, template-based, and ab initio USR approaches.

We observe that the binned method provides a competitive constraint on $\Omega_{\rm GW}h^2$ in the central frequencies close to the peak (barring oscillations induced by the poor resolution associated with choosing 15 bins). 
However, the constraint quickly degrades at both ends, due to the unconstrained curvature spectral amplitude there. 
The template-based method (assuming a priori a SIGW from a BPL scalar power spectrum) improves the reconstruction of the tails, but results in an overall loss of precision of a factor ${\cal O}(6)$ with respect to the posterior predictive derived assuming the USR scenario.
This is most probably due to the larger number of parameters in the BPL template compared to the USR model, and the known degeneracy between $\gamma$ and the two tilts around the peak. 

Also, $\Pz$, shown in the right panel, is most tightly constrained when using the USR model, showing the 
 effectiveness and robustness of our analysis pipeline in reconstructing the injected high SNR signal.
The BPL template gives an intermediate result on the IR tail, which, however, degrades much faster than  USR, due to the limited information on the tail of the SIGW, which is mostly controlled by the causality tail. 
Finally, the binned method gives a worse reconstruction of both the IR and UV tail, due to the small information within the LISA band about these regimes, and the independence assumed in this method between the central bins (best constrained) and the ones on the sides. 

Overall, 
the binned method proves to be a powerful approach to explore the interpretation of a primordial background at LISA within a more agnostic approach. The comparison with specific SIGW templates does not significantly outperform the binned method for the range of frequencies around the peak, which are the best constrained by LISA.
However, consistently with expectations, adopting the correct USR model provides greater accuracy in capturing the features of the signal, leading to a more precise reconstruction. 
These findings demonstrate the power of also adopting inference analyses based on explicit {\it ab initio} models (of which USR is just an example) that could outperform traditional template-based approaches.
This, of course, assumes one can identify the best early universe model through model comparison. 
We will come back to discussing how to compare different scenarios in Sec.~\ref{sec:tests}.

\subsection{Non standard thermal histories}\label{sec:reseMDRD}

Using information on the SIGW spectrum, LISA would be
able to challenge the vanilla assumption that the SIGW was emitted during a RD era. 
As discussed in Sec.~\ref{sec:computation}, the kernels entering the computation of the SIGW spectrum bring information about the equation of state around the epoch of SIGW emission.

\paragraph{A sudden transition from eMD era to the RD era.}
We exemplify this case by showing how LISA can constrain the SIGWs emitted within an alternative thermal history by considering an early period of matter domination (eMD).
We further assume sudden reheating, as introduced in Sec.~\ref{sec_tran}. 
As we discussed, during this eMD epoch  $\Phi$ does not decay, leading to an enhancement of the SIGW spectra around the scale $k \gtrsim 1/\eta_{\rm R}$. 
We take as a benchmark a nearly scale-invariant spectrum  $\mathcal{P}_\zeta$, with a cutoff at placed at $k_{\max}$. 
We simplistically describe the spectrum as
\begin{equation}\label{eq:Pzeta_eMDRD}
\mathcal{P}_\zeta(k)=A_s \Theta\left(k_{\max }-k\right),    
\end{equation}
with benchmark parameters
\begin{equation}
A_s=2.1 \times 10^{-9},
\qquad 
\eta_{\rm R} = 2000\,\mathrm{s},
\qquad 
k_{\max} = 0.06 \, \mathrm{s}^{-1}.
\end{equation} 
Therefore the free parameters to examine in this case are 
\begin{equation}
    {\bm \theta}_{\rm cosmo} =\{
    A_s,
    k_{\rm max},
    \eta_{\rm R}
    \}.
\end{equation}

Our phenomenological parametrization should be regarded as a toy model, with the UV cut-off scale $k_\mathrm{max}$ introduced to ensure perturbativity, as assumed when computing the SIGW. For this reason, given the fact that the energy density contrast grows linearly with the scale factor during a MD era, i.e.~${\delta\rho}/{\rho}\propto a$, one can associate $k_\mathrm{max}$ as the scale at which the power spectrum of density contrast becomes unity, i.e.~$\mathcal{P}_\delta(k_\mathrm{max})=1$~\cite{Assadullahi:2009nf,Inomata:2019ivs,Inomata:2019zqy}, although our actual choice is slightly more restrictive. 
One then can easily understand why $k_\mathrm{max}$ depends on the scale we are probing, as the source is largest for modes that spent the most time within the horizon during the eMD era.
While we do not model the non-linear part of the spectrum, it may lead to further observational signatures \cite{Jedamzik:2010hq,Eggemeier:2022gyo,Fernandez:2023ddy,Padilla:2024cbq}. 
Finally, the template \eqref{eq:Pzeta_eMDRD} can be made more realistic by introducing a smooth cut-off.  

\begin{figure}[t]
    \centering
    \includegraphics[width=0.52\linewidth]{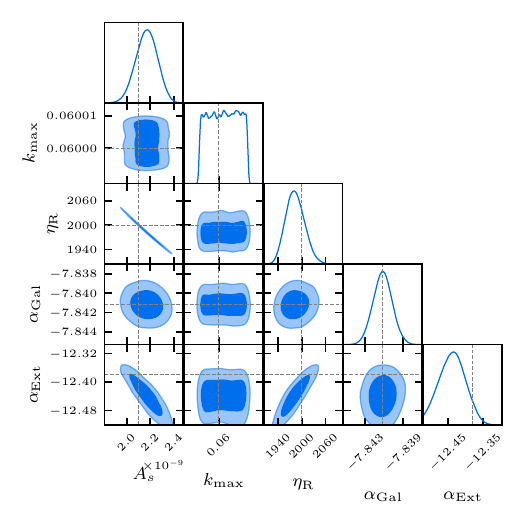}
    \includegraphics[width=0.47\linewidth]{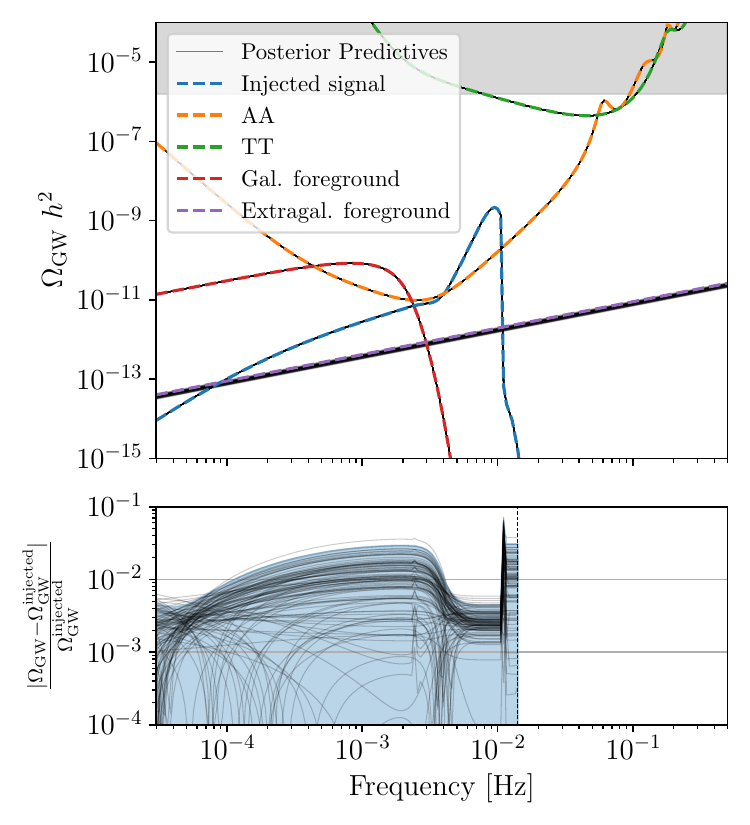}
    \caption{
    Left panel:
    Same as Fig.~\ref{fig:ln_corner}, but for the case of a nearly scale-invariant power spectrum and a sudden eMD to RD transition.
    $k_{\rm max}$ and $\eta_{\rm R}$ are expressed in units of $\mathrm{s}^{-1}$ and $\mathrm{s}$, respectively. 
    Right panel: 
    Corresponding posterior predictive distribution for $\Omega_{\mathrm{GW}}h^2$.
    The notation used matches the one in Fig.~\ref{fig:binned_no_injection_15}. }
    \label{fig:emd_corner}
\end{figure}

The parameters of this template are accurately reconstructed as shown in Fig.~\ref{fig:emd_corner} (left panel).
We notice that $\eta_{\mathrm R}$ and $A_s$ are strongly correlated, as the duration of the MD signal directly controls the growth of perturbations emitting SIGWs, which is therefore degenerate with the primordial amplitude. 
The cut-off scale is also strongly constrained, with a marginalised posterior distribution which is flat within a narrow range of scales corresponding to the resolution adopted in this forecast.
It should be kept in mind, however, that more realistic spectra would feature a smoother drop-off, alongside a contribution from non-linear scales not included here, thus jeopardizing the relevance of the constraining power on $k_{\rm kmax}$.
In the right panel of Fig.~\ref{fig:emd_corner}, we show the posterior predictive distribution for the SIGW. We find 
the reconstruction to be accurate up to the cutoff scale (better than a few $\%$). The SIGW spectrum is reconstructed well in the large-scale approximation, with the resonant amplification improving accuracy by an order of magnitude. 
The resonant peak is reconstructed with a larger accuracy due to its milder model dependence and due to its tilt being controlled by the resonant conditions (see discussion around Eq.~\eqref{eq:irdapp}).
Moreover, that part of the signal appears with a larger SNR in the LISA detector. 
The associated $\mathcal{P}_\zeta$ is accurately reconstructed as a flat spectrum with a maximum relative error of order $10 \%$. 

Our numerical pipeline can also be applied to other scenarios for the thermal history of the early Universe. Of particular interest would be the study of time-dependent EoS parameters on the SIGWs, like in the case of smooth-crossovers~\cite{Escriva:2023nzn,Escriva:2024ivo}, analogous to the QCD phase transition.

\subsection{Non-Gaussian effects on SIGWs}
\label{sec:NG}

As discussed in Sec.~\ref{sec:NG-SGWB}, the tensor power spectrum of SIGWs receives contributions from the four-point correlation function of curvature perturbations. This contribution can be split into disconnected and connected terms. While the disconnected one depends only on the scalar power spectrum, the connected part arises from the primordial trispectrum, 
i.e.~it is sensitive to primordial NG. 
And, as stressed in Sec.~\ref{sec:NG-SGWB}, for local NG, $\tau_{\rm NL}$ would be the key observable to extract a constraint on NG from SIGWs. However, 
when the curvature perturbation originates from a single fluctuating degree of freedom beyond the inflaton, the parameters $\tau_{\rm NL}$ and $f_{\rm NL}$ are connected as measures of higher-order correlations in the curvature perturbation, satisfying the relation  $\tau_{\rm NL}=\left(\frac{6}{5}f_{\rm NL}\right)^2$, which saturates the Suyama-Yamaguchi inequality \cite{Suyama:2007bg}. 
This relation has relevant implications for SIGWs, generated by models characterized by local-type NG, since observational constraints on one parameter can indirectly provide bounds on the other, assuming a given model. 
In the following analysis, we adopt the strategy of performing the analysis considering $f_{\rm NL}$ as a parameter of the model and assume the shape  \eqref{P-PG} of the full power spectrum.
We then discuss the implications for $\tau_{\rm NL}$ that derive from the constraints on $f_{\rm NL}$. In this case, we assume the 
curvature power spectrum to have a LN profile, (see Eq.~\ref{eq:PLN})\footnote{Note that our choice of the LN $\mathcal{P}_\zeta$ assumes that the dimensionless primordial curvature fluctuations -- including the higher order term coming from the trispectrum (the left-hand side of \cref{P-PG} multiplied by ${k^3}/{2\pi^2}$) -- describe a lognormal. This practically isolates the effect of computing $\mathcal{P}_\zeta$ including non-Gaussian contributions \textit{ab-initio} from the effect caused by a NG contribution on the computation of $\Omega_{\rm GW}$ given $\mathcal{P}_\zeta$. 
We will also compare our results to the LN case where only the Gaussian contribution is considered.} and as free parameters we use 
\begin{equation}\label{eq:LNNGs}
    {\bm \theta}_{\rm cosmo} =
    \{\log_{10} A_s,
    \log_{10}\Delta,
   \log_{10} \left(k_*/\mathrm{s}^{-1}\right),
    f_{\rm NL} \equiv 5/6 \sqrt{\tau_{\rm NL}}\}.
\end{equation}

\begin{figure}[t]
    \centering
    \includegraphics[width=0.49\textwidth]{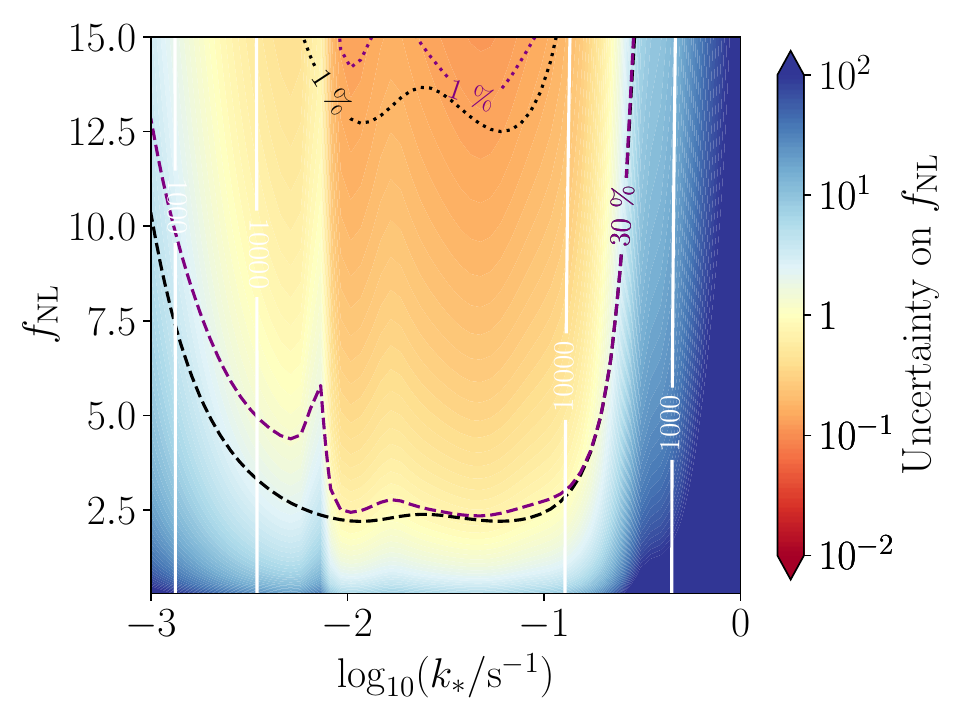}
    \includegraphics[width=0.49\textwidth]{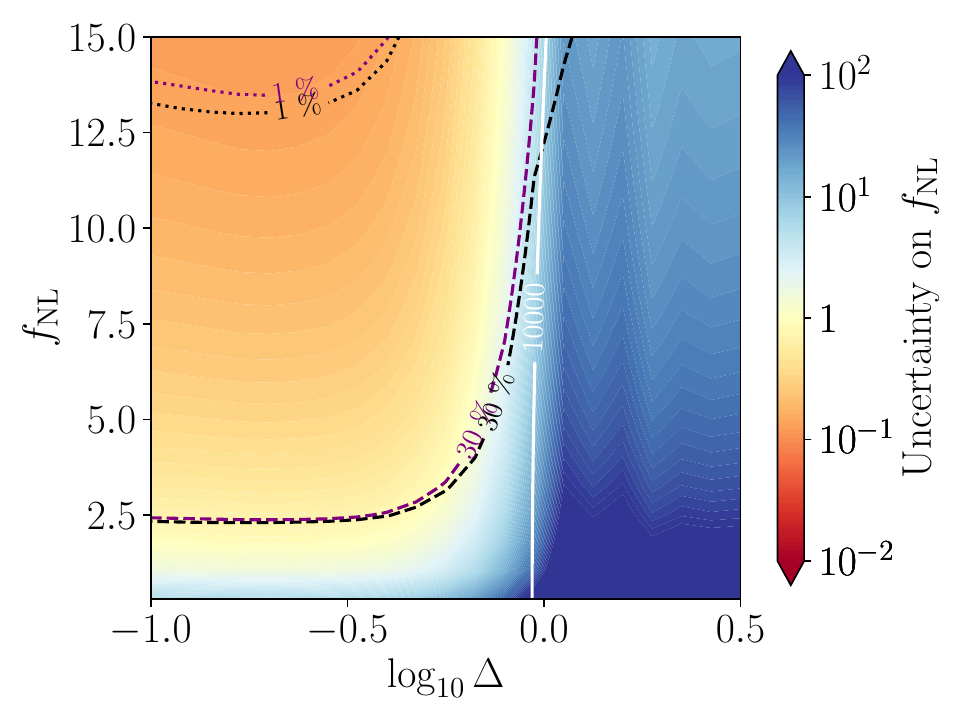}
    \caption{
    \emph{Left panel:} Absolute uncertainties of each of the parameters in the case of an injected signal which includes NG and assuming the LN template with parameters fixed to $\log_{10} A_s = -2 $ and $\log_{10}\Delta= -0.75$.
\emph{Right panel:} Same as the left panel, but varying $\Delta$ instead of $k_*$, for an injected signal which includes NG and assuming the LN template with parameters fixed to $\log_{10} A_s = -2 $ and $\log_{10}(k_*/s^{-1})= -1.4$.}
    \label{fig:scan_ng_k*_sigma}
\end{figure}

The left panel of Fig.~\ref{fig:scan_ng_k*_sigma} shows the absolute uncertainty associated with $f_{\rm NL}$ varying the parameter $f_{\rm NL}$ against $\log_{10}(k_*)$ computed using the FIM method. We fix the amplitude of $\mathcal{P}_\zeta$ to $\log_{10} A_s = -2 $ and the width to $\log_{10}\Delta= -0.75$. The dashed black (purple) contour lines represent the relative percentage error associated with $f_{\rm NL}$ when astrophysical foregrounds are not included (or are included). The white vertical line indicates the SNR. Notice that for small and large values of $k_*$, $f_{\rm NL}$ exhibits higher uncertainties, whereas the intermediate range of  $k_*\sim 10^{-2}-10^{-1}\,s^{-1}$ shows the minimal uncertainties for $f_{\rm NL}$. 
This suggests that 
tight constraints on $f_{\rm NL}$ can only be achieved within this specific range of $k_*$. Outside of this range, the errors increase notably, indicating less reliable measurements for $f_{\rm NL}$. Even in the optimal case, the reconstruction of $f_{\rm NL}$ only reaches the percent level for large $f_{\rm NL}\gtrsim 12.5$ . Notice that, in the presence of foregrounds, the accuracy on $f_{\rm NL}$ slightly degrades, in particular when $k_*$ coincides with the expected peak of the galactic foreground, i.e.~when $k_* \sim 10^{-2.2}\,$s$^{-1}$ .

Similarly, the right panel of Fig.~\ref{fig:scan_ng_k*_sigma} shows the FIM absolute uncertainty associated with $f_{\rm NL}$ varying the parameter $f_{\rm NL}$ against $\log_{10}\Delta$. We fix the amplitude of $\mathcal{P}_{\zeta}$ to $\log_{10} A_s = -2 $ and the peak scale to the optimal location  $\log_{10}(k_*/\mathrm{s}^{-1})= -1.4$.
In this case, the uncertainty when estimating $f_{\rm NL}$ is lower ($\lesssim 30\%$) in the region of $\log_{10}\Delta$ below $-0.3$, while it significantly degrades for larger widths.
This suggests that the most stringent measurements of $f_{\rm NL}$ will be obtained for relatively narrow curvature power spectra.

Given the relation between $f_{\rm NL}$ and $\tau_{\rm NL}$, improvements in the precision of the former directly translate into tighter constraints on the latter. The FIM analysis shows that percent-level accuracy on $f_{\rm NL}$ is achievable only for large $f_{\rm NL},$ which in turn would correspond to a percent-level constraint on $\tau_{\rm NL}$. 
In favorable scenarios --where the peak scale $k_*$ lies within $10^{-2}-10^{-1}\,\mathrm{s}^{-1}$ and the spectral width $\log_{10}\Delta$ is relatively narrow -- uncertainties in $f_{\rm NL}$ are minimal, restricting the allowed range of $\tau_{\rm NL}$. Conversely, when $f_{\rm NL}$ is less precisely determined, $\tau_{\rm NL}$ remains poorly constrained.

\begin{figure}[t!]
    \centering
\includegraphics[width=0.55\linewidth]{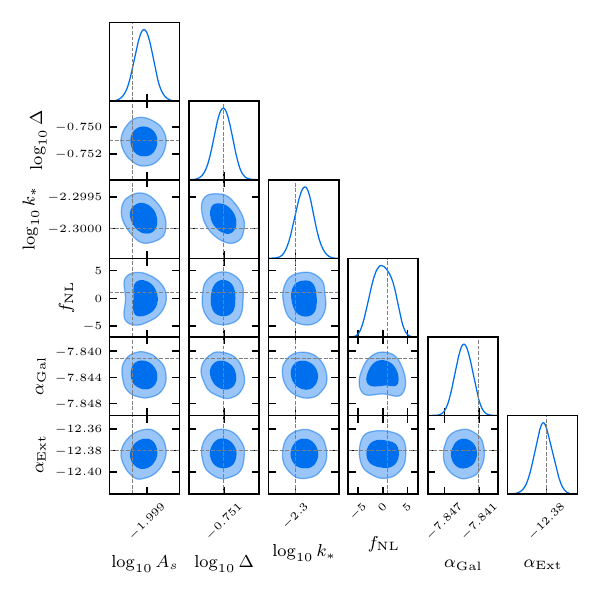} 
\includegraphics[width=0.44\linewidth]{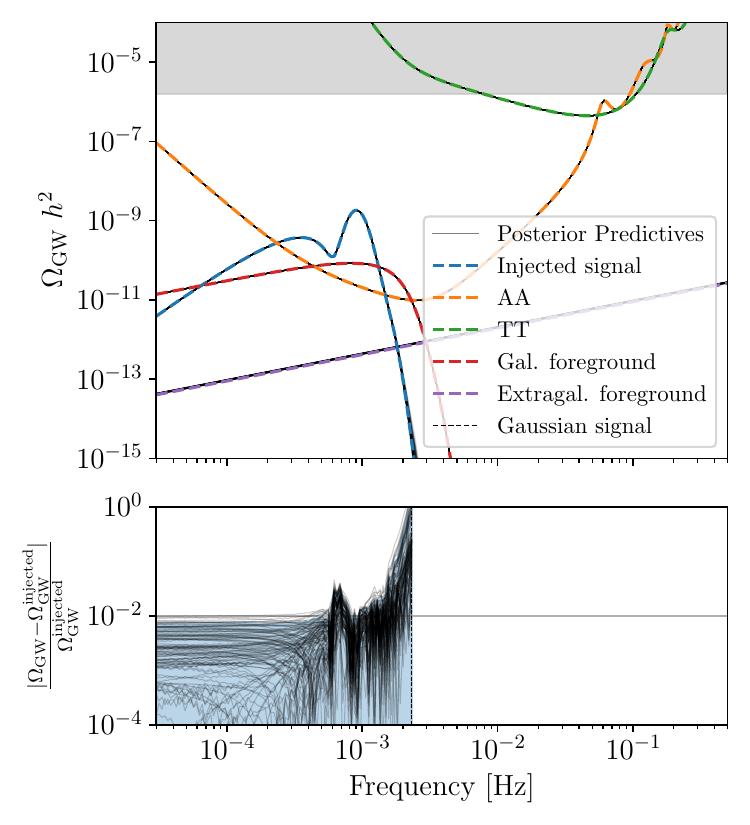}  
\caption{Posterior distribution for  
an injected signal which includes NG, $f_{\rm NL} = 1$ and assuming the LN with $\log_{10} A_{s}=-2$,  $\log_{10}\Delta= -0.75$, $\log_{10}(k_*/\mathrm{s}^{-1}) = -2.3$.}
    \label{fig:NGln1}
\end{figure}

The corner plot in Fig.~\ref{fig:NGln1} illustrates the posterior distributions of the SIGW signal parameters \{$\log_{10} A_{s}$, $\log_{10}(k_*/s^{-1})$, $\log_{10}\Delta$\}, including the primordial NG parameter $f_{\rm NL}$, alongside the extragalactic and galactic background amplitude energy densities $\log_{10}(h^2 \Omega_{\rm Ext}) \equiv \alpha_{\rm Ext}$, $\log_{10}(h^2 \Omega_{\rm Gal})\equiv \alpha_{\rm Gal}$.
As in the other cases, we omit the LISA noise parameters $A_{\rm noise}, P_{\rm noise}$ from the corner plot as they are tightly constrained and weakly correlated with the rest. As first benchmark, we injected a template with
$\log_{10} A_{s}=-2$,  $\log_{10}\Delta= -0.75$, $\log_{10}(k_*/\mathrm{s}^{-1})= -2.3$ and $f_{\rm NL} = 1$. 
On the right panel of Fig.~\ref{fig:NGln1}, we report the corresponding reconstructed $\Omega_{\rm GW}h^2$. For comparison, we also plot the GW energy density in the Gaussian case. Due to the low value of $f_{\rm NL}$ chosen, the reconstructed and Gaussian curves are almost superimposed, showing that the effects of NG are quite mild in this case.

The marginalized posterior for $f_{\rm NL}$ exhibits a broad posterior distribution, ranging from $-5$ to $5$. This distribution indicates a large uncertainty in $f_{\rm NL}$ spanning roughly $\pm 3$ at the $68\%$ C.L. and is compatible with $f_{\rm NL}=0$. This suggests a limited constraining power of LISA on $f_{\rm NL}$.
For such a signal, the posterior shows a bimodal structure, that arises because $f_{\rm NL}$ enters quadratically in the GW spectral energy density through the trispectrum. 
As the observed value suggests compatibility with a Gaussian primordial distribution, the broad posterior distribution also implies that ruling out moderate NG will be challenging, emphasizing the need for improved precision or additional data to refine these estimates.
The other parameters, such as the log-amplitude of the seed power spectrum ($\log_{10} A_s$), show a remarkable reconstruction, in line with the results of Sec.~\ref{sec:results_template_smooth}. Note however that we are injecting different values of 
\{$\log_{10} A_{s}$, $\log_{10}(k_*/s^{-1})$, $\log_{10}\Delta$\}. For a comparison to the fully Gaussian case with the same injection in $\mathcal{P}_\zeta$ see \cref{fig:ln_NG_gaussian_corner}.
The effect of adding the small NG correction $f_{\rm NL}=1$ on the recoverability of power spectral parameters \{$\log_{10} A_{s}$, $\log_{10}(k_*/s^{-1})$, $\log_{10}\Delta$\} and foreground parameters $\{\alpha_{\rm Gal}, \alpha_{\rm Ext}\}$ is small in this case, indicating that a small NG contribution does not spoil the reconstruction of curvature power spectra parameters. 
This also indirectly supports our choice of not including NG corrections in the benchmark USR scenario discussed in Sec.~\ref{sec:USR_pzeta_computation}, which is characterized by $f_{\rm NL}\simeq 0.09$. However, notice the visible non-zero correlation between $f_{\rm NL}$ and both $\{\log_{10} A_{s}, \alpha_{\rm Gal}\}$. 
The joint posterior $f_{\rm NL}$ - $\log_{10} A_{s}$ reflects the multimodality induced by the double peak structure of $f_{\rm NL}$.  Nevertheless, LISA can still strongly constrain the amplitude of the non-linear power spectrum. Hence the detection of the GWB is not strongly influenced by the primordial NG, which is beneficial for simplifying the analysis when considering models predicting small NGs.

The posterior distributions for astrophysical foreground amplitudes are broader and show limited correlation with signal parameters. In contrast, instrumental noise parameters are tightly constrained and largely independent of signal estimation. This decoupling ensures robust estimation of the signal amplitude, enhancing the reliability of SGWB detection despite noise uncertainties.

Overall, the joint distributions show only weak correlations, indicating that each of these parameters can be inferred with a good degree of independence from the others.

\begin{figure}[t!]
    \centering
\includegraphics[width=0.55\linewidth]{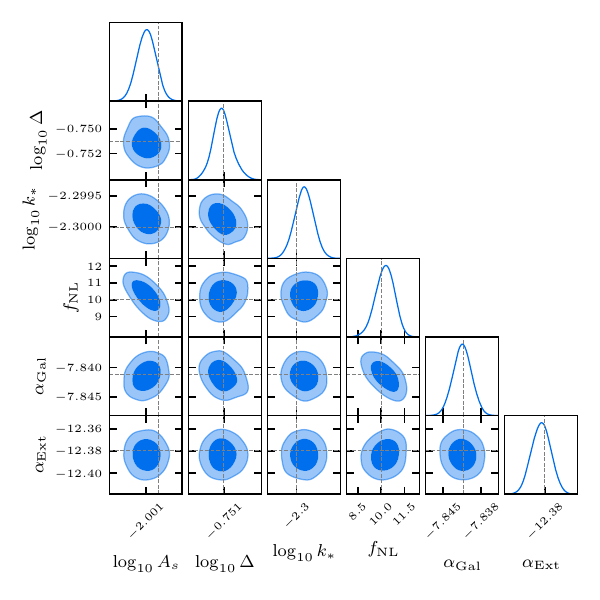}    
\includegraphics[width=0.44\linewidth]{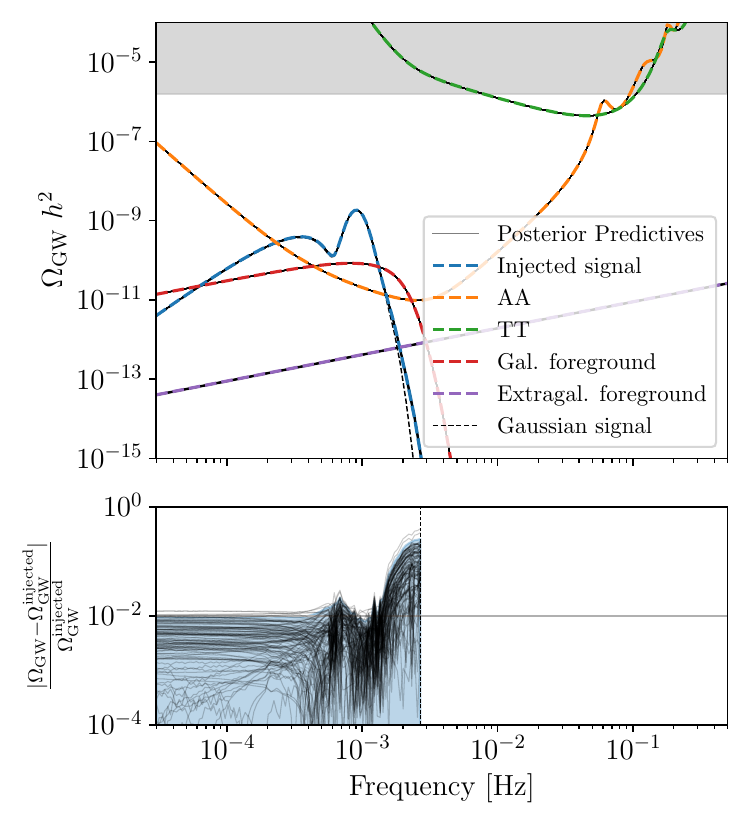} \caption{Same as Fig.~\ref{fig:NGln1} with $f_{\rm NL} = 10$. Note that there is a second, identical mode at $f_{\rm NL}=-10$ as only the square of $f_{\rm NL}$ enters $\Omega_{\rm GW}$. We omitted this mode.}
    \label{fig:NGln10}
\end{figure}

The situation changes when considering a larger value of  $f_{\rm NL} $, as shown in Fig.~\ref{fig:NGln10}. Specifically, we run the MCMC with the following injected template with
$\log_{10} A_{s}=-2$,  $\log_{10}\Delta= -0.75$, $\log_{10}(k_*/\mathrm{s}^{-1})= -2.3$ and $f_{\rm NL} = 10$, 
differing from the previous benchmark only in the choice of $f_{\rm NL}$.
On the right panel, we report the reconstructed $\Omega_{\rm GW}h^2$ as well as the Gaussian counterpart for comparison. Given the higher value of $f_{\rm NL}$, the differences are now more evident, resulting in an enhancement of the UV tail, but also in a slightly higher peak and more smoothed minimum (not visible in Fig.~\ref{fig:NGln10}). Now, the marginalized posterior distribution for $f_{\rm NL}$ shows only a narrow spread indicating that the estimate of $f_{\rm NL}$ is precise. 
The absolute value of $f_{\rm NL}$ has a mean reconstructed value that varies about $\pm 0.8$ at the $68\%$ C.L. which makes it incompatible with $f_{\rm NL}=0$. 

As with the previous case, the non-linear power spectral parameters 
are tightly constrained and no significant bias is observed with respect to the injected values. 
Both $k_*$ and $\Delta$  are very weakly correlated with $f_{\rm NL}$, suggesting that their reconstruction is not heavily influenced when $f_{\rm NL}$ is large. 
However, $\log_{10} A_{s}$ shows a larger anti-correlation with the $f_{\rm NL}$. An anti-correlation also appears between $f_{\rm NL}$ and $\alpha_{\text{Gal}}$, probably induced by the similar IR shape behavior.  
The tighter constraints associated with such a signal imply that higher levels of NG are easier to constrain, yielding clearer and more reliable effects. 

The independence between $f_{\rm NL}$ and other parameters (except for $\log_{10} A_{s}$), indicated by the weak correlations, suggests robustness in parameter estimation. This means that uncertainties in $f_{\rm NL}$ do not drastically affect the inference of other parameters, resulting in more precise parameter constraints compared to the $f_{\rm NL} = 1$ case. When $f_{\rm NL}$ is larger, the signal is more distinct, allowing for setting more stringent constraints on primordial NG. Note that the prior in Fig.~\ref{fig:NGln10} is restricted to positive values of $f_{\rm NL}$. Similarly to Fig.~\ref{fig:NGln1} the posterior distribution has a second mode at $f_{\rm NL}=-10$ since it only enters quadratically in the signal.

For the amplitude $A_s$ considered in this case, the imprints due to possible inaccuracies in accounting for the full non-Gaussian behavior for some models of inflation are expected to be negligible when $f_{\rm NL}=1$, but could be substantial when $f_{\rm NL}=10$, as recently argued by~\cite{Iovino:2024sgs}. For this analysis, we neglected those refinements.
We further stress that for the enhanced amplitude of the power spectrum considered here, $f_{\rm NL}=10$ represents much larger deviations from Gaussianity than on CMB scales, because the expansion parameter determining the relative size of the trispectrum versus power spectrum is $\tauNL \cdot {\cal P}_\zeta$. It is of order one in the current context, while less than $10^{-4}$ on CMB scales.

Concerning the implications for $\tau_{\rm NL}$, large uncertainties in $f_{\rm NL}$ directly translate into poor constraints on $\tau_{\rm NL}$. For example, for $f_{\rm NL}=1$ with a quite broad uncertainty of $\pm 3 $, $ \tau_{\rm NL} $ could span from values close to zero (if $f_{\rm NL} \approx 0$) up to $\simeq 23$ (if $f_{\rm NL}\approx 4$), making it challenging to clearly identify a primordial NG signal. In this range, even moderate NGs become difficult to distinguish from a Gaussian spectrum. Without improved precision on $f_{\rm NL}$, the corresponding $\tau_{\rm NL}$ will remain poorly determined, limiting our ability to discriminate between different levels of primordial NG.
When $f_{\rm NL}=10$, providing a more pronounced non-Gaussian signal, the corresponding $\tau_{\rm NL} = \left(\frac{6}{5}\cdot10\right)^2=144$ is now much more tightly constrained. Since the uncertainty in $ f_{\rm NL} $ is roughly $\pm 0.8$, $\tau_{\rm NL}$ varies up to a $\pm 15\%$ range. This tighter range is obviously better than the scenario with small $f_{\rm NL}$. Hence, larger $f_{\rm NL}$ values significantly improve our ability to determine $\tau_{\rm NL}$, allowing LISA to better distinguish between different levels of primordial NG in the SGWB.

Finally, it is important to highlight that while Planck provides constraints that are very close to zero~\cite{Akrami:2018odb}, indicating 
no evidence for primordial NG at large scales, the analysis we are performing for LISA focuses on NG at much smaller scales.
LISA's ability to provide tight constraints on $f_{\rm NL}$ suggests that GW detection could play a crucial role in refining our understanding of primordial NG, particularly in scenarios where the signal is expected to be strong. In addition, the sensitivity of LISA to different scales compared to Planck provides an important cross-check, helping to verify any scale dependence for $f_{\rm NL}$~\cite{Byrnes:2009pe,Byrnes:2010ft}. Overall, while Planck remains a benchmark for CMB-based constraints on $f_{\rm NL}$ at large scales, LISA shows the potential of GW detectors to significantly advance the search for, and the characterization of primordial NG.

\section{Testing the scalar-induced hypothesis}
\label{sec:tests}

In this section, we outline a procedure to test the compatibility of the SIGW hypothesis with a possible SGWB detection.
So far, our analyses have assumed that the cosmological contribution of the SGWB originates from SIGWs. There are, however, many alternative possible physical mechanisms for sourcing SGWBs in the early universe.
Our goal is to offer a practical approach for assessing the validity of the hypothesis explored in this work—namely, whether or not a hypothetically detected signal originates from enhanced scalar fluctuations of inflationary origin. To this end, we focus on two illustrative scenarios that are distinct in nature, %fundamentally different, 
leaving a detailed comparison of various early-universe signals—which is beyond the scope of this paper—for future work.
We use the evidence \eqref{eq:evidence} as an estimator for model selection. Specifically, given an injected signal, we consider different reconstruction techniques, for which we can compute the (log) evidence using the nested sampler \texttt{PolyChord}~\cite{polychord1,polychord2}. Then, we compare different hypotheses by computing the Bayes factors to test which one is favoured given the observed signal. 
It is important to note that the Bayes factor is a global estimator: it not only assesses the goodness of fit of a model to the data but also incorporates information about the prior volume and its compression as the prior transitions to the posterior. Therefore, given a similar fit to data, it naturally favours simpler models—those with fewer parameters—over more complex ones.

In the following analysis, we compare two different hypothesis: 
% follow two distinct philosophies, depending on whether they assume, or not, that the signal originates from scalar-induced GWs:
\begin{enumerate}
    \item {\bf 
    Hypothesis 1: The SGWB can be modelled directly in $\mathbf{\Omega_{\mathrm{GW}}(f)}$, independently of whether or not it is scalar-induced}.
    Thanks to the \texttt{SGWBinner} code~\cite{Caprini:2019pxz,Flauger:2020qyi}, we can use both {\it i)} a template-based \cite{LISACosmologyWorkingGroup:2024hsc}, as well as {\it ii)}  a model-agnostic approach to model $\Omega_{\rm GW}$ when reconstructing the signal. 
    Unlike the model-agnostic methods presented in this work, \texttt{SGWBinner}'s methods do not assume the underlying physics. The templates we use to fit the signal are of the same functional form as the injection template. Of course, with real LISA data, we will not necessarily know the correct template, and must either rely on model selection between several physically-motivated templates or, alternatively, model-agnostic methods.
    The binned approach of \texttt{SGWBinner} divides the frequency space into bins by SNR and then fits a power law within each bin. This results in an agnostic reconstruction of $\Omega_{\rm GW}$.

    \item {\bf Hypothesis 2: The SGWB is scalar-induced, and can be modelled at the level of $\mathbf{P_\zeta(k)}$.}
    By contrast, the various techniques presented in the previous sections assume that the signal that we are considering is due to SIGWs. 
    We can reconstruct the SGWB exploiting the \SIGWAY\ code in two different ways. One option is to use {\it i)} a template-based approach (see Sec. \ref{sec:analyticaltemplates}), in which a template for $\mathcal{P}_\zeta$ is specified. A second possibility is to use {\it ii)} the $\mathcal{P}_\zeta$-agnostic (still assuming SIGW to be the source of the SGWB) binned approach, as described in Sec.~\ref{sec:binnedPzeta}.\footnote{Let us note that there is a difference in the implementation of the binned $\mathcal{P}_\zeta$ approach, compared to the binned $\Omega_{\rm GW}$. In the latter approach, the \texttt{SGWBinner} code dynamically selects the optimal number of bins before the nested sampling, based on the Akaike information criterion~\cite{Akaike:1974vps} (we refer the interested reader to the discussion around Eq.~(3.6) of~\cite{Caprini:2019pxz} for more details). On the other hand, such a feature has not been implemented in the \SIGWAY\  code, as there are fundamental difficulties to attempting a similar approach in $\mathcal{P}_\zeta$-space (see App.~\ref{App:largeNbinning} for a detailed discussion) so the number of bins for $\mathcal{P}_\zeta(k)$ has to be chosen by hand. }
    We choose the number of bins $N_{\rm bins}=40$. We assume the SIGW to be produced during the radiation-dominated era.
\end{enumerate}

We simulate the following two qualitatively different signals:  
\begin{itemize}
\item {\bf Case 1. The injection is not SIGWs.} The first signal is a narrow lognormal in $\Omega_{\rm GW}(f)$. 
This injection serves as an example of a signal that cannot be produced by SIGWs, assuming the modes reenter during RD. In this case, regardless of how narrow the peak in $ \mathcal{P}_\zeta $ is, the generated SIGW will always exhibit the so-called ``causality tail" proportional to $f^3$. 
As a benchmark for this injection, we chose to reproduce the main peak of the double-peak background shown in the top-left panel of Fig.~11 in \cite{LISACosmologyWorkingGroup:2024hsc}, with a slightly lowered amplitude. This amounts to choosing the following parameters: $\log_{10}(h^2\Omega_*) = -9.5$, $\log_{10}(f_*/ {\rm Hz}) = -2.21$,
$\log_{10}(\rho) = -1.10$ in Eq.~(2.8) of~\cite{LISACosmologyWorkingGroup:2024hsc}. We will henceforth call this signal $\Omega_{\rm GW}^{\rm LN}$.

 \item {\bf Case 2. The injection is SIGWs.} The second injection is instead derived from a SIGW scenario. We inject a lognormal power spectrum of curvature perturbations, see Eq.~\eqref{eq:PLN} and compute the resulting SGWB numerically.
The power spectrum parameters used were: $\log_{10}A_s = -2.3$, 
$\log_{10}\Delta = -0.70$,
$\log_{10}(k_*/{\rm s^{-1}}) = -1.5$. 
The resulting shape in the SGWB can be described through the double-peak template in Eq.~(2.10) of \cite{LISACosmologyWorkingGroup:2024hsc}, with parameters $\{\log_{10}(h^2\Omega_*),\log_{10}(f_*/\mathrm{Hz}),\beta,\kappa_1,\kappa_2,\rho,\gamma\}=\{-9.5,-5,0.242, 0.456, 1.234, 0.08, 6.91\}$.
With this choice, the main peak of the SIGW coincides with the injection of {\bf Case 1}. 

\end{itemize}

\begin{figure}[t!]
    \centering
    \includegraphics[width=0.49\linewidth]{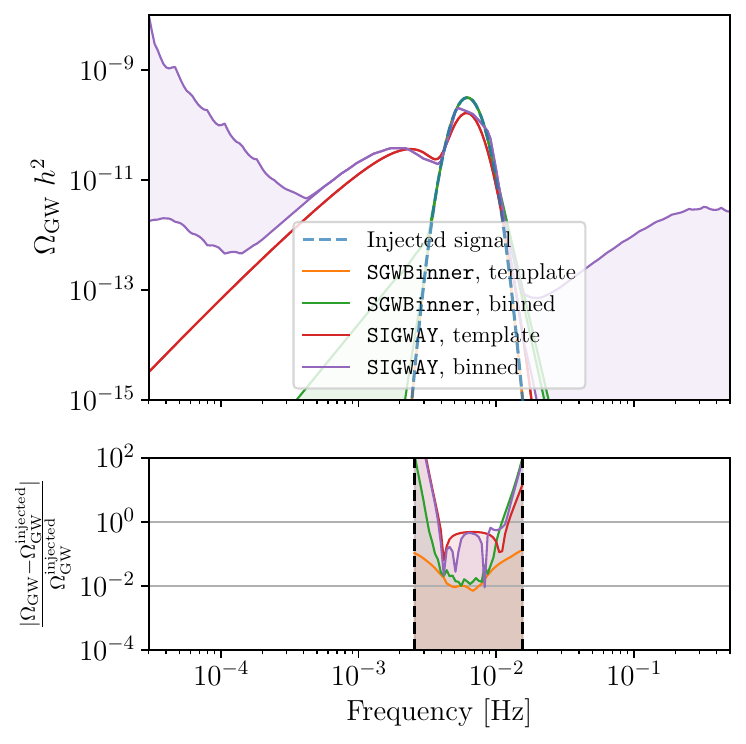}
    \includegraphics[width=0.49\linewidth]{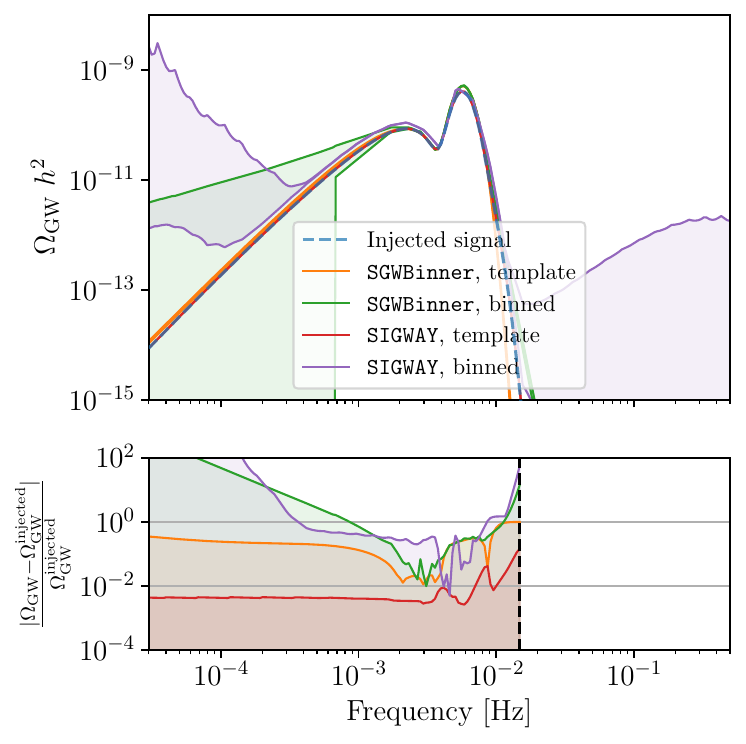}
    \caption{Reconstruction of $\Omega_{\rm GW}$ for case 1 (left) and case 2 (right). We only show the reconstruction of the injected cosmological contribution to the SGWB. It is clearly visible that the two models that assume SIGWs cannot reconstruct the narrow log-normal peak in case 1 and therefore lead to much lower evidence.
    }
    \label{fig:evidence_reconstruction_plots}
\end{figure}
We fit each of the two injections using the four models specified above. The results of our analysis are summarized in Fig.~\ref{fig:evidence_reconstruction_plots} (see also Fig.~\ref{fig:evidence_reconstruction_plots_full} in App.~\ref{app:additionalfigs} showing the reconstruction including foregrounds), and Tab.~\ref{tab:Bayes_factors}, which we now describe in order.

In {\bf Case 1}, the models that perform the worst in terms of model selection are the two based on the SIGW hypothesis (marked by the row \SIGWAY{}). As shown in the left panels of Fig.~\ref{fig:evidence_reconstruction_plots}, both the lognormal and binned $ \mathcal{P}_\zeta(k) $ models attempt to fit the lognormal SGWB using the primary peak of the SIGW. However, the secondary peak at lower frequencies severely undermines the fit to the data, resulting in a very poor evidence value. According to Jeffreys' scale (see Sec.~\ref{sec:data_analysis}), these models are decisively ruled out when compared to the two alternative hypotheses. As expected, the binned $ \mathcal{P}_\zeta(k) $ performs worst, due to its significantly larger number of parameters.

We also fit the same injection using a lognormal template for $\Omega_{\rm GW}$ (which corresponds to the true injection) and a binned $\Omega_{\rm GW}$ reconstruction with \texttt{SGWBinner}. As illustrated in Fig.~\ref{fig:evidence_reconstruction_plots}, both models reconstruct the signal very well, closely matching the injection. Furthermore, they yield very similar best-fit likelihood values with a preference for the lognormal template (matching the injected signal) due to its smaller number of parameters.

The main takeaway from Case 1 is that if a similar signal was detected, we could conclude with very high statistical significance that the signal does not have a scalar-induced origin.

In {\bf Case 2}, all the models considered successfully recover the injected signal. Unlike in the previous case, the Bayes factors are closer together, while still being orders of magnitude apart (we would like to stress that the differences quoted in Tab.~\ref{tab:Bayes_factors} are $\ln(Z)/10^{4}$). Again, the model-agnostic reconstructions -- whether in $ \Omega_{\rm GW} $ or $ \mathcal{P}_\zeta(k) $ -- perform the worst. However, their flexibility and agnostic nature still makes them useful in the real data analysis to pre-constrain the shape of $\Omega_{\rm GW}$ and $\mathcal{P}_\zeta$. The Bayes factors are shown in Tab.~\ref{tab:Bayes_factors}.

The accuracy of the recovery of the injected signal is sufficient to show a strong preference for the injected \SIGWAY template in $\mathcal{P}_\zeta$ when compared to the empirical template for $\Omega_{\rm GW}$ developed in~\cite[Eq.~(2.10)]{LISACosmologyWorkingGroup:2024hsc}. This outcome arises because the template in~\cite{LISACosmologyWorkingGroup:2024hsc} is described by seven free parameters compared to three for the lognormal $\mathcal{P}_\zeta(k)$, and because there is a slight mismatch between the numerical computation and the phenomenological fit. This highlights the need for accurate models for the SGWB. This behaviour can be seen in the right column of Fig.~\ref{fig:evidence_reconstruction_plots}.

%It is also interesting to stress that, in case 2, the template-based \SIGWAY method performs better than the template-based \texttt{SGWBinner} in reconstructing $\Omega_{\rm GW}$, as can be seen in the right column of Fig.~\ref{fig:evidence_reconstruction_plots}. This improvement arises because the assumption of SIGWs enforces a more restricting shape for $\Omega_{\rm GW}$, characterized by a smaller number of parameters. In contrast, the \texttt{SGWBinner} imposes weaker restrictions in the reconstruction.
On the other hand, when comparing the agnostic approaches, we see that the former gives slightly better $\Omega_{\rm GW}$ reconstructions around the peak of the signal, with an oscillatory behaviour of the residuals due to the finite resolution of the binned approach (40 bins).
This result stresses the importance of adopting optimal modelling of the eventual cosmological scenario when reconstructing the signal (see also App.~\ref{app:additionalfigs}).

\begin{table}[t]\centering
\begin{tabular}{c|c|c|c}
\toprule
\multirow{2}{*}{Package} & \multirow{2}{*}{Method} 
 & Case 1: not SIGW signal & Case 2: SIGW signal\\
& & $\ln \Big[ Z(\Omega_{\rm GW}^{\rm LN})\Big]/10^4$& $\ln\Big[Z({\cal P}_\zeta^{\rm LN})\Big]/10^4$\\
\midrule
\multirow{2}{*}{\texttt{SGWBinner}} & Template 
& $ \mathbf{0.5444} $ & $-0.5278  $ \\
& Binned 
& $-0.5625  $ & $-0.5479  $  \\ 
 \midrule
\multirow{2}{*}{\SIGWAY}  & Template 
& $-25.6078$ & $ \mathbf{-0.5203} $\\
 & Binned  % (40 bins)
 & $-25.7934  $ & $-2.712  $ \\
\bottomrule
\end{tabular}
\caption{
$\ln$ Bayes factors (normalized to a reference value $10^4$) comparing the SGWB reconstruction (either using the injected model as a template, or a model-agnostic binned method) with the SIGW reconstruction for two signals: {\it i)}
a log-normal power spectrum in $\Omega_{\rm GW}$ which cannot be generated by SIGW (within the assumptions we are working in), {\it ii)} a log-normal power spectrum in $\mathcal{P}_\zeta$ which generates detectable SIGW. 
In bold, we show the Bayes factors for the recovery which assumes the injected template. As expected, they are the best reconstructions for each injection.}
\label{tab:Bayes_factors}
\end{table}

All in all, the results of this section, although based on two illustrative examples, demonstrate that LISA has the potential to confirm or rule-out the scalar-induced nature of the SGWB with high statistical significance.
These results confirm that the true models achieve the best Bayes factors when appropriately matched. Alternative and binned models consistently show inferior fits, highlighting the distinctiveness of the injected signals.

However, is important to stress that not all SGWB that may appear in LISA lead to such a clear difference between signals that can or cannot be generated by SIGW. 
The characteristic double-peak structure that we observe with the injected $\mathcal{P}_\zeta^{\rm LN}$ signal is only measurable by LISA if \textit{(a)} the peak in $\mathcal{P}_\zeta$ is sufficiently narrow, and \textit{(b)} both peaks in $\Omega_{\rm GW}$ happen to fall within the sensitivity of LISA. On the other hand, there are many potential shapes for $\mathcal{P}_\zeta$ where these conditions are not met. In these cases the SIGW signal can easily mimic one expected from other cosmological sources, potentially making it much harder to rule out models. We will leave a more detailed discussion of this for future work.

\section{Conclusions}
\label{sec:conclusions}

In this work, we investigated the potential of the LISA detector for reconstructing the SGWB sourced by second-order scalar perturbations. 
Three approaches were explored: 
A binned spectrum reconstruction, 
template-based methods, 
and a direct modeling approach rooted in first-principles scenarios (taken to be the single-field USR inflationary model for presentation purposes). 

Our results demonstrate that the direct modeling approach yields the tightest constraints on the primordial curvature power spectrum $\mathcal{P}_\zeta$, particularly capturing both the IR and UV tails of the signal with better precision than alternative methods, due to the stronger prior information inevitably included in the fit. This highlights the power of incorporating {\it ab initio} physics into signal reconstruction pipelines to leverage the constraining power of LISA observations at their best.

The binned spectrum reconstruction approach is complementary and proved effective in providing model-independent upper bounds on $\mathcal{P}_\zeta$ when the cosmological contribution to the SGWB is below the sensitivity. Capturing the overall shape of the SIGW spectrum with this approach proved to be difficult, due to a combination of missing SNR towards the edges of the LISA window and strong degeneracies between the bins when choosing a large number of them. Despite these shortcomings, it is possible -- if a cosmological contribution to the SGWB is detected -- to tell apart signals that can be SIGW from those that cannot, by Bayesian model selection. 

In comparison, the template-based methods provided a more consistent reconstruction across frequencies, though they inherently rely on prior assumptions about the shape of the spectrum. The complementary strengths of these approaches suggest that an optimal reconstruction strategy would involve their combined use, with model-dependent templates guiding reconstructions and binned methods offering flexibility in capturing unanticipated spectral features or in setting bounds that are agnostic on the spectral shape.

We also examined the impact of going beyond the simplest vanilla cosmological scenarios on the SIGW reconstruction investigating the sensitivity of SIGW signals to early-universe physics. In particular, we included in our analysis the study of the effect of the transition from early matter-dominated to radiation-dominated eras, as well as the role of non-Gaussianity in the SIGW spectrum. Future research will include the study of the effect of a time-dependent equation-of-state parameter on the SIGW spectrum, as would be generated in the case of a smooth crossover \cite{Escriva:2023nzn,Escriva:2024ivo}, such as in the QCD phase transition.
Overall our analysis demonstrated how SIGW searches in LISA will provide constraints that vastly outperform those deduced from the effective number of relativistic species $\Delta N_\text{eff}$ and PBH overproduction bounds. In this regard, SIGW searches will also be an invaluable tool for probing the asteroid mass window of PBH dark matter.

Looking forward, several key avenues remain open for future work. 
On the phenomenological side, it remains an open question how well LISA will be able to constrain primordial non-Gaussianity or non-standard thermal histories while allowing for a fully non-parametric curvature power spectrum. In this work we have only quoted template-based constraints on these effects, e.g.~in Figures \ref{fig:emd_corner} and \ref{fig:NGln1}, which do not consider possible degeneracies with the shape of the spectrum.

Concerning the binned approach to reconstructing $\mathcal{P}_\zeta$, a more mature method that incorporates assumptions about the smoothness of the spectrum of scalar perturbations and addresses the computational difficulties with the binning will be needed in the future. 
It is likely that -- even allowing for non-Gaussianities and non-standard thermal histories -- some shapes of the SGWB cannot be scalar-induced and can be confirmed as signatures of directly sourced tensor perturbations, even without identifying a specific early-universe source.

On the theory side, future work should consider expanding the scope of single-field scenarios by exploring more general actions in the Jordan frame, incorporating non-minimal coupling and non-canonical kinetic terms (e.g.~Eq.~2.1 of \cite{Kannike:2017bxn}). Also, going beyond single-field USR models, first-principle multi-field inflationary scenarios merit investigation, as discussed in Sec.~\ref{sec:models_multi-field}. In some cases, multi-field models can effectively be reduced to single field descriptions, making some of the techniques developed here already applicable, and enabling simpler parameter space scans as done in \cite{Geller:2022nkr,Qin:2023lgo}.
A reverse engineering approach could be particularly valuable, where inflationary dynamics are modeled based on a minimal set of parameters, and the corresponding inflationary potential is reconstructed within single- or multi-field frameworks, as demonstrated in Refs.~\cite{Franciolini:2022pav,Autieri:2024wye}. 
Expanding the framework for computing the non-Gaussian signatures predicted in most SIGW models beyond the lognormal template in Sec.~\ref{sec:NG-SGWB} could serve as a diagnostic tool for breaking degeneracies in cases where the SGWB spectrum alone is insufficient. Also, implementing the binned approach in scenarios with NGs could also allow us to reduce the computational costs of these analyses.
While in this work we only considered the monopole signal, as non-Gaussianities may impact the large scale SIGW anisotropies, it would be interesting to include information from higher order in the multiple expansion of power in the sky \cite{Bartolo:2019zvb,LISACosmologyWorkingGroup:2022kbp,Li:2023qua,Ruiz:2024weh}.
Finally, integrating these advanced modeling and reconstruction techniques into the global fit pipeline of LISA, as well as incorporating measurements from other experiments, will be essential for unlocking the mission’s full potential in probing the early universe's cosmological landscape.

\vspace{.2cm}
\noindent
{\bf Acknowledgments.}
We thank Nicola Bartolo, Gaetano Luciano, Marco Merchand, 
Sabino Matarrese, and Toni Riotto for discussions and interactions in an early stage of this project.
We acknowledge the LISA Cosmology Working Group members for seminal discussions. We especially thank the authors of refs.~\cite{Caprini:2024hue,Blanco-Pillado:2024aca,LISACosmologyWorkingGroup:2024hsc,Kume:2024sbu} for the collaborative developments of the \texttt{SGWBinner} code upon which we built to produce many of the forecasts we presented in this work. 
The research of JF is supported by the grant PID2022-136224NB-C22, funded by MCIN\allowbreak/\allowbreak AEI\allowbreak/\allowbreak 10.13039\allowbreak/501100011033\allowbreak/\allowbreak FEDER, UE, and by the grant\allowbreak/ 2021-SGR00872.
RR was supported by an appointment to the NASA Postdoctoral Program at the NASA Marshall Space Flight Center, administered by Oak Ridge Associated Universities under contract with NASA.
The work of EM is supported by the Italian Ministry of University and Research grant Rita Levi-Montalcini ``New directions in axion cosmology''.
EM acknowledges the support of Istituto Nazionale di Fisica Nucleare (INFN) through the \textit{iniziativa specifica} TAsP.
JE and GN acknowledge support by the ROMFORSK grant project no.~302640.
TP acknowledges the contribution of the COST Action CA21136 ``Addressing observational tensions in cosmology with systematics and fundamental physics (CosmoVerse)'' and of the INFN Sezione di Napoli \textit{iniziativa specifica} QGSKY. He acknowledges as well financial support from the Foundation for Education and European Culture in Greece. 
DR is supported by the UZH Postdoc Grant 2023 Nr.\,FK-23-130. 
During the course of this work, S.RP and DW
were supported by the European Research Council under
the European Union’s Horizon 2020 research and innovation programme (grant agreement No 758792, Starting Grant project GEODESI).
The work of AG is supported by the UKRI AIMLAC CDT, funded by grant EP/S023992/1.
The work of AG, GT, and IZ is partially funded by the STFC grants ST/T000813/1 and ST/X000648/1.
The work of HV was supported by the Estonian Research Council grants PSG869 and RVTT7 and the Center of Excellence program TK202.
M.Pe acknowledges support from Istituto Nazionale di Fisica Nucleare (INFN) through the Theoretical Astroparticle Physics (TAsP) project and from the MIUR Progetti di Ricerca di Rilevante Interesse Nazionale (PRIN) Bando 2022 - grant 20228RMX4A, funded by the European Union - Next generation EU, Mission 4, Component 1, CUP C53D23000940006.
JK acknowledges support from the JSPS Overseas Research Fellowships and the INFN TAsP project.
GP acknowledges partial financial support by ASI Grant No. 2016-24-H.0.
AR acknowledges support from Istituto Nazionale di Fisica Nucleare (INFN) through the \textit{iniziativa specifica} TEONGRAV and by the project BIGA - ``Boosting Inference for Gravitational-wave Astrophysics" funded by the MUR Progetti di Ricerca di Rilevante Interesse Nazionale (PRIN) Bando 2022 - grant 20228TLHPE - CUP I53D23000630006.

\vspace{.2cm}
\noindent
{\bf Authors' contribution.}
JE: 
Main developer of \SIGWAY. Implementation and optimization of techniques discussed in App. \ref{app:SIGWBinner}, coding the interface to \texttt{SGWBinner}. 
Co-coding different thermal histories. Running analyses and producing figures for all results except Secs.~\ref{sec:NG} and \ref{sec:tests}. 
Writing Secs.~\ref{sec:results} and Appendices.
Reviewing all draft.
AG: 
Co-coding different thermal histories. 
Writing Secs.~\ref{sec_tran} and \ref{sec:reseMDRD}.
Reviewing Sec.~\ref{sec:modellingPzeta}.
GF: 
Proposing and coordinating the project with RR, including defining goals and managing tasks. 
Co-coding parts of \SIGWAY, mainly on SIGW computations and the USR module. 
Writing and Reviewing all sections of the draft.
TP:
Co-coding kernel functions and different thermal histories.
Writing Sec.~\ref{sec:reseMDRD} and parts of Secs.~\ref{sec:models_inflation}, \ref{sec:models_multi-field},
\ref{sec:other_models}, \ref{sec:analyticaltemplates} and 
\ref{sec:conclusions}.
Reviewing Secs.~\ref{sec_resBM}, \ref{sec:template_method} and \ref{sec:USR_inference}.
MPe: 
Devising methods for computations in Sec.~\ref{sec:USR_pzeta_computation} and developing the algorithm for the binned spectrum approach, 
contributing to the code implementation. Writing Secs.~\ref{sec:binnedPzeta} and \ref{sec:OmegaGWijk}. Reviewing Secs.~\ref{sec:intro}, \ref{sec:models}, \ref{sec:tempbinnerpipe}, and \ref{sec:conclusions}.
GP: Co-coding SIGW computations with NGs in \SIGWAY. 
Running and producing Figs. for Sec.~\ref{sec:NG}. 
Writing App.~\ref{App:NG_tec} and Sec.~\ref{sec:computation}. Reviewing Sec.~\ref{sec:computation}.
MPi: Co-coding SIGW computation in the \SIGWAY\ and interfacing with \texttt{SGWBinner}.
Writing Sec.~\ref{sec:tempbinnerpipe}.
AR: Co-coding SIGW computations with NGs in \SIGWAY. 
Running and producing Figs. for Sec.~\ref{sec:NG}. 
Writing Sec.~\ref{sec:NG}. 
Reviewing Sec.~\ref{sec:intro}.
RR: Proposing and coordinating the project with GF, including defining goals and managing tasks. 
Co-coding the USR module. Writing and Reviewing all sections of the draft.
GT: Developing the algorithm for the binned spectrum approach, 
contributing to the code implementation. 
Writing Secs.~\ref{sec:binnedPzeta} and \ref{sec:OmegaGWijk}. 
Reviewing the draft.
MB: Running the analysis and writing of Sec.~\ref{sec:tests}.
JF: Running the analysis and writing of Sec.~\ref{sec:tests}. 
Reviewing Secs.~\ref{sec:tests}, \ref{sec:temposc}, \ref{sec:models_multi-field} and Sec.~\ref{sec:models}.
JK: Coding \texttt{SGWBinner}.
Writing Sec.~\ref{sec:tempbinnerpipe}. Reviewing Sec.~\ref{sec:modellingPzeta}.
EM: Contributing to an early version of the USR code. Writing Secs.~\ref{sec:models_inflation},
\ref{sec:models_multi-field}, 
\ref{sec:benchmodelUSR}, 
\ref{sec:analyticaltemplates}, 
\ref{sec:USR_pzeta_computation}, \ref{sec:template_method}, and \ref{sec:USR_inference}. Reviewing all draft.
GN: Writing Sec.~\ref{sec:intro}. Reviewing all draft.
DR: Writing Secs.~\ref{sec:modellingPzeta} (intro), \ref{sec:SIGW derivation}, \ref{sec_RDE}, and \ref{sec:results_template_smooth}. Reviewing Secs.~\ref{sec:computation} and \ref{sec:tests}.
SRP: Devising and interpreting Sec.~\ref{sec:NG-SGWB} and \ref{sec:NG}.
Writing Sec.~\ref{sec:NG-SGWB} and related elements in Secs.~\ref{sec:computation} and \ref{sec:NG}.
Reviewing all draft. 
HV: Devising Secs.~\ref{sec:models} and \ref{sec:modellingPzeta}. 
Writing Secs.~\ref{sec:models}, \ref{sec:modellingPzeta}, and \ref{sec:template_method}. Reviewing  Secs.~\ref{sec:intro}, \ref{sec:reseMDRD}, \ref{sec:NG}, and \ref{sec:conclusions}.
DW: 
Devising and interpreting Sec.~\ref{sec:NG-SGWB}.
Reviewing Secs.~\ref{sec:NG-SGWB}, \ref{sec:intro}, and \ref{sec:conclusions}.
IZ: Writing parts of Sec.~\ref{sec:computation} and
reviewing Secs.~\ref{sec:models}, \ref{sec:modellingPzeta}, and \ref{sec:computation}.

\appendix
\section{SIGWAY code: technicalities }\label{app:SIGWBinner}

In this appendix, we describe some technical aspects of the \SIGWAY\  code developed for the analysis performed in this work. 

\subsection{Perturbations in USR scenarios: code structure}
Given a potential $V(\phi)$ inducing a USR phase, the curvature power spectrum is computed in three steps that are described below.
$\mathcal{P}_\zeta(k)$ is then interpolated and $\Omega_{\mathrm{GW}} h^2$ is computed as described in \ref{appendix:computation_of_omega}.

Using the notation described in Sec.~\ref{sec:USR_pzeta_computation}, the inputs in the code are:
\begin{itemize}
    \item the inflaton potential $V(\phi)$;
    \item the number of $e$-folds from when CMB modes exit the Hubble horizon to the end of inflation $N_{\rm CMB\to end}$.
    \item The initial conditions $\phi_0$ and $\pi_0=\phi'_0$. 
\end{itemize}
The code then automatically defines the dimensionless variables in Eq.~\eqref{eq:adim_phi_V}.
For definiteness, in this paper, we have fixed $N_{\rm CMB\to end} = 58$.
Fixing the number of $e$-folds from the CMB scale to the end of inflation effectively allows us to set the correspondence between $N$ and wavenumbers $k$. 
Fixing $N_{\rm CMB}$ also implicitly fixes the thermal history of reheating and subsequent phases. We do not model these eras for simplicity, but they would be fixed in a complete USR+reheating model.
Also, we checked that the initial SR attractor would quickly pull the field space trajectory to the background evolution, and thus one could also assume negligible initial velocity for simplicity.

\begin{figure}[h]
    \centering
    \includegraphics[width=0.65\textwidth]{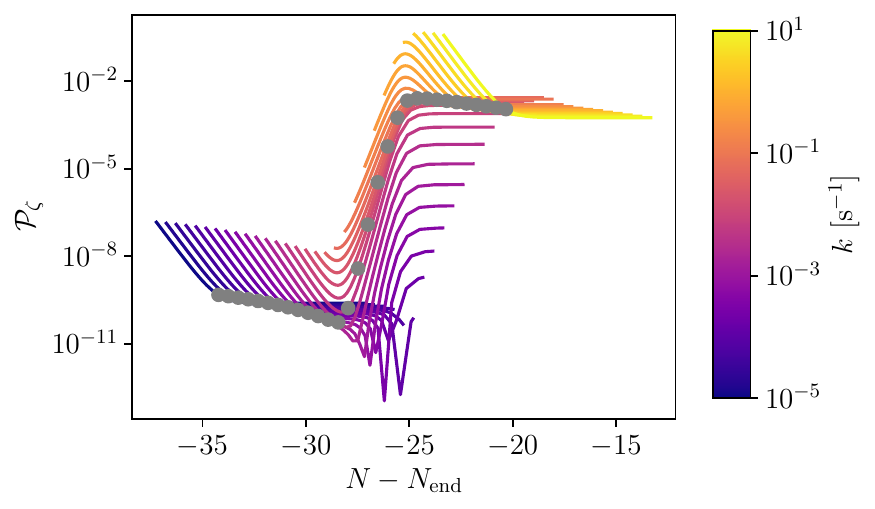}
    \caption{
    Power spectrum as a function of $N$ for different modes $k$ indicated by the color palette, close to the onset of the USR phase $N- N_{\rm end}\simeq -28$.
    This plot shows the evolution and freezing out of modes for different values of $k$ for the potential from Eq.~\eqref{eq:Germani-potential}.
    The evolution of each mode is traced for around $\Delta N \simeq 10$ $e$-folds from within the Hubble sphere to sufficiently after Hubble crossing and freezing. The time of Hubble crossing is marked with gray dots.}
    \label{fig:usr_k_mode_evolution}
\end{figure}

The computation then proceeds as follows: 
\begin{enumerate}
\item 
Solve for the background evolution using Eqs.~\eqref{eq:adim_EomBkgUSR}. We evolve the dynamics until the first SR condition is violated ($\epsilon_H=1$), 
collecting $x(N), y(N)$ and $h(N)$. 
We denote the number of $e$-folds at the end of inflation $N_{\rm end}$.
\item Compute the relation between the wave number $k$ and the number of $e$-folds $N$ at Hubble crossing using
\begin{align}
    k = k_{\rm CMB}
    \frac{h(N)}{h_{\rm CMB}} 
    \exp(N-N_{\rm CMB}),
\end{align}
where $k_{\rm CMB}=0.05/\,\mathrm{Mpc}$, $N_{\rm CMB}=N_{\rm end}-N_{\rm CMB\to end}$ 
and $h_{\rm CMB} = h(N_{\rm CMB})$. With this, we can compute $\mathcal{P}_\zeta(k)$ in the SR approximation according to Eq.~\eqref{eq:PzetaSR}.
\item We are only interested in computing the spectrum of curvature perturbations beyond the SR approximation for modes of relevance for LISA.
Once we have computed $k(N)$, a set of modes covering the LISA frequency band $\sim (10^{-5} - 10^{-1})\,\mathrm{Hz}$ is selected and those modes are evolved according to Eq.~\eqref{eq:MSequationadim} from $N_{\rm in}$ to $N_{\rm out}$, where $N_{\rm in}$ and $N_{\rm out}$ can be set by the user. We found $N_{\rm in}=N-3$ and $N_{\rm out}=N+7$, where $N$ is the horizon crossing time of the mode, to be long enough for the modes to freeze out as there is no super horizon evolution in these models once USR has ended. Notice that the USR can only last about ${\cal O}(3)$ $e$-folds without making perturbations grow beyond the validity of perturbation theory.
Additionally, for simple shapes of the potential, we find that evolving $\sim100$ modes in $k$ and interpolating between them yields sufficient precision. 
\item Lastly, $\mathcal{P}_\zeta(k)$ is computed without resorting to the SR approximation using Eq.~\eqref{eq:Pzeta_from adim}. This quantity is then passed to the algorithm computing $\Omega_{\rm GW} h^2$ which can then be passed to the LISA likelihood.
\end{enumerate}
Fig.~\ref{fig:usr_k_mode_evolution} shows the evolution of a number of modes in $\mathcal{P}_\zeta$ as a function of $N$ across the USR phase. As this algorithm is called many times when sampling the LISA likelihood, we are using the package \texttt{diffrax}~\cite{diffrax} to solve the inflationary perturbation equation of motion.

\subsection{Computation of SIGWs from the spectrum of curvature perturbations}\label{appendix:computation_of_omega}
For a given shape of $\mathcal{P_\zeta}(k)$, computing $\Omega_{\rm GW} h^2$ is relatively straightforward by evaluating the double integral in Eq.~\eqref{eq:P_h_ts}. The main concern here is making this computation as fast as possible, as this dominates the time it takes for each likelihood evaluation. To take full advantage of vectorization, we compute $s, t, k$ on a grid, 
where  $s$ is linearly spaced, and $t, k$ are logarithmically spaced. 
The integration in $t$ is convergent for any realistic shape of $\Pz$ 
as both the value of $\overline{I^2(k, s, t)}$ tends towards 0 for large $t$ 
and there needs to be some cutoff controlling the amplitude of $\Pz$ at large momenta not to violate BBN bounds. 
If $\mathcal{P}_\zeta(k)$ features a scale after which it drops rapidly, it can be advantageous to define a custom $t(k)$ and compute the grid as $s, t(k), k$.

After evaluating the integrand in Eq.~\eqref{eq:P_h_ts} on the grid, the integral is computed with Simpson-integration. Depending on the shape of $\mathcal{P}_\zeta$, we found a number of points in the grid around $N_s\sim 10-100$, $N_t\sim 300-1000$ and $N_k\sim 100$ would give sufficient precision.

To take full advantage of threading, we use \texttt{JAX} and just-in-time compilation~\cite{jax2018github} for computing the integrand and performing the integration itself. Altogether, this results in a significant speedup over other publicly available codes (e.g.~\cite{Witkowski:2022mtg}). We record wall-clock times of $\sim 10^{-2}\,\mathrm{s}$ for calculating $100$ values of $\Omega_{\rm GW}$ for $\mathcal{P}_\zeta$ containing only \texttt{JAX}-native functions. In the case discussed in Sec.~\ref{sec:rapid_transitions} this time rises to about $1\,\mathrm{s}$. Crucially, this speedup allows us to sample the posterior distribution efficiently and obtain good MCMC convergence with a laptop on timescales of ${\cal O}$(hours).

\subsection{Computation of SIGWs using binned coefficients}
In the case where we bin $\mathcal{P}_{\zeta}(p)$ in momentum space, $\Omega_{\rm GW}(k)$ can be computed in a straightforward manner through Eq.~\eqref{eq:OmegaMatrix}. The most computationally expensive part of this is the computation of the coefficients $\Omega_{\rm GW}^{(i,j)}(k)$. Luckily we can precompute these coefficients on a grid and save them as a $N_k\times N_p\times N_p$ tensor $K_{ij}^k \equiv \Omega_{\rm GW}^{(i,j)}(k)$. At runtime, we then compute a $N_p\times N_p$ tensor of $B = A\otimes A$, with which we can conveniently re-write Eq.~\eqref{eq:OmegaMatrix} as
\begin{align}
    \Omega_{\rm GW}^k = \Omega_{ij}^{k}B^{ij}\,,
\end{align}
where we used Einstein sum convention. The reason for doing this arguably very simple conversion is that the matrix equation is fully vectorizable with \texttt{JAX}. 

Using this trick we are able to compute $\Omega_{\rm GW}$ from $\mathcal{P}_\zeta$ in $\lesssim 10^{-3}\,\mathrm{s}$ for $N_k=N_p=50$, thus making inference possible despite a large number of parameters to sample.

\subsection{Computation of SIGWs including primordial NGs}
\label{App:NG_tec}
In this appendix, we provide additional technical details regarding the MCMC analyses that resulted in Figures \ref{fig:NGln10} and \ref{fig:NGln1}, as well as the parameter scan leading to Fig.~\ref{fig:scan_ng_k*_sigma}. As reported in Sec.~\ref{sec:NG-SGWB}, the trispectrum arising from the local expansion Eq.~\eqref{eq:zeta-local-anzatz} leads to additional contributions to the SIGW spectrum. In particular, from the connected part, one obtains the following two terms
\begin{equation}
\begin{aligned}
    \Omega_{\rm GW}(k,\eta)|_{\tc} =& \hspace{0.1cm}\frac{1}{12\pi} \left(\frac{k}{aH}\right)^2 \left(\frac{3}{5}f_{\rm NL}\right)^2 \int_{0}^{\infty} d t_1 \int_{-1}^{1} ds_1 \int_{0}^{\infty} d t_2 \int_{-1}^{1} ds_2\\
    &\hspace{1.8cm}\times\int_{0}^{2\pi}d\varphi_{12} \cos 2\varphi_{12}\frac{u_1v_1}{(u_2v_2)^2}\frac{1}{w_{a,12}^3}\overline{\Tilde{J}(u_1,v_1,x)\Tilde{J}(u_2,v_2,x)}\\
    &\hspace{1.8cm}\times\calP_{\zeta_g}(v_2k)\calP_{\zeta_g}(u_2k)\calP_{\zeta_g}(w_{a,12} k)\,,
    \label{eq:ng_t_comp}
\end{aligned}
\end{equation}
and

\begin{equation}
\begin{aligned}
    \Omega_{\rm GW}(k,\eta)|_{\uc} =& \hspace{0.1cm}\frac{1}{12\pi} \left(\frac{k}{aH}\right)^2 \left(\frac{3}{5}f_{\rm NL}\right)^2 \int_{0}^{\infty} d t_1 \int_{-1}^{1} ds_1 \int_{0}^{\infty} d t_2 \int_{-1}^{1} ds_2\\
    &\hspace{1.8cm}\times\int_{0}^{2\pi}d\varphi_{12} \cos 2\varphi_{12} \frac{u_1u_2}{(v_1v_2)^2}\frac{1}{w_{b,12}^3}\overline{\Tilde{J}(u_1,v_1,x)\Tilde{J}(u_2,v_2,x)}\\
    &\hspace{1.8cm}\times\calP_{\zeta_g}(v_1k)\calP_{\zeta_g}(v_2k)\calP_{\zeta_g}(w_{b,12} k)\,,
    \label{eq:ng_u_comp}
\end{aligned}
\end{equation}
were the integration variables $t_i$ and $s_i$ are defined as in Eq. \eqref{eq:t-s}. To keep the equation concise, some terms have been left expressed as functions of $u_i$ and $v_i$, but they have to be intended as depending on the integration variables $t_i$ and $s_i$. Moreover, we introduced $\Tilde{J}(u_i,v_i,x) = v_i^2k^2\sin^2\theta{I}(u_i,v_i,x)$, with ${I}(u,v,x)$ the integration kernel defined in the main text and $w_{a,12}$ and $w_{b,12}$, defined as
\begin{equation}
\begin{aligned}
    w_{a,12} =& [v_1^2+v_2^2-2v_1v_2(\cos\theta_1\cos\theta_2+\sin\theta_1\sin\theta_2\cos\varphi_{12})]^{1/2} \,,
\end{aligned}
\end{equation}
and 
\begin{equation}
    \begin{aligned}
    w_{b,12} =&\big[1+v_1^2+v_2^2+2v_1v_2(\cos\theta_1\cos\theta_2+\sin\theta_1\sin\theta_2\cos\varphi_{12})\\
    &\hspace{1cm}-2v_1\cos\theta_1-2v_2\cos\theta_2\big]^{1/2}\,.
\end{aligned}   
\end{equation}
The sine and cosine functions are related to the integration variables by
\begin{align}
    \cos\theta_i = \hspace{0.1cm}\frac{1-s_i(1+t_i)}{t_i-s_i+1}\,,\qquad
    \sin^2\theta_i =\hspace{0.1cm} \frac{(1-s_i^2)t_i(2+t_i)}{(t_i-s_i+1)^2}\,.
\end{align}
As shown in Eqs.~\eqref{eq:ng_t_comp} and \eqref{eq:ng_u_comp}, the evaluation of the NG corrections requires a 5 dimensional integration for each of the frequencies at which the final GW spectrum is evaluated. However, $f_{\rm NL}$ and $A_s$ are multiplicative parameters and the effect of $k_*$ just results in a shift of the spectrum along the $k$-axis. Hence, once the spectrum is evaluated for a fixed width $\Delta$, it can be used for different values of the parameters reported above, without requiring further evaluation. When $\Delta$ is varied, instead, a new evaluation of the spectrum is required each time. Hence, just a single evaluation of the spectrum would require relatively little time, but the evaluation of the whole spectrum for each point of the MCMC would notably slow down the run, making it difficult to get the final posterior in a reasonable time, also considering the presence of other parameters in the MCMC evaluation. 

For this reason, to speed up the evaluation we proceed as follows:
we numerically pre-compute a grid of NG contributions to the GW spectra as a function of frequency for different widths, in order to explore the range $\log_{10}\Delta \in [-1,1]$. This grid is then used to obtain the NG corrections to the spectrum corresponding to any value of $\Delta$ in the range considered, by interpolating them from the pre-computed ones. In detail, to get the spectrum corresponding to those values of $\bar{\Delta}$ not present in the grid, we first search for ${\Delta}_{\rm max}$ and ${\Delta}_{\rm min}$, respectively immediately above and below $\bar{\Delta}$. Then we compute the interpolated spectrum by Taylor expanding around these values, obtaining
\begin{align}
    \Omega_{\rm GW}(\bar{\Delta}) =  \Omega_{\rm GW}(\Delta_{\rm max})w_{\rm min} +  \Omega_{\rm GW}(\Delta_{\rm min})(1-w_{\rm min})\,,
\end{align}
with
\begin{equation}
    w_{\rm min} = \frac{(\bar{\Delta}-\Delta_{\rm min})}{\left(\Delta_{\rm max}-\Delta_{\rm min}\right)}\,.
\end{equation}
For the evaluation of the scans that require a Fisher forecast and hence the derivatives with respect to the parameters, we proceed in a similar way. We pre-compute a grid of derivatives\footnote{Note that when taking the derivative with respect to $A_s$ and $f_{\rm NL}$ the integrals remain unchanged, hence we consider the same pre-computed NG contributions used in the MCMC runs. When taking the derivative with respect to $k_*$ or $\Delta$, instead, since the integrand is varied, we pre-compute a grid for each of these two derivatives.} in the range $\log_{10}\Delta \in [-1,1]$ and then we interpolate as explained above.

\subsection{Inference}
\begin{figure}[h!]
    \centering
    \includegraphics[width=0.85\linewidth]{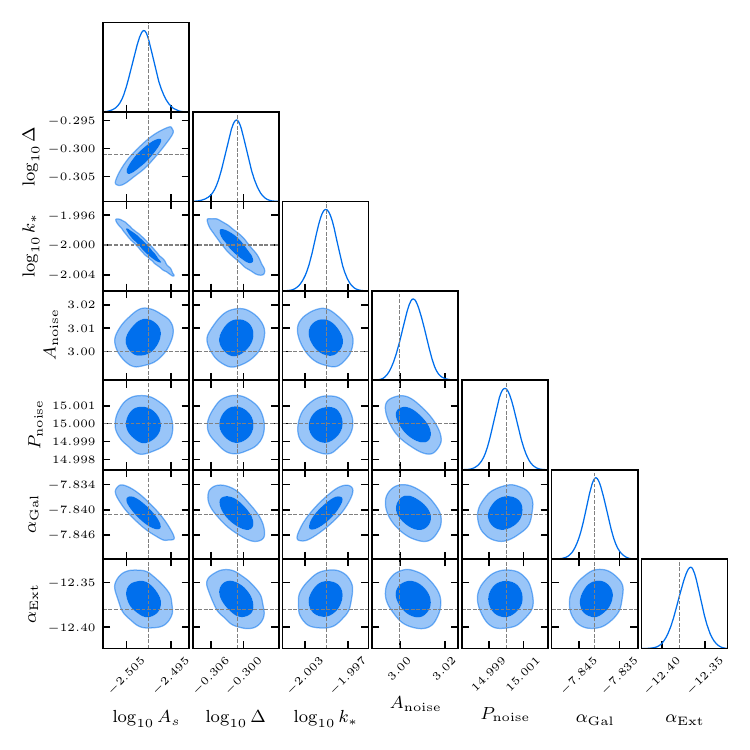}
    \caption{Same as Fig.~\ref{fig:ln_corner} but showing the marginalised contours for all sampled parameters including the noise parameters $A_{\rm noise}, P_{\rm noise}$. There is a relatively mild degeneracy between $A_{\rm noise}$ and $\alpha_{\rm Gal}$ but no correlations between the signal parameters $A_s, \Delta, k_*$ and the noise parameters. We found similar correlations (or the lack thereof) for all other injections.}
    \label{fig:ln_corner_full}
\end{figure}

Once $\Omega_{\rm GW}$ has been computed by the \SIGWAY, the resulting spectrum is interpolated in log-space and passed to the \texttt{SGWBinner} which computes the posterior distribution according to Eq.~\eqref{eq:full_post}. \texttt{Cobaya}~\cite{Torrado:2020dgo,2019ascl.soft10019T} is used as an inference-framework. We use different samplers for Monte Carlo sampling depending on the dimensionality, requirements, and structure of the posterior surface:
\begin{itemize}
    \item The inferences in Figs.~\ref{fig:binned_no_injection_15},
    \ref{fig:binned_injection_15},
    \ref{fig:usr_bpl_corner},
    \ref{fig:usr_bpl_posterior_predictives},
    \ref{fig:oscillations_multi_field_corner},
    \ref{fig:oscillations_multi_field_posterior_predictives},
    and \ref{fig:emd_corner} have been run using the nested sampler \texttt{nessai}~\cite{nessai,Williams:2021qyt,Williams:2023ppp}.
    \item Figs.~\ref{fig:ln_corner},
    \ref{fig:ln_posterior_predictives},
    \ref{fig:usr_posterior_predictives_V},
    \ref{fig:NGln1},
    and \ref{fig:NGln10} have been obtained using \texttt{Cobaya}'s \texttt{CosmoMC}~\cite{Lewis:2002ah,Lewis:2013hha} MCMC, where we started the chains at the injected values and injected FIM estimates of the covariance matrix to speed up convergence.
    \item The evidences in Sec.~\ref{sec:tests} have been computed with the nested sampler \texttt{PolyChord}~\cite{polychord1,polychord2}
    \item The inference for Figs.~\ref{fig:oscillations_sharpfeature_corner} and
    \ref{fig:oscillations_sharpfeature_posterior_predictives} was performed using the active learning algorithm \texttt{GPry}~\cite{gpry1,gpry2} due to the prohibitively slow speed of computing $\mathcal{P}_\zeta$ stemming from the Bessel functions in its equation.
\end{itemize}
All corner plots have been created with \texttt{GetDist}~\cite{Lewis:2019xzd}. In the corner plots, we omitted showing the marginalised constraints on $A_{\rm noise}$ and $P_{\rm noise}$, as in all cases they were well constrained and showed weak degeneracies with the signal parameters. To give an idea of how tightly these parameters tend to be constrained, Fig.~\ref{fig:ln_corner_full} shows a corner plot including the constraints on $A_{\rm noise}$ and $P_{\rm noise}$ for the injected lognormal $\mathcal{P}_\zeta$ (see Eq.~\eqref{eq:PLN}).

\section{Challenges with binned analyses and a large number of bins}\label{App:largeNbinning}
The binned approach to performing the double integration going from $\mathcal{P}_\zeta$ to $\Omega_{\rm GW}$ introduced in Sec.~\ref{sec:OmegaGWijk} -- while in principle extremely powerful at reconstructing any SIGW spectrum without model-dependence -- unfortunately suffers from some crucial shortcomings as will be explained in this section.

For the sake of illustration, we will only consider the case where all modes reenter during radiation domination (see Sec.~\ref{sec_RDE}) where the kernel is $k$-independent. The situation changes a bit if the kernel has a $k$-dependence such as is the case during an early matter domination era (see Sec.~\ref{sec_tran}), however, our main arguments remain unchanged.

\begin{figure}
    \centering
    \includegraphics[width=0.6\linewidth]{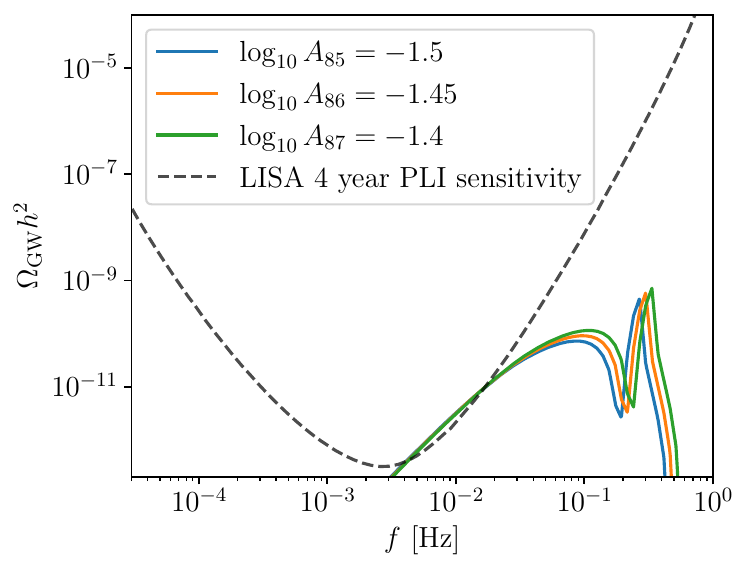}
    \caption{Three different spectra in $\Omega_{\rm GW}$ generated with the binned approach with 100 bins where for each spectrum only one of the bins is non-zero. The black dashed line shows the approximate power law integrated sensitivity of LISA assuming a 4-year mission. It is clear from this picture that the three adjacent bins shown are entirely degenerate when trying to resolve them with LISA as the peaks are well outside the sensitivity and the causality tails generated are exactly the same.}
    \label{fig:degenerate_bins_illustration}
\end{figure}

It is clear from the structure of the integral in Eq.~\eqref{eq:P_h_ts}, that a single wavenumber $k$ in $\mathcal{P}_\zeta$ affects multiple frequencies in $\Omega_{\rm GW}$. An easy way to understand this is to consider a $\mathcal{P}_\zeta$ that is sufficiently close to a monochromatic source $\mathcal{P}_\zeta (k) = A_s \delta(k-k_*)$. In our binned approach this translates to one single bin $A_*$ being non-zero. Fig.~\ref{fig:degenerate_bins_illustration} shows three such spectra with 100 bins, where each one contains a single non-zero bin $A_\star$ (bin nr. 85, 86, 87 in this case). It is evident from this figure that if the peak towards $k_*$ is not resolved, and only the causality tail in the IR regime enters the LISA sensitivity, these spectra become entirely degenerate. This means that there is not necessarily a unique mapping $\Omega_{\rm GW}\mapsto\mathcal{P}_\zeta$. In other words, the power from the bins is ``leaking" into adjacent bins.

\begin{figure}[h]
    \centering
    \includegraphics[width=0.49\textwidth]{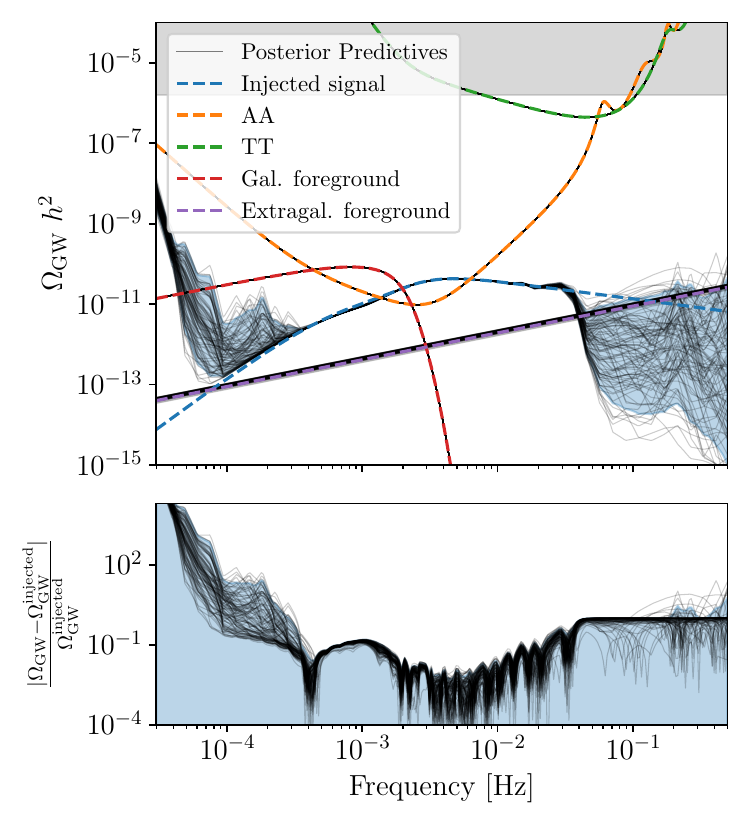}
    \includegraphics[width=0.49\textwidth]{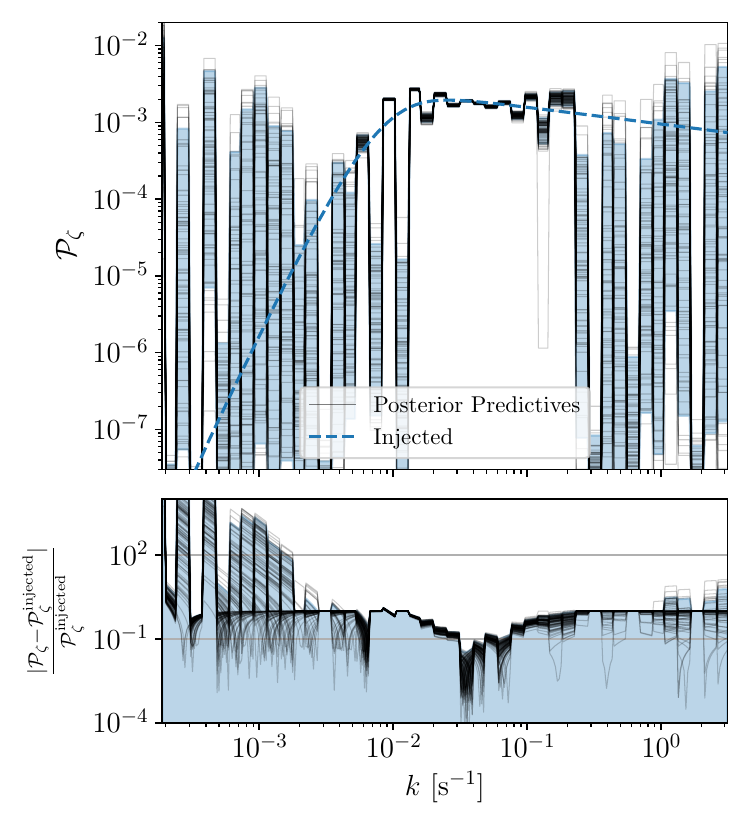}
    \caption{
   Same as Fig.~\ref{fig:binned_no_injection_15}, but for an injected BPL spectrum following the benchmark USR model and using $50$ bins.}
    \label{fig:binned_injection_50x}
\end{figure}

In reality, this means that, for many bins and towards low SNR, the binned reconstructed $\mathcal{P}_{\zeta}$ is highly degenerate, and valid reconstructions include ``oscillations" between the bins as lower power in one bin can be compensated by higher power in an adjacent one. 

The aforementioned degeneracies are not unexpected and are really just a feature of the physical properties of the process of scalar-induced gravitational waves.
However, they do induce some practical complications.
In an ideal template-agnostic pipeline, we would want to perform inference on the $N_{\rm bins}$ bins in $\mathcal{P}_\zeta$ to reconstruct the signal with the highest evidence parameterization.
Due to the large (non-linear) degeneracies between the bins, the likelihood is far from Gaussian and the FIM approximation is invalid, making MC-sampling necessary. Sampling over this space is very computationally challenging due to \textit{(a)} the very narrow degeneracies \textit{(b)} the resulting large number of posterior modes and \textit{(c)} the high dimensionality of the parameter space. In practice, this leads to overconfidence in the reconstruction, as some posterior modes are inevitably missed or underexplored by the MC sampler. Fig.~\ref{fig:binned_injection_50x} shows the binned reconstruction of the USR model injection from Sec.~\ref{sec:benchmodelUSR} with 50 bins. The oscillation effect is clearly visible in the low SNR region, where $\mathcal{P}_\zeta$ oscillates between high power and low power, thus overconstraining certain bins. These degeneracies are partially broken by fewer bins, as visible in Fig.~\ref{fig:binned_no_injection_15}.

This leaves us in a dilemma: we would like to bin $\mathcal{P}_\zeta$ as finely as possible to increase the frequency resolution of the template, but as one increases $N_{\rm bins}$ the posterior becomes much more difficult to sample. In our tests, we found $N_{\rm bins}=15$ to be reliable in terms of convergence for the BPL signal (Fig.~\ref{fig:binned_no_injection_15}) and $N_{\rm bins}=40$ for the injected lognormal signals in $\Omega_{\rm GW}$ and $\mathcal{P}_\zeta$ (Fig.~\ref{fig:evidence_reconstruction_plots}) that occupy less of the frequency range. However, it is clear that this low number of bins cannot reconstruct the shape of $\mathcal{P}_\zeta$ with high fidelity.

Luckily, this problem does not appear when no signal is present, as the posterior distribution in $A_i$ becomes a simple upper bound, an unimodal structure that is easy to map by a nested sampler, even in high dimensions. We can therefore conclude that the upper bounds obtained by this method are reliable even with many bins.

Future work on this approach could include studying improved bases for the reconstructed bins (a basis of Gaussians or other wide kernels in $P_\zeta$ may be less multi-modal), or improving sampling by manually adding jump proposals to the degenerate modes of the posterior in a given basis (e.g.~as is done for LISA black hole binary sampling in \texttt{BBHx} \cite{Katz:2024oqg}).

\begin{figure}[t]
    \centering
    \includegraphics[width=0.6\linewidth]{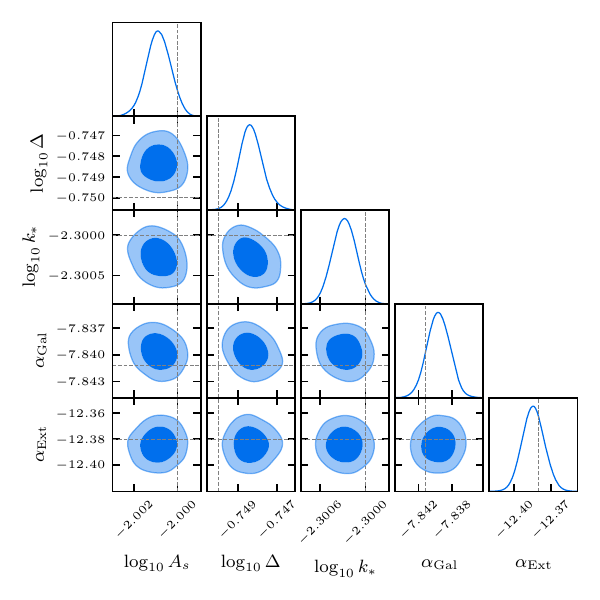}
    \caption{Same as \cref{fig:ln_corner} for the same injected parameters as in Sec.~\ref{sec:NG}.}\label{fig:ln_NG_gaussian_corner}
\end{figure}

\section{Testing the resolvability of Non-Gaussian corrections: Additional plots}

Fig.~\ref{fig:ln_NG_gaussian_corner} shows a corner plot that was obtained by injecting a purely Gaussian SIGW signal with a lognormal shape in $\mathcal{P}_\zeta$ (see \cref{eq:PLN}) that is equivalent to the cases discussed in \cref{fig:NGln1,fig:NGln10} with $f_{\rm NL}=\tau_{\rm NL}=0$. The remaining injected parameters are the same as in \cref{sec:NG}: $\log_{10} A_{s}=-2$,  $\log_{10}\Delta= -0.75$, $\log_{10}(k_*/\mathrm{s}^{-1}) = -2.3$. By comparing these results to the one obtained for $f_{\rm NL}=1$ shown in \cref{fig:NGln1} and $f_{\rm NL}=10$ shown in \cref{fig:NGln10}, we see that the NG contribution neither significantly improves, nor worsens the constraints on the signal and foreground parameters.

\section{Testing the scalar-induced hypothesis: Additional plots}\label{app:additionalfigs}

\begin{figure}[h!]
    \centering
    \includegraphics[width=0.49\linewidth]{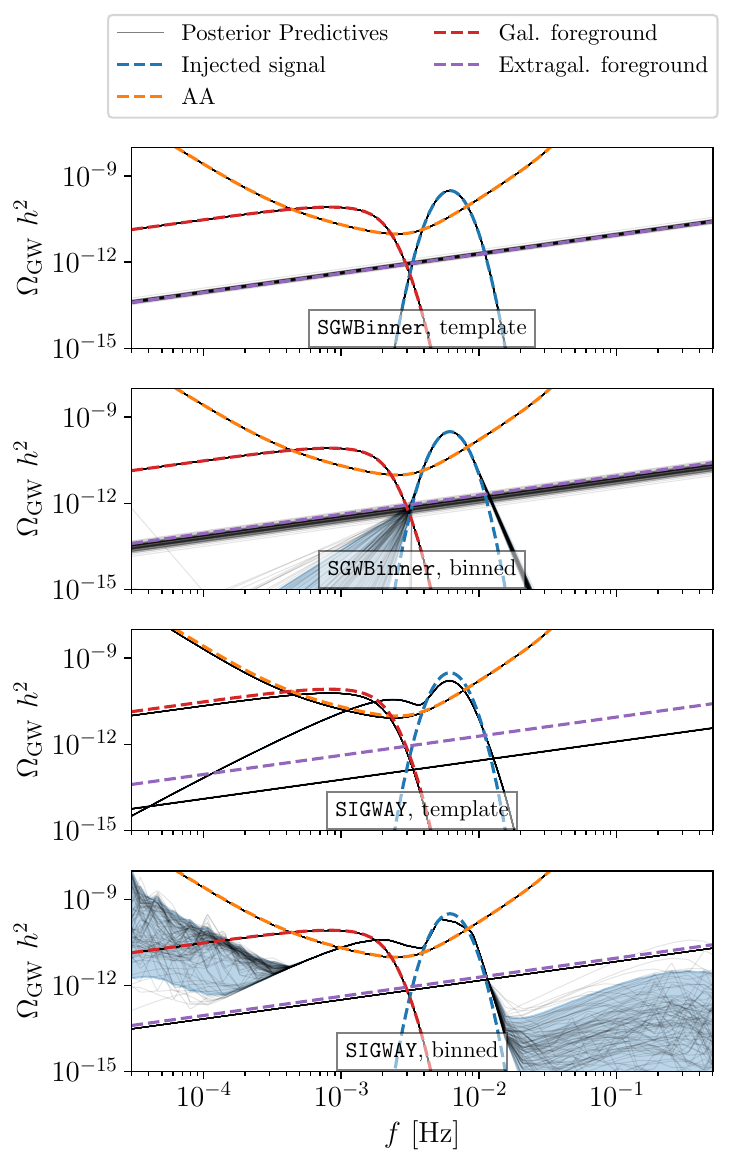}
    \includegraphics[width=0.49\linewidth]{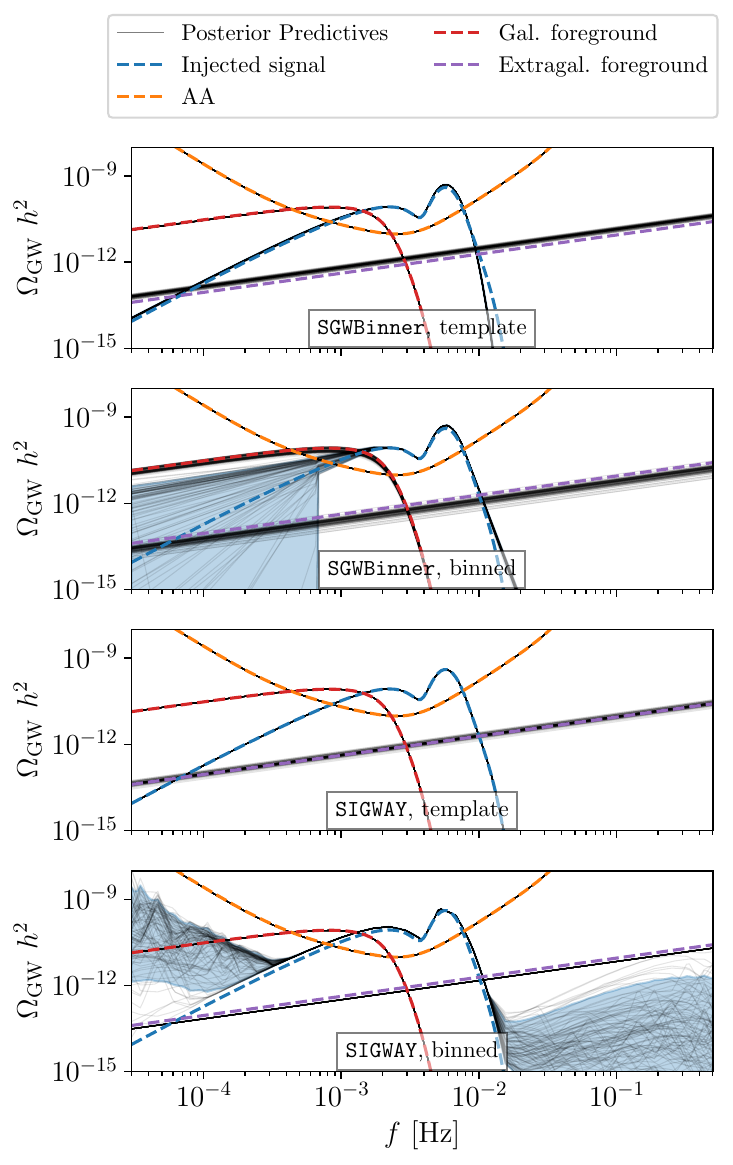}
    \caption{Reconstruction of $\Omega_{\rm GW}$ for case 1 (left) and case 2 (right) including the noise and foreground. See Sec.~\ref{sec:tests} for more details. The TT-channel component of the noise is above $10^{-7}$ and we therefore omit it. }
    \label{fig:evidence_reconstruction_plots_full}
\end{figure}

Figure \ref{fig:evidence_reconstruction_plots_full} provides further insight into the quality of reconstruction when comparing the \texttt{SGWBinner} and the \SIGWAY\  methods. The noise is accurately reconstructed in all cases, with only a slight underestimation in case 1 (left panel) using the \SIGWAY\  template reconstruction. In contrast, the extragalactic background is reconstructed less accurately with the \texttt{SGWBinner} compared to the other methods.

A particularly notable observation is that in case 1 (left column), the \SIGWAY\  template method significantly underestimates both the extragalactic and galactic foregrounds. This underestimation can be attributed to the model compensating for excess power in the causality tail by reducing the power allocated to the foregrounds, due to the limited flexibility in the shape of $\Omega_{\rm GW}$ provided by the template. Interestingly, in case 2 (right column), the \SIGWAY\  method -- using the template or not -- performs better than the \texttt{SGWBinner} in reconstructing the foregrounds. This improvement arises because the assumption of SIGWs enforces a specific shape for $\Omega_{\rm GW}$, which cannot be easily mimicked by the foregrounds. In contrast, the \texttt{SGWBinner} does not impose such restrictions during reconstruction.
This result stresses the importance of adopting optimal modeling of the eventual cosmological signal even when reconstructing the astrophysical properties of the foreground sources. 

\bibliographystyle{JHEP}
\bibliography{biblio}

\end{document}